\documentclass[twocolumn,aps,prd,preprintnumbers,superscriptaddress,nofootinbib,10pt]{revtex4-2}

\usepackage{amsmath,amsfonts,amssymb}
\usepackage{epsfig,epstopdf}
\usepackage{graphicx}
\usepackage{subfigure}
\usepackage[dvipsnames]{xcolor}
\usepackage[colorlinks,linkcolor=blue,anchorcolor=black,citecolor=blue]{hyperref}
\usepackage[all]{hypcap}

\newcommand\dd{\ensuremath{\mathrm{d}}}

\newcommand\qperp{\ensuremath{q_\perp}}
\newcommand\uperp{\ensuremath{u_\perp}}

\newcommand\vperp{\ensuremath{v_\perp}}

\newcommand\bperp{\ensuremath{b_\perp}}

\newcommand\rperp{\ensuremath{r_\perp}}

\newcommand\Sperp{\ensuremath{S_\perp}}
\newcommand\alphas{\ensuremath{\alpha_s}}

\newcommand\qa{\ensuremath{ q_{1\perp}}}
\newcommand\qb{\ensuremath{ q_{2\perp}}}
\newcommand\qc{\ensuremath{ q_{3\perp}}}
\allowdisplaybreaks[4]

\begin{document}

\title{Forward Inclusive Jet Productions in \texorpdfstring{$pA$}{pA} Collisions}

\author{Lei Wang} \email{leiwang@mails.ccnu.edu.cn}
\affiliation{Key Laboratory of Quark and Lepton Physics (MOE) and Institute of Particle Physics, Central China Normal University, Wuhan 430079, China}

\author{Lin Chen}  \email{raymondchen@cuhk.edu.cn}
\affiliation{School of Science and Engineering, The Chinese University of Hong Kong, Shenzhen 518172, China} 
\affiliation{University of Science and Technology of China, Hefei, Anhui, 230026, P.R.China}

\author{Zhan Gao} \email{gaozhan2@foxmail.com}
\affiliation{Key Laboratory of Quark and Lepton Physics (MOE) and Institute of Particle Physics, Central China Normal University, Wuhan 430079, China}

\author{Yu Shi} \email{yu.shi@sdu.edu.cn} 
\affiliation{Key Laboratory of Particle Physics and Particle Irradiation (MOE), Institute of frontier and interdisciplinary science, Shandong University, Qingdao, Shandong 266237, China}

\author{Shu-Yi Wei}  \email{shuyi@sdu.edu.cn}
\affiliation{Key Laboratory of Particle Physics and Particle Irradiation (MOE), Institute of frontier and interdisciplinary science, Shandong University, Qingdao, Shandong 266237, China}

\author{Bo-Wen Xiao}  \email{xiaobowen@cuhk.edu.cn}
\affiliation{School of Science and Engineering, The Chinese University of Hong Kong, Shenzhen 518172, China}

\begin{abstract}
Motivated by recent experimental LHC measurements on the forward inclusive jet productions and based on our previous calculations on forward hadron productions, we calculate single inclusive jet cross-section in $pA$ collisions at forward rapidity within the color glass condensate framework up to the next-to-leading-order. Moreover, with the application of jet algorithm and proper subtraction  of the rapidity and collinear divergences, we further demonstrate that the resulting  next-to-leading-order hard coefficients are finite. In addition, in order to deal with the large logarithms that can potentially spoil the convergence of the perturbative expansion and improve the reliability of the numerical predictions, we introduce the collinear jet function and the threshold jet function and resum these large logarithms hidden in the hard coefficients. 
\end{abstract}

\maketitle

\section{Introduction}

Due to the Bremsstrahlung radiation, the gluon field strength and density inside a hadron rise rapidly with the hadron energy. Generally, large-$x$ quarks and gluons inside fast moving hadrons can be viewed as color sources from which small-$x$ gluons~\cite{Kuraev:1977fs,Balitsky:1978ic} are emitted, where $x$ is the longitudinal momentum fraction of the gluon w.r.t. the parent hadron. The increase in gluon density can be described by the well-known Balitsky-Fadin-Kuraev-Lipatov (BFKL) evolution equation~\cite{Balitsky:1978ic}, which resums large logarithms in the form of $\alpha_s\ln\frac{1}{x}$.  When more and more gluons are packed in a confined hadron, these gluons start to overlap and recombine~\cite{Gribov:1984tu, Mueller:1985wy}. This can lead to the nonlinear QCD evolution well described by the Balitsky-Kovchegov and Jalilian-Marian-Iancu-McLerran-Weigert-Leonidov-Kovner (BK and JIMWLK) equation~\cite{Balitsky:1995ub, JalilianMarian:1996mkd, JalilianMarian:1997jx, JalilianMarian:1997gr,Kovchegov:1999yj, Iancu:2000hn, Ferreiro:2001qy, Kovner:2000pt, Iancu:2001ad}. Eventually, the radiation and reabsorption of gluons tend to balance, which leads to so-called gluon saturation~\cite{Gribov:1984tu,Mueller:1985wy,McLerran:1993ni,McLerran:1993ka,arXiv:1002.0333, CU-TP-441a}. As common practice, one usually introduces the saturation momentum $Q_s(x)$ to characterize the typical size of the soft gluons and separate the non-linear dynamics from the linear BFKL evolution. 

One of the major goals of high energy QCD studies is to search for the compelling evidence for the gluon saturation phenomenon. In the past few years, tremendous contributions~\cite{Marquet:2009ca, Benic:2016uku, Iancu:2020jch, Hatta:2022lzj, Caucal:2021ent, Caucal:2022ulg, Taels:2022tza, Bergabo:2022tcu, Bergabo:2022zhe, Tong:2022zwp} have been made to the search for such an intriguing phenomenon. Relativistic heavy ion collider (RHIC) ~\cite{Arsene:2004ux, Adams:2006uz, Braidot:2010ig, Adare:2011sc} and the large hadron collider (LHC)~\cite{ALICE:2012xs, ALICE:2012mj, Hadjidakis:2011zz} have provided us with a large amount of experimental data~\cite{Arsene:2004ux, Adams:2006uz, Braidot:2010ig, Adare:2011sc, Hadjidakis:2011zz, ALICE:2012xs, ALICE:2012mj, ATLAS:2016xpn, LHCb:2021abm, LHCb:2021vww}. Quantitative and precise phenomenological tests of saturation physics in heavy-ion collisions have been a hot topic for the past decades. Early attempts include the measurement of structure-function at HERA and the measurement of the production of forward single inclusive jet (or hadron) in $pA$ collisions at RHIC and LHC. Also, studying the onset of gluon saturation is also one of three physics pillars of the upcoming electron-ion collider (EIC)~\cite{Boer:2011fh, Accardi:2012qut, Proceedings:2020eah, AbdulKhalek:2021gbh}. 

Forward inclusive hadron and jet productions in $pA$ collisions have attracted many theoretical interests in recent years~\cite{Kovchegov:1998bi,Dumitru:2002qt, Dumitru:2005kb, Dumitru:2005gt, Albacete:2010bs, Levin:2010dw, Dominguez:2010xd, Altinoluk:2011qy, Fujii:2011fh, Chirilli:2011km, Albacete:2012xq, Albacete:2013ei, Stasto:2013cha, Lappi:2013zma, vanHameren:2014lna, Stasto:2014sea, Altinoluk:2014eka, Watanabe:2015tja, Stasto:2016wrf, Iancu:2016vyg, Ducloue:2016shw, Ducloue:2017dit, Kang:2019ysm, Liu:2020mpy, Shi:2021hwx, Liu:2022ijp}. For example, since the projectile proton (or deuteron) can be treated as a dilute probe in comparison with the ultra-dense gluon fields in the nuclear target~\cite{Chirilli:2011km, Chirilli:2012jd, Dominguez:2010xd, Dominguez:2011wm}, the forward hadron (or jet) productions have been widely used to study the gluon saturation. Moreover, the experimental studies of the evolution of the nuclear modification factor $R_{dAu}$~\cite{Arsene:2004ux, Adams:2006uz} have provided strong hints for gluon saturation~\cite{Kharzeev:2003wz, Kharzeev:2004yx, Albacete:2003iq, Iancu:2004bx, Albacete:2013ei}. The forward inclusive mini-jet cross-section in $pA$ collisions within the color glass condensate (CGC) framework~\cite{Mueller:1993rr, Mueller:1999wm, McLerran:1993ni, McLerran:1993ka, McLerran:1994vd, Jalilian-Marian:1997ubg, Weigert:2000gi, Iancu:2003xm, Gelis:2010nm, Kovchegov:2012mbw, Albacete:2014fwa, Blaizot:2016qgz, Morreale:2021pnn} was first studied in Ref.~\cite{Dumitru:2002qt}. Subsequently, thanks to the abundant data made available by RHIC and the LHC, the research attention on the theoretical side was mostly focused on hadron productions. In addition, a lot of progress has been made on the calculation of the one-loop diagrams and next-to-leading order (NLO) corrections for hadron productions. In particular, the full NLO contributions of single hadron productions include the one-loop contributions computed in Ref.~\cite{Chirilli:2011km, Chirilli:2012jd} and the additional kinematic corrections~\cite{Watanabe:2015tja}. 

The study of forward jet productions~\cite{Abazov:2004hm,Abelev:2007ii,Marquet:2007vb,Khachatryan:2011zj,daCosta:2011ni,Aad:2010bu,Chatrchyan:2011sx,Adamczyk:2013jei,Aaboud:2019oop,Liu:2022xsc} provides us with another channel besides the hadron probe. Usually, one views the leading hadron in a jet as the surrogate for the full jet. The theoretical calculation for the inclusive jet production has many aspects in common with hadron productions with a few notable differences. Experimentally, the inclusive very forward jet production in proton-lead collisions has been measured by the CMS experiment at the LHC~\cite{CMS:2018yhi}. The comparison between the CMS data and the LO CGC calculation is later carried out in Ref.~\cite{Mantysaari:2019nnt}.

The main objective of this paper is to compute the NLO corrections to the forward jet production in the CGC framework based on the previous progress in the hadron case. It is worth mentioning that the results presented in this manuscript is akin to those in Ref.~\cite{Liu:2022ijp}, while the detailed computation and the resummation approach employed in this paper are different.

In forward $pA$ collisions, the active partons with longitudinal momentum fraction $x=\frac{\qperp}{\sqrt{s}}e^y$ in the proton projectile can be treated as dilute probes. Here $\qperp$ and $\sqrt{s}$ are the measured final state parton transverse momentum and the total energy in the center-of-mass frame for $pp$ collisions, respectively.\footnote{Note that the notations for the kinematic variables $x$ and $q_\perp$ in this work are $x_p$ and $k_\perp$, respectively, in the forward hadron paper~\cite{Chirilli:2012jd}.} Meanwhile, the active partons in the target nucleus with longitudinal momentum fraction $x_A=\frac{\qperp}{\sqrt{s}}e^{-y}$ formed a dense gluon background. When these partons traverse the ultra-dense gluonic medium of the target, they can accumulate a typical transverse momentum of the order of the saturation momentum $Q_s(x)$ through multiple interactions with the nuclear target. For positive and sufficiently large rapidity $y$, the active parton from the proton projectile is from the large $x$ region while the active parton from the nucleus target is deeply in the low $x$ region. In $pA$ collisions, since the target nucleus is large enough we can integrate over the impact parameter to get the transverse area of the target. Therefore, we can neglect the impact parameter dependence and greatly simplify the calculations in $pA$ collisions. Compared with other physical processes such as $pp$ collisions, the production of the forward single jet in $pA$ collisions is an ideal process for observing the saturation phenomena. Compared with hadron productions, the advantage of measuring jet productions is that jets provide more direct transverse momentum $q_\perp$ information without involving fragmentation functions (FFs). However, the saturation effects are expected to be small for high $p_T$ jets. It may be challenging to measurement jets with relatively low transverse momenta around a few times of $Q_s(x)$. 

The physical picture of the forward single inclusive jet production in $pA$ collisions can be understood as follows,
\begin{align}
p+A\to \text{jet}+X.
\end{align}
where a parton from the right-moving proton (with momentum $q$) scatters off the nucleus target (with momentum $P_A$), and becomes a final sate jet with momentum $P_J$ and rapidity $\eta$. The kinematics at NLO are the similar as in Refs.~\cite{Chirilli:2011km,Chirilli:2012jd}. One needs to resum multiple interactions as the gluon density of the target becomes high. In this paper, we follow the factorization formalism (color-dipole or CGC) as in the Refs.~\cite{Chirilli:2011km,Chirilli:2012jd} to evaluate this process up to one-loop order.

The leading order (LO) calculation is straightforward, and it has been studied extensively in Refs.~\cite{Blaizot:2004wu, Blaizot:2004wv, Albacete:2010bs, Levin:2010dw, Fujii:2011fh, Albacete:2012xq, Lappi:2013zma, vanHameren:2014lna, Bury:2017xwd, Mantysaari:2019nnt}. We first outline the LO results in the following section. Then, to evaluate NLO corrections, we calculate the gluon radiation contributions, including both real and virtual diagrams at the one-loop order. When one integrates over the phase space of the additional gluon, one finds various divergences in both real and virtual contributions~\cite{Chirilli:2011km, Chirilli:2012jd}. For example, there are collinear divergences associated with the incoming parton distribution. With a proper jet definition, final state collinear singularities cancel between real and virtual diagrams. When the final state partons form a jet, there are no collinear singularities anymore after summing real diagrams and virtual diagrams. To tackle the calculation more efficiently, we use the narrow jet approximation (NJA)~\cite{Jager:2004jh, Mukherjee:2012uz, Sun:2014gfa,Sun:2014lna} (also known as the small cone approximation) to simplify the calculation. NJA allows one to simplify calculations and neglect small contributions of order $R^2$ with $R$ defined as the jet cone size. In addition, there are also rapidity divergences associated with the small-$x$ multiple-point correlation function~\cite{Blaizot:2004wv,Dominguez:2008aa, Dominguez:2011wm, Dominguez:2012ad,Shi:2017gcq, Zhang:2019yhk}. These rapidity divergences allow one to reproduce the BK equation~\cite{Balitsky:1995ub,Kovchegov:1999yj}. After solving BK evolution equations, one resum $\ln \frac{1}{x_A}$ type logarithms automatically. 

Furthermore, there are additional large logarithms from the one-loop corrections which require further theoretical treatments. In forward jet productions, one enters an extremely asymmetric kinematic region. In this region, one finds $x\to 1$ and $x_A\to 0$, which can maximize the saturation effect to the greatest extent. Meanwhile, the longitudinal momentum fraction of active partons in the proton goes to $1$, indicating that this process has reached the kinematic boundary of the phase space. Therefore, the logarithm, such as $\alpha_s\ln(1-x)$, can become large and cause an issue for the perturbative expansion. As shown in Ref.~\cite{Stasto:2013cha}, the NLO corrections for hadron productions start to become large and negative. This indicates that additional theoretical technique is required to ensure the reliability of the NLO calculation. Early attempts have been devoted to solving this issue~\cite{Altinoluk:2014eka, Kang:2014lha, Stasto:2014sea, Watanabe:2015tja, Stasto:2016wrf, Ducloue:2016shw, Iancu:2016vyg, Ducloue:2017mpb, Ducloue:2017dit, Xiao:2018zxf, Liu:2019iml, Kang:2019ysm, Liu:2020mpy}. We believe that the origin of the negativity issue stems from the threshold logarithms due to soft gluon radiations near the threshold region. The resummation of such logarithms is known as the threshold resummation, which is also called Sudakov resummation in some literature. To deal with the remaining final state collinear logarithms, we introduce the collinear jet functions (CJFs) $\mathcal{J}_{i}(z)$ which is similar to the usual FFs $\mathcal{D}_{h/i}(z)$ with $z$ the longitudinal momentum fraction of the parton carried by the final state measured hadron or jet. The CJFs $\mathcal{J}_{i}(z)$ satisfy the well-known Dokshitzer–Gribov–Lipatov–Altarelli–Parisi (DGLAP) evolution equations equivalently. 

In addition, we introduce the jet threshold resummation for the threshold logarithms, which arise from integrating over the soft and collinear regions of soft gluon emissions near the kinematic boundary. By identifying the soft (and collinear) part of the phase space, one can develop the corresponding counting rule for the threshold logarithms and resum them in terms of Sudakov factors. The threshold resummation can help restore the predictive power of the CGC NLO calculation and extend its applicable window to larger transverse momentum regions. Two different formulations of the threshold resummation within the CGC framework have been proposed in Refs.~\cite{Xiao:2018zxf, Shi:2021hwx} and Refs.~\cite{Kang:2019ysm, Liu:2020mpy}, respectively. After choosing the appropriate initial condition and semi-hard scales, it was shown in Ref.~\cite{Shi:2021hwx} that the resummed NLO results can describe the experimental data from both RHIC and the LHC well. In this study, we follow the similar framework developed in Refs.~\cite{Xiao:2018zxf, Shi:2021hwx} by introducing the CJFs for the collinear logarithms and the jet threshold resummation for Sudakov type single and double logarithms, while we put the rest of the NLO contributions into the NLO hard factor. By choosing proper scales, we ensure that large logarithms are taken care of by various evolution 
(or renormalization) equations and the NLO hard factors only bring small corrections numerically. It appears to us that the threshold and collinear resummations are universal and indispensable to many high energy processes. 

At last, we have been assuming the eikonal approximation for the interaction between the quark or gluon from the projectile proton and the target nucleus. Lately, there have also been efforts made beyond the eikonal approximation for $pA$ collisions~\cite{Altinoluk:2014oxa, Altinoluk:2015gia, Chirilli:2018kkw, Altinoluk:2020oyd}. Recently, there have been many other NLO CGC calculations~\cite{Mueller:2012bn, Hentschinski:2014esa, Benic:2016uku, Boussarie:2016bkq, Roy:2018jxq, Roy:2019cux, Roy:2019hwr, Roy:2019hwr, Iancu:2020mos, Caucal:2021ent, Caucal:2022ulg, Taels:2022tza, Bergabo:2022tcu, Bergabo:2022zhe, Mantysaari:2022kdm, vanHameren:2022mtk, Beuf:2022kyp, Hanninen:2022gje, Iancu:2022gpw} for various processes. This resummation technique may be applied to other small-$x$ calculations as well.   

The rest parts of this paper are organized as follows. To be self-contained, we briefly present the leading order results for inclusive jet production in $pA$ collision in Sec.~\ref{section2}. Sec.~\ref{section3} is devoted to the NLO calculations which are divided into four parts. In subsection~\ref{section31}, we first evaluate the $q\to q$ and set up the framework of the calculation for the NLO forward jet cross-section, present the cross-section in the coordinate space, then transform the results into the momentum space to extract the large threshold logarithms. Following the same strategy, we compute the $g\to g$, $q\to g$, and $g\to q$ channels in subsections ~\ref{section32},~\ref{section33}, and~\ref{section34}, respectively. In Sec.~\ref{section4}, various kinds of large logarithms extracted from Sec.~\ref{section3} are identified and resummed. Firstly, we discuss the special plus function contributions which stems from the final state gluon radiations. We show that the resummation of collinear logarithms can be achieved with two slightly different methods, i.e., the DGLAP evolution and renormalization-group equation in subsections~\ref{collinear1} and~\ref{collinear2}, respectively. In the subsection~\ref{collinear1}, we resum $\ln R^2$ and collinear logarithms with the help of the DGLAP evolution by setting scale $\mu_J$ to the scale $\Lambda$. In Sec.~\ref{soft}, we take care of the threshold logarithms and derive the final resummation results. The summary and further discussions are given in Sec.~\ref{section5}.

\section{The leading order single inclusive jet cross-section}
\label{section2}
\begin{figure}[ht]
\includegraphics[width=6.6cm]{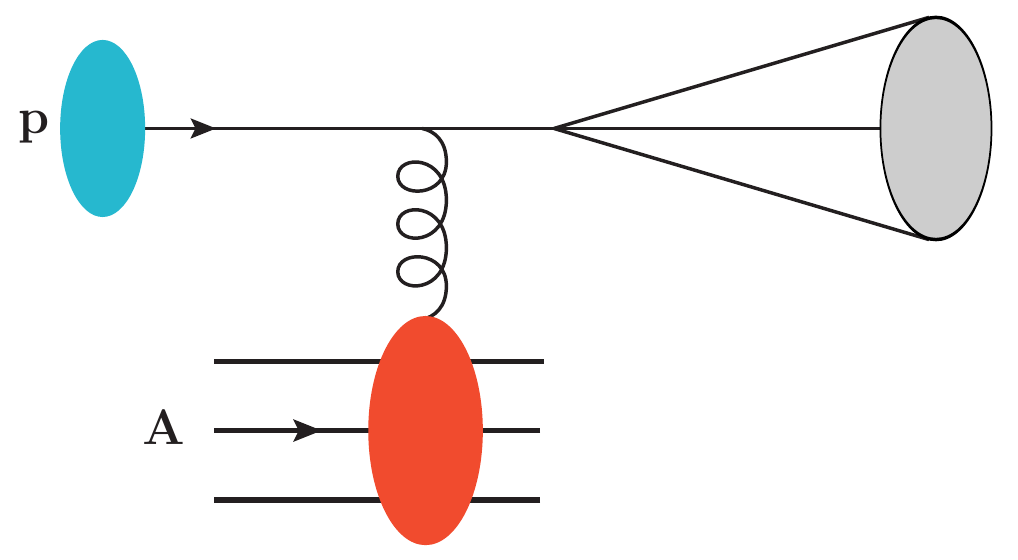} 
\caption[*]{A schematic diagram for the quark jet production at LO.}
\label{qqLO}
\end{figure}
As illustrated in Fig.~\ref{qqLO}, the forward single inclusive jet production in $pA$ collisions at leading order can be viewed as a probe to the saturation. In this process, a collinear parton (either a quark or a gluon) with momentum fraction $x$ from the proton projectile scatters off the dense nuclear target $A$ and subsequently fragments into a final state jet which is measured at forward rapidity $\eta$ with transverse momentum $P_J=z q_\perp$. The LO results for the forward quark jet cross-section in $pA$ collisions was first derived in Ref.~\cite{Dumitru:2002qt}. The LO calculation for jet productions is the same as the hadron production case at the parton level. We first take the quark channel in $pA$ collisions as an example, then the gluon channel can be done similarly. The leading-order cross-section for producing a quark with transverse momentum $\qperp$ at rapidity $\eta$ can be expressed as follows 

\begin{widetext}

\begin{equation}
\frac{\dd\sigma ^{\text{LO}}_{p+A\rightarrow q+X}}{\dd\eta \dd^{2}\qperp}%
=\sum_{f}xq_{f}(x)\int \frac{\dd^{2}x_{\perp }\dd^{2}y_{\perp }}{(2\pi )^{2}}%
e^{-i\qperp\cdot (x_{\perp }-y_{\perp })}\frac{1}{N_{c}}\left\langle
\text{Tr}U(x_{\perp })U^{\dagger }(y_{\perp })\right\rangle _{Y},
\end{equation}
where $q_f(x)$ is the quark distribution function with the longitudinal momentum fraction $x$ and the flavor $f$. $U(x_\perp)$ is the Wilson line in the fundamental representation which contains the multiple interaction between the quark and the dense gluon field of the target nucleus. The notation $\langle \dots \rangle _{Y}$ represents the CGC average of the color charges over the nuclear wave function with $Y\simeq\ln 1/x_g$. As to the gluon initiated channel, one finds
\begin{equation}
\frac{\dd\sigma ^{\text{LO}}_{p+A\rightarrow g+X}}{\dd\eta \dd^{2}\qperp}%
=xg(x)\int \frac{\dd^{2}x_{\perp }\dd^{2}y_{\perp }}{(2\pi )^{2}}
e^{-i\qperp\cdot (x_{\perp }-y_{\perp })}\frac{1}{N_{c}^2-1}\left\langle \text{Tr}%
W(x_{\perp})W^{\dagger }(y_{\perp })\right\rangle _{Y},
\end{equation}%
where $W(x_\perp)$ is the Wilson line in the adjoint representation. By using the usual convention, one can rewrite the cross-section in a compact form with the Fourier transform of the dipole scattering amplitude $\mathcal{F}({\qperp})$ and $\tilde{\mathcal{F}}(\qperp)$ in the fundamental and adjoint representations, respectively. Therefore, the full LO cross-section for jet productions reads 
\begin{eqnarray}
\frac{\dd\sigma^{\text{LO}}_{p+A\to \text{jet}+X}}{\dd\eta \dd^2P_J }=\int_\tau^1\frac{\dd z}{z^2}\left[\sum_fxq_f(x)\mathcal{F}(\qperp )\mathcal{J}_q(z)+xg(x)\tilde{\mathcal{F}}(\qperp )\mathcal{J}_g(z)\right],
\end{eqnarray}
where $\tau=P_Je^\eta/\sqrt{s}$ is the longitudinal fraction of the final state jet with $P_J=z\qperp$ is the transverse momentum of the jet. Note we introduce the CJFs $\mathcal{J}_f(z)$ which represent the probability of final state partons becoming a jet with the momentum fraction $z$, where the label $f=q, g$ for quark and gluon jets, respectively. In particular, the leading order CJFs $\mathcal{J}_{q}^{(0)}(z)$ and $\mathcal{J}_{g}^{(0)}(z)$ are trivial since partons are identified as jets at LO
\begin{align}
\mathcal{J}_{q}^{(0)}(z)=\delta(1-z), \quad \quad 
\mathcal{J}_{g}^{(0)}(z)=\delta(1-z).
\label{initial2}
\end{align} 
The dipole gluon distributions follow the definitions
\begin{align}
\mathcal{F}(\qperp )=\int \frac{\dd^{2}x_{\perp }\dd^{2}y_{\perp }}{(2\pi )^{2}}%
e^{-i\qperp\cdot (x_{\perp }-y_{\perp })}S_Y^{(2)}(x_\perp,y_\perp),\\
\tilde{\mathcal{F}}(\qperp )=\int \frac{\dd^{2}x_{\perp }\dd^{2}y_{\perp }}{(2\pi )^{2}}%
e^{-i\qperp\cdot (x_{\perp }-y_{\perp })}\tilde{S}_Y^{(2)}(x_\perp,y_\perp),
\end{align}
with $S_Y^{(2)}(x_\perp,y_\perp)=\frac{1}{N_{c}}\left\langle
\text{Tr}U(x_{\perp })U^{\dagger }(y_{\perp })\right\rangle _{Y}$ and $\tilde{S}_Y^{(2)}(x_\perp,y_\perp)=\frac{1}{N_{c}^2-1}\left\langle \text{Tr}%
W(x_{\perp})W^{\dagger }(y_{\perp })\right\rangle _{Y}$ are the quark and the gluon dipole amplitude, respectively. By utilizing the Fierz identity and the large-$N_c$ limit, one can rewrite the scattering amplitude $\tilde{S}_Y^{(2)}(x_\perp,y_\perp)$ as $S_Y^{(2)}(x_\perp,y_\perp)S_Y^{(2)}(y_\perp,x_\perp)$. As shown in previous studies~\cite{Chirilli:2011km,Chirilli:2012jd, Shi:2021hwx}, the large $N_c$ limit can greatly simplify both the analytic and numerical computations. Thus, we will take large $N_c$ limit throughout this paper for simplicity and only keep the leading $N_c$ contributions.

\section{The next-to-leading order cross section}
\label{section3}

In this section, we aim to present the detailed evaluations for the NLO corrections. In principle, there are four partonic channels need to be considered: $q\to qg$, $g\to gg$, $q\to gq$, $g\to q\bar{q}$. We first take $q\to qg$ as an example to illustrate our calculation strategy and set up the framework for the NLO calculations of jet production. We mainly concentrate on the final state radiation since this is the most difficult part compared to hadron productions. Meanwhile, the initial state radiations, interference contributions, and virtual contributions are akin to the calculations of the hadron production case once we take the small cone limit. In the end, we list the final results in order to be self-contained. 

\subsection{The \texorpdfstring{$q\to q$}{q->q} channel}
\label{section31}

For the $q\to q$ channel, the NLO real diagrams of this channel have an additional gluon radiation. This process includes both the initial state gluon radiation and the final state gluon radiation. One measures the final state quark jet after the multiple scattering with the nucleus target. The $q\to q$ channel has been studied widely in Ref.~\cite{hep-ph/0405266, Dominguez:2010xd, Dominguez:2011wm,Chirilli:2011km,Chirilli:2012jd}. We take Eq.(11) of Ref.~\cite{Chirilli:2012jd} as our starting point since the partonic cross-section is the same. According to the previous studies \cite{Dominguez:2011wm,Chirilli:2012jd}, the partonic cross-section reads as follows
\begin{eqnarray}
\frac{\dd\sigma _{qA\rightarrow qg X}}{\dd^3l\dd^3k}&=&\alpha
_{S}C_F\delta(q^+-l^+-k^+) \int \frac{\text{d}^{2}x_{\perp}}{(2\pi)^{2}}%
\frac{\text{d}^{2}x_{\perp}^{\prime }}{(2\pi )^{2}}\frac{\text{d}%
^{2}b_{\perp}}{(2\pi)^{2}}\frac{\text{d}^{2}b_{\perp}^{\prime }}{(2\pi )^{2}} e^{-il_\perp\cdot(x_\perp-x_\perp')}e^{-ik_\perp\cdot(b_\perp-b_\perp')}\sum_{\lambda\alpha\beta}
\psi^{\lambda\ast}_{\alpha\beta}(u^{\prime}_{\perp})\psi^\lambda_{\alpha%
\beta}(u_{\perp})  \notag \\
&&\times \left[S^{(6)}_{Y}(b_{\perp},x_{\perp},b^{\prime}_{\perp},x^{%
\prime}_{\perp})+S^{(2)}_{Y}(v_{\perp},v^{\prime}_{\perp}) - S^{(3)}_{Y}(b_{\perp},x_{\perp},v^{\prime}_{%
\perp})-S^{(3)}_{Y}(v_{\perp},x^{\prime}_{\perp},b^{\prime}_{\perp})\right],
\label{partqqg}
\end{eqnarray}
with $l$ and $k$ being the momenta of the final state gluon and quark, respectively. $\xi=\frac{k^+}{q^+}$ is the longitudinal momentum fraction carried by the final state quark with $q$ being the momentum of the incoming quark. $x_\perp$ and $b_\perp$ are the transverse coordinates of gluon and quark in the amplitude, respectively. $x_\perp'$ and $b_\perp'$ are the transverse coordinates of gluon and quark in the conjugate amplitude, respectively. For convenience, we have also defined $u_{\perp}=x_{\perp}-b_{\perp}$, $u^{\prime}_{\perp}=x^{\prime}_{%
\perp}-b^{\prime}_{\perp}$, $v_{\perp}=(1-\xi)x_{\perp}+\xi b_{\perp}$, $%
v^{\prime}_{\perp}=(1-\xi)x^{\prime}_{\perp}+\xi b^{\prime}_{\perp}$. The correlators are 
\begin{align}
S^{(6)}_{Y}(b_{\perp},x_{\perp},b^{\prime}_{\perp},x^{\prime}_{\perp})&=%
\frac{1}{C_FN_c}\left\langle\text{Tr}\left(U(b_{\perp})U^\dagger(b^{%
\prime}_{\perp})T^dT^c\right)\left[W(x_{\perp})W^\dagger(x^{\prime}_{\perp})%
\right]^{cd}\right\rangle_{Y}, \\
S^{(3)}_{Y}(b_{\perp},x_{\perp},v^{\prime}_{\perp})&=\frac{1}{C_FN_c}%
\left\langle\text{Tr}\left(U(b_{\perp})T^dU^\dagger(v^{\prime}_{\perp})T^c%
\right)W^{cd}(x_{\perp})\right\rangle_{Y}.
\end{align}
In addition, we also include virtual diagrams. The calculations of virtual diagrams are straightforward in the dipole picture. It eventually leads to 
\begin{eqnarray}
\frac{\dd\sigma_{\text{virt}}}{\dd^3k}&=&-2\alpha _{s}C_{F}\int \frac{\dd^{2}v_\perp}{(2\pi )^{2}}\frac{%
\dd^{2}v^{\prime }_\perp}{(2\pi )^{2}}\frac{\dd^{2}u_\perp}{(2\pi )^{2}}%
e^{-i\qperp\cdot (v_\perp-v^{\prime
}_{\perp})}\sum_{\lambda\alpha\beta}\psi^{\lambda\ast}_{\alpha\beta}(u_%
\perp)\psi^{\lambda}_{\alpha\beta} (u_\perp) \left[ S_Y^{(2)}(v_\perp,v_\perp^{\prime
})-S^{(3)}_{Y}(b_{\perp},x_{\perp},v^{\prime}_{\perp})\right].  \label{v1}
\end{eqnarray}
The square sum of the splitting wave function (splitting kernel) can be written as follows
\begin{eqnarray}
\sum_{\lambda\alpha\beta} \psi^{\lambda\ast}_{\alpha\beta}(\xi,
u^{\prime}_{\perp})\psi^\lambda_{\alpha\beta}(\xi, u_{\perp})=2(2\pi)^2\frac{1}{p^+}
\left[
\frac{1+\xi^2}{1-\xi}
-\epsilon (1-\xi) \right]
\frac{%
u_{\perp}^{\prime}\cdot u_{\perp}}{u_{\perp}^{\prime 2} u_{\perp}^{ 2}},
\end{eqnarray}
which is consistent with the splitting function in $D=4-2\epsilon$ dimension given by Ref.~\cite{Catani:1998nv}.
In the above function, the contribution from the second correction term $-\epsilon (1-\xi)$ is expected to be small. This correction term only affects the contributions of the initial state radiation, since the corresponding contribution has the collinear divergence. In contrast, the final results of the jet production are free of the collinear singularity, thus there is no finite contributions from the $-\epsilon (1-\xi)$ term when final state gluon radiations are considered.

\subsubsection{The final state radiation}

\begin{figure}[ht]
\centering
\subfigure[]{\includegraphics[width=4cm]{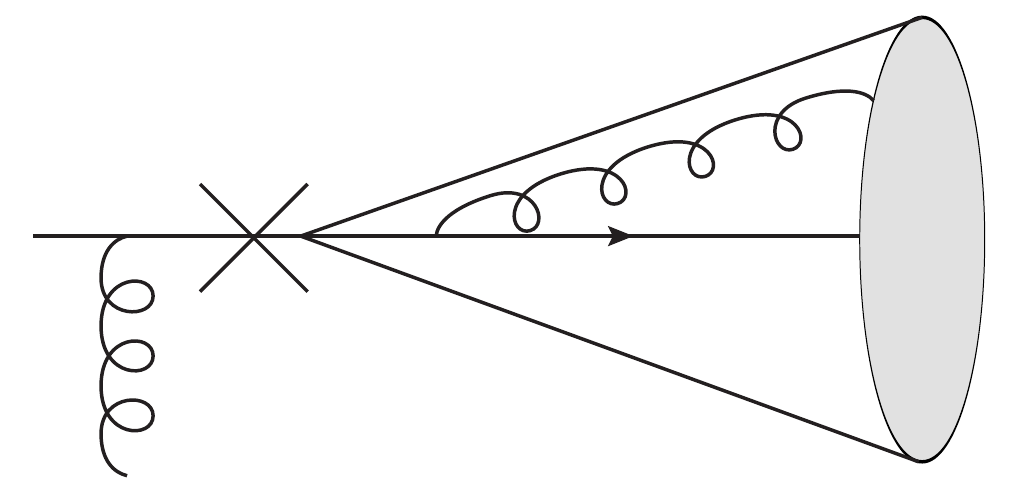}}
~~\raisebox{27pt}[0pt][0pt]{--}~~
\subfigure[]{\includegraphics[width=4cm]{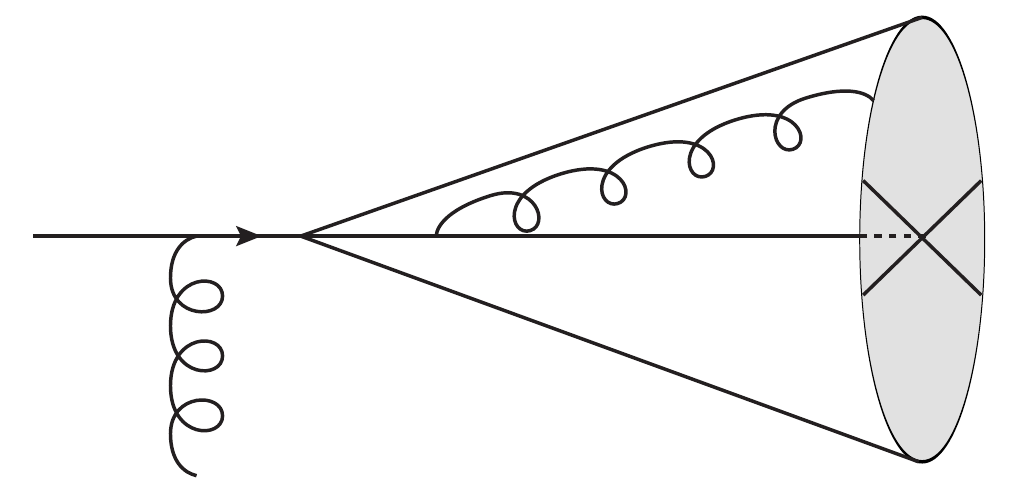}}
~~\raisebox{27pt}[0pt][0pt]{+}~~
\subfigure[]{\includegraphics[width=4cm]{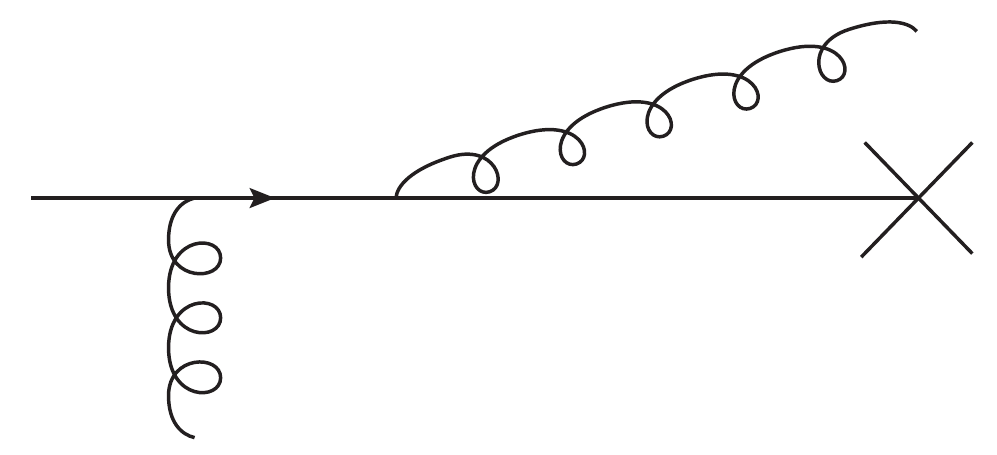}}
\caption[*]{The definition of the jet-cross section from partonic cross-section where the symbol $\times$ indicates the measured jet with transverse momentum $P_J$.}
\label{qqfinal}
\end{figure}

In this subsection, let us elaborate the calculation on the final state gluon radiation since this is the part which differs the most from the hadron production case. The $S_Y^{(2)}(v_\perp,v_\perp')$ term in Eq. (\ref{partqqg}) corresponds to the final state gluon radiation contribution. It resums only the multiple interactions with nucleus target before the quark splitting. We use the narrow jet approximation as described in Refs.~\cite{Jager:2004jh, Mukherjee:2012uz, Sun:2014gfa, Sun:2014lna}. Let us explain the procedure and the results by using the Feynman diagrams as depicted in Fig.~\ref{qqfinal}. Depending on whether the radiated gluon is inside the jet cone or not, there are three different cases. Firstly, Fig.~\ref{qqfinal} (a) represents the so-called in-cone contribution which indicates the radiated gluon and its parent quark are almost collinear. In this case, the final state quark and gluon are combined together and treated as a single jet. Therefore, the momentum of the measured quark jet is equal to the sum of the momenta of these two particles. Secondly, Fig.~\ref{qqfinal} (b) stands for the false identification of the jet as the final state quark when the radiated gluon is inside the quark jet cone. This part should be subtracted from the total contribution. In addition, Fig.~\ref{qqfinal} (c) denotes the quark plus the gluon radiation without any constraint. Therefore, $\sigma_\text{c}-\sigma_\text{b}$ yields the out-cone contribution. Therefore, the corresponding quark jet cross-section can be expressed as follows,  
\begin{eqnarray}
\sigma_{qq}^{\text{final}}=\sigma_\text{a}+(\sigma_\text{c}-\sigma_\text{b})+\sigma_{qq}^{\text{virt}(\text{jet})} , \label{com}
\end{eqnarray}
where $\sigma_{qq}^{\text{virt}(\text{jet})} $ is the virtual contribution. It is straightforward to determine the momentum constraints for the above three cases via the light cone perturbation formalism. To proceed, we need to define the jet cone. Firstly, we define the momenta of the radiated gluon and the final state quark as $l^\mu\equiv(l^+=\frac{1}{\sqrt{2}}l_\perp e^{y_1},l^-=\frac{1}{\sqrt{2}}l_\perp e^{-y_1},l_\perp)$ and $k^\mu\equiv(k^+=\frac{1}{\sqrt{2}}k_\perp e^{y_2},k^-=\frac{1}{\sqrt{2}}k_\perp e^{-y_2},k_\perp)$, respectively. Then, the relative distance between the quark and the radiated gluon is characterized by their invariant mass
\begin{align}
(l+k)^2= l_\perp k_\perp \left(e^{y_1-y_2}+e^{y_2-y_1}\right)-2 l_\perp k_\perp \cos(\phi_1-\phi_2) \equiv l_\perp k_\perp R_{qg}^2,
\end{align}
where $R_{qg}\equiv\sqrt{\Delta y^2+\Delta \phi^2}$ when their rapidity difference $\Delta y\equiv y_1-y_2$ and azimuthal angle difference $\Delta\phi\equiv \phi_1-\phi_2$ are very small. Furthermore, the virtuality of the quark-gluon pair can also be expressed as 
\begin{equation}
(l+k)^2=\frac{(\xi l_\perp-(1-\xi)k_\perp)^2}{\xi(1-\xi)}.
\end{equation}
Here $\xi$ is the longitudinal momentum carried by the final state quark, and $p_\perp=\xi l_\perp-(1-\xi)k_\perp$ stands for the relative transverse momentum of the quark-gluon pair. Once we define the jet cone size as $R$, then the requirement $R_{qg}\leq R$ indicates that the final state quark and the radiated gluon are located within one jet cone.

In the following, we derive the momentum constraints for the above three cases. First, the radiated gluon and the final state quark are put inside the same jet cone. Therefore, the momentum of the measured jet is equal to the momentum summation of the quark-gluon pair that is $ P_J=z\qperp=z(l_\perp+k_\perp)$ with $z=1$. In order to get the differential cross-section of the transverse momentum of jet, we need to integrate the relative momentum $p_\perp$. By requiring that both the quark and gluon are inside the same jet cone, we have the kinematic constraint as follows
\begin{align}
\frac{p_\perp^2}{\xi(1-\xi)}\leq l_\perp k_\perp R^2\simeq\frac{\xi(1-\xi)}{z^2}P_J^2R^2=\xi(1-\xi)\qperp^2R^2,
\label{constraint1}
\end{align} 
where in the last step we approximately write $zl_\perp=(1-\xi)P_J$ and $zk_\perp=\xi P_J$. Taking all considerations into account, the kinematic constraint becomes $p_\perp^2\leq \xi^2(1-\xi)^2\qperp^2R^2$.

The second diagram indicates the contribution from the false identification of tagged quark when the emitted gluon is also inside the jet cone. In this case, the transverse momentum of the measured final state quark jet is $P_J=zk_\perp=z\qperp$. Note that the in-cone constraint is slightly different from the constraint of $\sigma_\text{a}$. We still have the approximate relation $\frac{l_\perp}{k_\perp}=\frac{1-\xi}{\xi}$. Therefore, the constraint becomes
\begin{eqnarray}
\frac{p_\perp^2}{\xi(1-\xi)}=l_\perp k_\perp R_{qg}^2\leq l_\perp  k_\perp R^2=\frac{1-\xi}{\xi z^2} P_J^2R^2.
\label{constraint2}
\end{eqnarray}
In this case, the kinematic constraint changes to $p_\perp^2\leq(1-\xi)^2 \qperp^2R^2$. This contribution should be subtracted since it comes from the false tagging of an individual parton inside a jet cone. 

For the last part, we do not impose any jet cone constraints, and then integrate the momentum of radiated gluon over the full phase space. In this case, the calculation is identical to that of the hadron production.

To proceed, we apply the dimensional regularization~\cite{Ciafaloni:1998hu} and the modified minimal subtraction scheme ($\overline{\text{MS}}$) to evaluate the remaining part of the integral. Thus, we can write $\sigma_\text{a}$, $\sigma_\text{b}$ and $\sigma_\text{c}$ as follows
\begin{eqnarray}
\sigma_\text{a}
&=&\frac{\alpha_sC_F}{2\pi}\left[\frac{1}{\epsilon^2}+\frac{3}{2\epsilon}-\frac{1}{\epsilon}\ln\frac{q_\perp^2R^2}{\mu^2}-\frac{3}{2}\ln\frac{q_\perp^2R^2}{\mu^2}+\frac{1}{2}\ln^2\frac{q_\perp^2R^2}{\mu^2}+6-\frac{3}{4}\pi^2   +\frac{1}{2} \right]\sigma_{\text{LO}}(x,q_\perp), \label{sigmaa}\\
\sigma_\text{b}&=&\frac{\alpha_sC_F}{2\pi}\int_x^1\dd\xi~\sigma_{\text{LO}}\left(\frac{x}{\xi},\frac{\qperp}{\xi}\right)\frac{1+\xi^2}{(1-\xi)_+}\frac{1}{\xi^{2}}\left[-\frac{1}{\epsilon}-\ln\frac{\xi^2\mu^2}{q_\perp^2R^2}\right]+\frac{\alpha_sC_F}{2\pi}\int_x^1\dd\xi~\sigma_{\text{LO}}\left(\frac{x}{\xi},\frac{\qperp}{\xi}\right)\left[\frac{\ln(1-\xi)^2}{(1-\xi)}\right]_+\frac{1+\xi^2}{\xi^2} \label{sigmab} \nonumber \\
&&+\frac{\alpha_sC_F}{2\pi}\left[\frac{1}{\epsilon^2}-\frac{1}{\epsilon}\ln\frac{q_\perp^2R^2}{\mu^2}-\frac{\pi^2}{12}+\frac{1}{2}\ln^2\frac{q_\perp^2R^2}{\mu^2}\right]\sigma_{\text{LO}}(x,q_\perp)+\frac{\alpha_sC_F}{2\pi}\int_x^1\dd\xi~\sigma_{\text{LO}}\left(\frac{x}{\xi},\frac{\qperp}{\xi}\right)\frac{1-\xi}{\xi^{2}}
,\\
\sigma_\text{c}&=&\frac{\alpha_sC_F}{2\pi}\int_x^1\dd\xi~\sigma_{\text{LO}}\left(\frac{x}{\xi},\frac{\qperp}{\xi}\right)\frac{1+\xi^2}{(1-\xi)_+}\frac{1}{\xi^2}\left[-\frac{1}{\epsilon}+\ln\frac{c_0^2}{\mu^2r_\perp^2}\right] +\frac{\alpha_sC_F}{2\pi}\int_x^1\dd\xi~\sigma_{\text{LO}}\left(\frac{x}{\xi},\frac{\qperp}{\xi}\right)\frac{1-\xi}{\xi^{2}},
\end{eqnarray}
where $c_0=2e^{-\gamma_E}$ with the Euler's constant $\gamma_E \simeq 0.577$. 
The splitting function $\mathcal{P}_{qq}(\xi)$ is defined as
\begin{eqnarray}
\mathcal{P}_{qq}(\xi)=\left(\frac{1+\xi^2}{1-\xi} \right)_+=\frac{1+\xi^2}{(1-\xi)_+}+\frac{3}{2}\delta(1-\xi).
\end{eqnarray}
In arriving at the above result, we have defined 
\begin{equation}
\sigma_{\text{LO}}\left(\frac{x}{\xi},\frac{\qperp}{\xi}\right)=S_\perp\int_\tau^1\frac{\dd z}{z^2}\mathcal{J}_q^{(0)}(z)\int\frac{\dd^2r_\perp}{(2\pi)^2}S^{(2)}(r_\perp)\frac{x}{\xi}q\left(\frac{x}{\xi}\right)e^{-i\frac{\qperp\cdot r_\perp}{\xi}},
\end{equation}
with $r_\perp=x_\perp-y_\perp$ and $S_\perp$ being the transverse area of the target nucleus. We can see that the cone-size dependence comes from $\sigma_\text{a}$ and $\sigma_\text{b}$. Meanwhile, the rapidity divergence only comes from $\sigma_\text{c}$ when the radiated gluon almost is back-to-back with the parent quark. 
\begin{figure}[ht]
\centering
\subfigure[]{
\includegraphics[width=5cm]{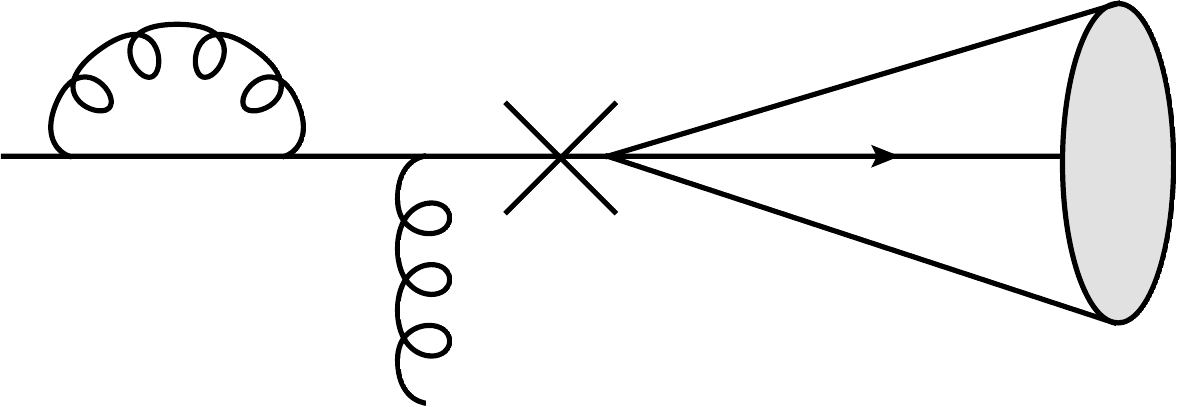}
}
\subfigure[]{
\includegraphics[width=5cm]{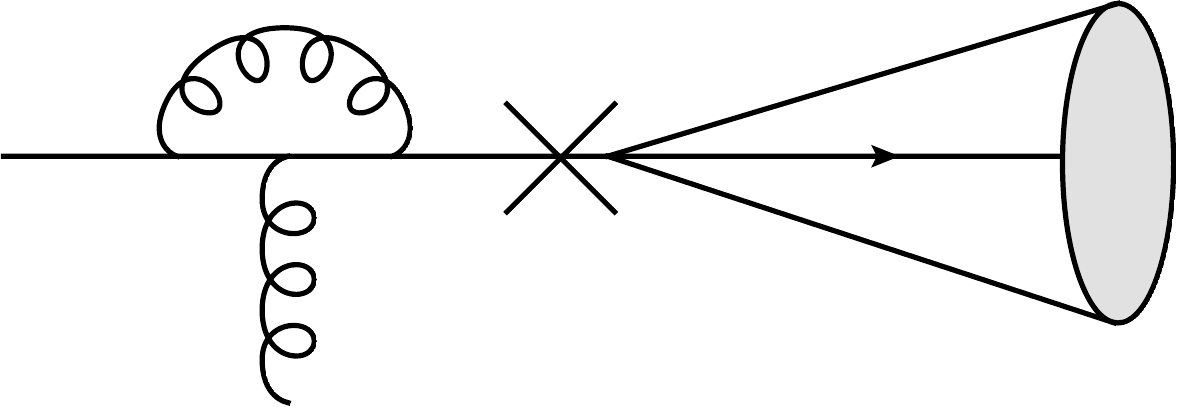}
}
\subfigure[]{
\includegraphics[width=5cm]{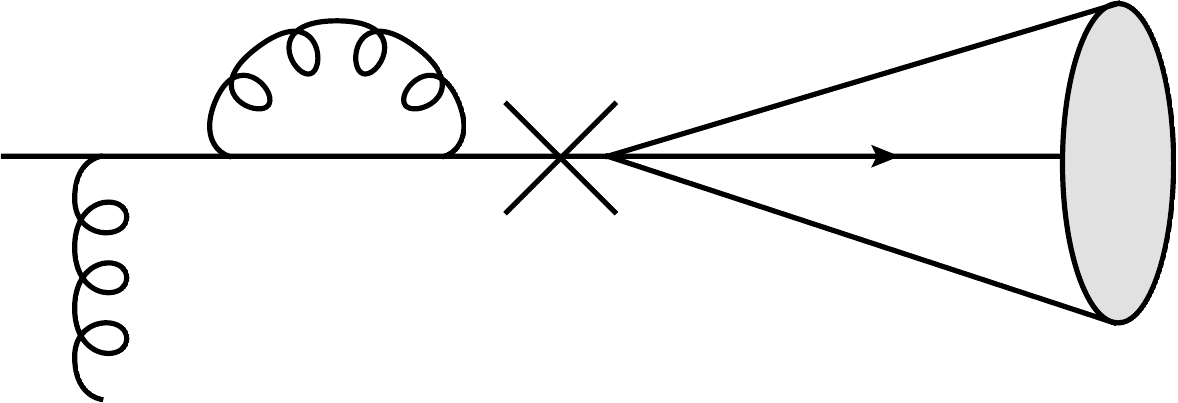}
}
\caption[*]{The virtual contributions to the quark jet from the $q\to q$ channel.}
\label{qqvirtual0}
\end{figure}
In addition, Fig.~\ref{qqvirtual0} contains all virtual contributions of $q\rightarrow qg$ channel. To illustrate the cancellation of final state singularities, we single out part of final virtual contributions associated with the jet, and demonstrate the complete cancellation. Comparing to the hadron production, this case is new since there is a residual collinear divergence associated with the hadron fragmentation function. The remaining virtual contributions will also be taken into account later and they are combined with other real contributions in the following sections. The virtual jet contribution $\sigma_{qq}^{\text{virt}(\text{jet})}$ is given as 
\begin{eqnarray}
 \sigma_{qq}^{\text{virt}(\text{jet})} =   S_\perp\frac{3}{2}\frac{\alpha_sC_F}{2\pi}\int_\tau^1\frac{\dd z}{z^2}\mathcal{J}_q^{(0)}(z)\int\frac{\dd^2r_\perp}{(2\pi)^2}e^{-i\qperp\cdot r_\perp}S^{(2)}(r_\perp)\left[ - \frac{1}{\epsilon}+\ln\frac{\qperp^2}{\mu^2} -\frac{1}{2} \right].
\label{qqvirtual}
\end{eqnarray}
By combining all these contributions together as indicated in Eq.~(\ref{com}), we find that all the divergences cancel. The final results read as
\begin{eqnarray}
\frac{\dd\sigma_{qq}^{\text{final}}}{\dd\eta \dd^2P_J}&=&-\frac{\alpha_sC_F}{2\pi}\int_x^1\dd\xi\frac{1}{\xi^2}\sigma_{\text{LO}}\left(\frac{x}{\xi},\frac{\qperp}{\xi}\right)\left[-\frac{1+\xi^2}{(1-\xi)_+}\ln\frac{c_0^2}{\qperp^2r_\perp^2}-\mathcal{P}_{qq}(\xi)\ln\frac{1}{R^2}\right.\notag\\
&&\left.\quad+(1+\xi^2)\left(\frac{\ln\frac{(1-\xi)^2}{\xi^2}}{1-\xi}\right)_+-\left(6-\frac{4}{3}\pi^2\right)\delta(1-\xi)\right].
\label{qqfinalcross-section}
\end{eqnarray}
Note here we have terms proportional to $\left(\frac{\ln(1-\xi)^2}{1-\xi}\right)_+$ and $\left(\frac{\ln\xi^2}{1-\xi}\right)_+$ in $\sigma_\text{b}$. By implementing the definition of the plus function $\int_a^1\dd\xi(f(\xi))_+g(\xi)=\int_a^1\dd\xi f(\xi)g(\xi)-g(1)\int_0^1\dd\xi f(\xi)$ with $g(\xi)$ being a non-singular function and $f(\xi)$ being singular at $\xi=1$, we can combine these two terms as follows 
\begin{eqnarray}
-\int_x^1\dd\xi\left(\frac{\ln(1-\xi)^2}{1-\xi}\right)_+\frac{1+\xi^2}{\xi^2}+\int_x^1\dd\xi\left(\frac{\ln\xi^2}{1-\xi}\right)_+\frac{1+\xi^2}{\xi^2}=-\int_x^1\dd\xi\left[\frac{\ln\frac{(1-\xi)^2}{\xi^2}}{1-\xi}\right]_+\frac{1+\xi^2}{\xi^2}-\frac{2}{3}\pi^2 ,
\end{eqnarray}
which gives rise to an additional constant factor of $-\frac{2\pi^2}{3}$. In comparison with the hadron production case, there are several terms which are unique to the jet production. First, the terms in the second line of Eq.~(\ref{qqfinalcross-section}) are new. Moreover, the cone-size dependent term is akin to the collinear divergence in the hadron production. Only the first term inside the square brackets in Eq.~(\ref{qqfinalcross-section}) is identical to the corresponding one in the hadron production. 

\subsubsection{Other contributions}

This section is devoted to the discussion of the remaining contributions. As seen in the following, the computations for these contributions are almost the same as those for the hadron production case.

\begin{figure}[ht]
\centering
\includegraphics[width=4cm]{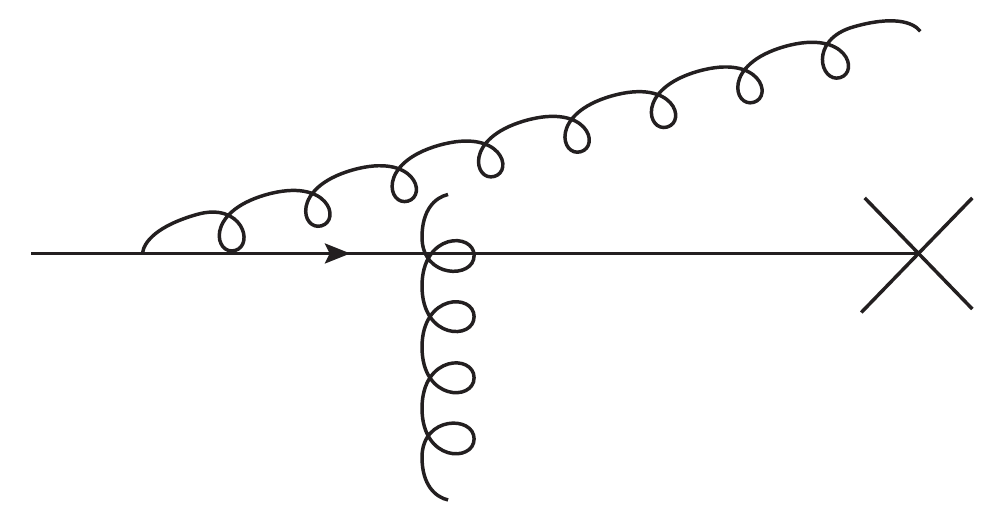}
\caption[*]{The real contribution to the quark jet due to initial state gluon radiations.}
\label{qqinitial}
\end{figure}

The Feynman diagram of the initial state gluon radiation is illustrated in Fig.~\ref{qqinitial}, where the multiple interactions occur not only in the amplitude but also in the conjugate amplitude. Both the multiple scattering of the quark and gluon with the target nucleus should be resumed, which corresponds to the $S_Y^{(6)}(b_{\perp},x_{\perp},b^{\prime}_{\perp},x^{\prime}_{\perp})$ term in Eq. (\ref{partqqg}). 

Fig.~\ref{qqinitial} shows the initial state radiation. In principle, the radiated gluon has a finite probability that it goes into the jet cone of the final state quark. However, it is expected that this probability is proportional to $R^2$, and thus it is negligible in the narrow jet approximation. Therefore, we can approximately integrate the momentum of the initial state radiated gluon over the entire phase space. In this sense, the calculation is the similar as the one in the hadron production. By integrating over the unobserved gluon momentum, we identify the transverse coordinate of the gluon from $x_\perp$ to $x_\perp'$, which simplifies the correlator $S_Y^{(6)}(b_{\perp},x_{\perp},b^{\prime}_{\perp},x^{\prime}_{\perp})$ and reduces it to $S^{(2)}_Y(\bperp,\bperp')$.

The next step is to use dimensional regularization and $\overline{\text{MS}}$ subtraction scheme again together with the above momentum constraints to evaluate the integration of $S_Y^{(2)}(v_\perp,v_\perp')$. The details can be found in Ref.~\cite{Chirilli:2012jd}. We list the final results here for completeness 
\begin{eqnarray}
\frac{\dd\sigma^{\text{initial}}}{\dd\eta \dd^2P_J}&=&S_\perp\frac{\alpha_sC_F}{2\pi}\int_\tau^1\frac{\dd z}{z^2}\mathcal{J}_q^{(0)}(z)\int\frac{\dd^2r_\perp}{(2\pi)^2}e^{-i\qperp\cdot r_\perp}S(r_\perp)\int_x^1\dd\xi\frac{1+\xi^2}{(1-\xi)_+}\frac{x}{\xi}q\left(\frac{x}{\xi}\right)\left[-\frac{1}{\epsilon}+\ln\frac{c_0^2}{\mu^2r_\perp^2}\right]\notag\\
&&+S_\perp\frac{\alpha_sC_F}{2\pi}\int_\tau^1\frac{\dd z}{z^2}\mathcal{J}_q^{(0)}(z)\int\frac{\dd^2r_\perp}{(2\pi)^2}e^{-i\qperp\cdot r_\perp}S(r_\perp)\int_x^1\dd\xi\frac{x}{\xi}q\left(\frac{x}{\xi}\right)\left[1-\xi \right].
\label{qqreal}
\end{eqnarray}
Note here in Eq.~(\ref{qqreal}) we have used the usual subtraction scheme of the rapidity divergence which changes the splitting function $\int\dd\xi\frac{1+\xi^2}{1-\xi}$ into $\int\dd\xi\frac{1+\xi^2}{(1-\xi)_+}$. In addition, as mentioned before, by considering the full four-momentum conservation before and after scattering, additional exact kinematical constraints will occur, which is equivalent to modifying the dipole splitting function. Therefore, for the full rapidity subtraction, several new terms emerge~\cite{Watanabe:2015tja}. 

In Eq. (\ref{partqqg}), $S^{(3)}_{Y}(b_{\perp},x_{\perp},v^{\prime}_{\perp})$ and $S^{(3)}_{Y}(v_{\perp},x_{\perp}',b_{\perp}')$ are the interference contributions. By taking the narrow jet approximation, we can simplify the evaluation of interference diagrams. We will list the final results in the next subsection. By combining the collinear singularities in Eq.~(\ref{qqreal}) and Eq.~(\ref{qqvirtual}), the coefficient of the collinear singularities becomes 
\begin{eqnarray}
-C_F\frac{1}{2\pi}\frac{1}{\epsilon}\left[\frac{3}{2}xq(x)+\int_{x}^1\dd\xi\frac{1+\xi^2}{(1-\xi)_+}\frac{x}{\xi} q\left(\frac{x}{\xi}\right)\right]=-C_F\frac{1}{2\pi}\frac{1}{\epsilon}\int_{x}^1\dd\xi\mathcal{P}_{qq}(\xi)\frac{x}{\xi} q\left(\frac{x}{\xi}\right).
\label{collinear}
\end{eqnarray}
In arriving at Eq. (\ref{collinear}), we have rewritten $\frac{3}{2}=\int_x^1\dd\xi\frac{3}{2}\delta(1-\xi)$ to change the first term in the left hand side. At the end of the day, by redefining the quark distribution function we can remove the collinear singularities as follows
\begin{equation}
q(x,\mu)=q^{(0)}(x)-\frac{1}{\epsilon}\frac{\alpha_s}{2\pi}\int_x^1\frac{\dd\xi}{\xi}C_F\mathcal{P}_{qq}(\xi)q\Big(\frac{x}{\xi}\Big),
\end{equation}
where $q^{(0)}(x)$ is the bare quark distribution.

\subsubsection{The complete one-loop cross-section in the coordinate space}

After removing all the divergences by renormalizing the quark distribution functions and the subtraction of the rapidity divergences, the final contributions should be finite. To proceed, we assemble all the finite terms together. For the quark channel: $qA\to \text{jet}+X$, we have the differential cross-section in the coordinate space as the following two parts 
\begin{align}
\frac{\dd \sigma_{qq}}{\dd \eta \dd^2 P_J}
= 
\frac{\dd \sigma_{qq}^{\rm LO}}{\dd \eta \dd^2 P_J}
+ \frac{\dd \sigma^{\rm NLO}_{qq}}{\dd \eta\dd^{2}P_J} 
= 
\frac{\dd \sigma_{qq}^{\rm LO}}{\dd \eta \dd^2 P_J}
+ 
\sum_{i=a}^m
\frac{\dd \sigma_{qq}^{i}}{\dd \eta \dd^2 P_J},
\end{align}
where the LO and NLO parts read
\begin{align} 
\frac{\dd \sigma_{qq}^{\rm LO}}{\dd \eta \dd^2 P_J}
= &
\Sperp \int_\tau^1\frac{\dd z}{z^2}\mathcal{J}_q^{(0)}(z)x q(x,\mu^2) \int \frac{\dd^2\rperp}{(2\pi)^2} e^{-i\qperp \cdot \rperp} S^{(2)}(\rperp),
\\ 
\frac{\dd \sigma_{qq}^{a}}{\dd \eta \dd^2 P_J}
= &
\frac{\alpha_s}{2\pi} \Sperp C_F\int_\tau^1\frac{\dd z}{z^2}\mathcal{J}_q^{(0)}(z)
 \int_{x}^1 \dd\xi \frac{x}{\xi} q\left(\frac{x}{\xi},\mu^2\right)
\int \frac{\dd^2\rperp}{(2\pi)^2} 
 e^{-i\qperp \cdot \rperp}  S^{(2)}(\rperp)  
\left[
\mathcal{P}_{qq} (\xi) 
 \ln \frac{c_0^2}{\rperp^2 \mu^2}
 +(1-\xi)
 \right],
\\
\frac{\dd \sigma_{qq}^{b}}{\dd \eta \dd^2 P_J}
= &
- \frac{\alpha_s}{2\pi} \Sperp C_F\int_\tau^1\frac{\dd z}{z^2}\mathcal{J}_q^{(0)}(z)x q(x,\mu^2)  \int \frac{\dd^2\rperp}{(2\pi)^2} e^{-i\qperp\cdot \rperp}S^{(2)}(\rperp) 
\left[
 \frac{3}{2}  \ln \frac{c_0^2}{\rperp^2\qperp^2} +\frac{1}{2}
\right] 
   \, ,
\\
\frac{\dd \sigma_{qq}^{c}}{\dd \eta \dd^2 P_J}
= &
-8\pi\frac{\alpha_s}{2\pi} \Sperp C_F \int_\tau^1\frac{\dd z}{z^2}\mathcal{J}_q^{(0)}(z)
 \int_{x}^1 \dd\xi \frac{x}{\xi} q\left(\frac{x}{\xi},\mu^2\right)  \int \frac{\dd^2\uperp\dd^2\vperp}{(2\pi)^4}   e^{-i \qperp \cdot (\uperp-\vperp)}  e^{-i \frac{1-\xi}{\xi}\qperp \cdot \uperp} 
 \nonumber \\
& \times \frac{1+\xi^2}{(1-\xi)_+} \frac{1}{\xi} \frac{\uperp \cdot \vperp}{\uperp^2 \vperp^2}
S^{(2)}(\uperp)S^{(2)}(\vperp),
\\
\frac{\dd \sigma_{qq}^{d}}{\dd \eta \dd^2 P_J}
= & 
8\pi\frac{\alpha_s}{2\pi} \Sperp C_F \int_\tau^1\frac{\dd z}{z^2}\mathcal{J}_q^{(0)}(z)
 x q(x,\mu^2) \int \frac{\dd^2\uperp\dd^2\vperp}{(2\pi)^4}  e^{-i \qperp \cdot ( \uperp -\vperp)} S^{(2)}(\uperp)S^{(2)}(\vperp)
\nonumber\\
& \times 
\int_0^1 \dd\xi' \frac{1+\xi'^2}{(1-\xi')_+}
\left[
e^{-i(1-\xi') \qperp \cdot \vperp} \frac{1}{\vperp^2} - \delta^2 (\vperp) \int \dd^2 \rperp' e^{i\qperp \cdot \rperp'}\frac{1}{\rperp'^2} 
\right],
\\
\frac{\dd \sigma_{qq}^{e}}{\dd \eta \dd^2 P_J}
= & \frac{\alpha_s}{\pi^2} \Sperp C_F\int_\tau^1\frac{\dd z}{z^2}\mathcal{J}_q^{(0)}(z)x q(x,\mu^2) \int \frac{\dd^2 \uperp\dd^2\vperp}{(2\pi)^2} e^{-i\qperp\cdot (\uperp-\vperp)} [S^{(2)}(\uperp)S^{(2)}(\vperp) - S^{(2)}(\uperp-\vperp)] 
\nonumber\\
& \times \Bigg[
\frac{1}{\uperp^2} \ln \frac{\qperp^2 \uperp^2}{c_0^2} +
\frac{1}{\vperp^2} \ln \frac{\qperp^2 \vperp^2}{c_0^2} - \frac{2\uperp\cdot\vperp}{\uperp^2 \vperp^2} \ln \frac{\qperp^2 |\uperp| |\vperp|}{c_0^2}
\Bigg],\\
\frac{\dd \sigma_{qq}^{ f}}{\dd \eta \dd^2 P_J}
= &
\frac{\alpha_s}{2\pi} \Sperp C_F\left(6-\frac{4}{3}\pi^2\right)\int_\tau^1\frac{\dd z}{z^2}\mathcal{J}_q^{(0)}(z)x q(x) \int \frac{\dd^2\rperp}{(2\pi)^2} e^{-i\qperp \cdot \rperp} S^{(2)}(\rperp),
\\ 
\frac{\dd \sigma_{qq}^{g}}{\dd \eta \dd^2 P_J}
= &-\frac{\alpha_s}{2\pi}\Sperp C_F\int_\tau^1\frac{\dd z}{z^2}\mathcal{J}_q^{(0)}(z)\int_x^1\dd\xi \frac{x}{\xi}q\left(\frac{x}{\xi}\right)\left[\frac{\ln\frac{(1-\xi)^2}{\xi^2}}{1-\xi}\right]_+\frac{1+\xi^2}{\xi^2} \int \frac{\dd^2\rperp}{(2\pi)^2} e^{-i\frac{\qperp\cdot \rperp}{\xi} } S^{(2)}(\rperp),\\
\frac{\dd \sigma_{qq}^{h}}{\dd \eta \dd^2 P_J}
= & \frac{\alpha_s}{2\pi} \Sperp C_F \int_\tau^1\frac{\dd z}{z^2}\mathcal{J}_q^{(0)}(z)\int_x^1\dd\xi \frac{x}{\xi}q\left(\frac{x}{\xi}\right)\frac{1+\xi^2}{(1-\xi)_+} \frac{1}{\xi^2}\int \frac{\dd^2\rperp}{(2\pi)^2} e^{-i\frac{\qperp\cdot \rperp}{\xi} }S^{(2)}(\rperp)  \ln \frac{c_0^2}{\rperp^2\qperp^2}  ,\\
\frac{\dd \sigma_{qq}^{m}}{\dd \eta \dd^2 P_J}
= &\frac{\alpha_s}{2\pi}\Sperp C_F\int_\tau^1\frac{\dd z}{z^2}\mathcal{J}_q^{(0)}(z)\int_x^1\dd\xi \frac{x}{\xi}q\left(\frac{x}{\xi}\right)\frac{1}{\xi^2}\mathcal{P}_{qq}(\xi)\int \frac{\dd^2\rperp}{(2\pi)^2} e^{-i\frac{\qperp\cdot \rperp}{\xi}} S^{(2)}(\rperp)\ln\frac{1}{R^2}.
\end{align}
To compare our one-loop results with those in Ref.~\cite{Liu:2022ijp}, we need to set $q_\perp$ to $p_{J\perp}$. Firstly, the LO results $d\sigma^{(0)}$ in Ref.~\cite{Liu:2022ijp} is the same as our $\sigma_{qq}^{\text{LO}}$. Secondly, our results from the initial state gluon radiations agree with Eqs.(40) and (41) in Ref.~\cite{Liu:2022ijp}. The $\eta$-pole induced BK logarithmic term $\mathcal{H}_{q,\text{BK}}$ in their calculation is subtracted and put into the BK evolution equation. The remaining unresolved term $d\sigma_{R+V,soft}^{un-resolv.}$ or $\mathcal{H}_{q,kin.}$ coincides with our $\sigma_{qq}^{e}$ term which arises from the kinematic constraint. Therefore, our one-loop results are consistent with those in Ref.~\cite{Liu:2022ijp}. Nevertheless, as we will present in the later discussion, our resummation strategy is different.

Since the splitting function reads as $\mathcal{P}_{qq}(\xi)=\frac{1+\xi^2}{(1-\xi)_+}+\frac{3}{2}\delta(1-\xi)$, we rewrite $\sigma^h$ as
\begin{align}
\frac{\dd \sigma_{qq}^{h}}{\dd \eta \dd^2 P_J}
= & \frac{\alpha_s}{2\pi} \Sperp C_F\int_\tau^1\frac{\dd z}{z^2}\mathcal{J}_q^{(0)}(z)x q(x) \int_x^1\dd\xi \frac{x}{\xi}q\left(\frac{x}{\xi}\right) \frac{1}{\xi^2}\mathcal{P}_{qq}(\xi)\int \frac{\dd^2\rperp}{(2\pi)^2} e^{-i\frac{\qperp \cdot \rperp}{\xi}}S^{(2)}(\rperp)  \ln \frac{c_0^2}{\rperp^2\qperp^2}\notag \\
&- \frac{3}{2} \frac{\alpha_s}{2\pi} \Sperp C_F\int_\tau^1\frac{\dd z}{z^2}\mathcal{J}_q^{(0)}(z)x q(x)  \int \frac{\dd^2\rperp}{(2\pi)^2} e^{-i\qperp \cdot \rperp}S^{(2)}(\rperp)  \ln \frac{c_0^2}{\rperp^2\qperp^2}.
\end{align}
Note here the first term of the above equation should be combined with $\sigma^m$, therefore $\sigma^m$ becomes
\begin{align}
\frac{\dd \sigma_{qq}^{m'}}{\dd \eta \dd^2 P_J}
= &\frac{\alpha_s}{2\pi}\Sperp C_F\int_\tau^1\frac{\dd z}{z^2}\mathcal{J}_q^{(0)}(z)\int_x^1\dd\xi \frac{x}{\xi}q\left(\frac{x}{\xi}\right)\frac{1}{\xi^2}\mathcal{P}_{qq}(\xi)\int \frac{\dd^2\rperp}{(2\pi)^2} e^{-i\frac{\qperp\cdot \rperp}{\xi} } S^{(2)}(\rperp) \ln\frac{c_0^2}{\rperp^2\qperp^2R^2}.   
\end{align}

To summarize what we have done so far, let us compare our calculations to the cross-section of the hadron production but without the FFs~\cite{Chirilli:2012jd}. By comparison, we find an interesting relation between the cone size logarithm of the forward jet production and the collinear singularity in the hadron production
\begin{eqnarray}
\ln\frac{1}{R^2} \Leftrightarrow -\frac{1}{\epsilon}+\ln\frac{\qperp^2}{\mu^2}.
\label{relation}
\end{eqnarray}
The above replacement can be understood as follows: by taking the $R\to 0$ limit, and replacing the transverse momentum $\qperp$ by the transverse momentum of the produced parton $k_\perp$ for inclusive hadron productions, one can reproduce the collinear singularity associated with the final state gluon radiation. As a common practice in forward hadron production calculations, one usually remove such collinear singularities by redefining FFs. Therefore, we find the corresponding relationship between the inclusive jet production and the hadron production given by Eq.~(\ref{relation}). Moreover, note here our $\sigma_a$ and $\sigma_h$ originate from the initial state gluon radiation and the final state gluon radiation, respectively. Also, the sum of $\sigma_a$ and $\sigma_h$ are corresponds to $\sigma_a$ in the supplemental material of the Ref.~\cite{Shi:2021hwx}. $\sigma_b$ is the virtual contribution. $\sigma_c$ and $\sigma_d$ are the interference contributions. $\sigma_e$ is the additional term that comes from kinematic constraint correction~\cite{Watanabe:2015tja}. $\sigma_b$, $\sigma_c$, $\sigma_d$, and $\sigma_e$ are the same as these in Ref.~\cite{Shi:2021hwx}. Note here that $\sigma^f$ and $\sigma^g$ are unique which are from jet productions, there are no such corresponding terms in the hadron production case. Therefore, Eq.~(\ref{relation}) can also be the consistency check of our results. By using the same procedure above one can do the calculation of the other three channels accordingly. As we will see in the following calculation, the relation Eq.~(\ref{relation}) holds for other channels too. The above relation can help us to compare our jet calculation to the previous calculation for hadron productions.

\subsubsection{The complete one-loop cross-section in the momentum space}

This subsection is devoted to improve the accuracy of the numerical calculations. Since the phase factor $e^{-i\qperp\cdot \rperp}$ results in an oscillatory integral, it is well-known that numerical calculations are easier to carry out in the momentum space~\cite{Shi:2021hwx}. Therefore, we perform the Fourier transform and convert the cross-section into the momentum space in order to make it more suitable for numerical calculations. More detailed discussions of this problem and Fourier transform tricks can be found in Ref.~\cite{Shi:2021hwx}. After Fourier transformation, we get the cross-section in the momentum space as follows 
\begin{align}
\frac{\dd\sigma_{qq}}{\dd \eta \dd^2P_J} = 
\frac{\dd \sigma^{\rm LO}_{qq}}{\dd \eta\dd^{2}P_J} 
+ \frac{\dd \sigma^{\rm NLO}_{qq}}{\dd \eta\dd^{2}P_J} 
=
\frac{\dd \sigma^{\rm LO}_{qq}}{\dd \eta\dd^{2}P_J} 
+ \sum_{i=1}^{11} \frac{\dd \sigma^{i}_{qq}}{\dd \eta\dd^{2}P_J} .
\end{align}
The corresponding LO and NLO contributions are 
\begin{align}
\frac{\dd \sigma^{\rm LO}_{qq}}{\dd \eta\dd^2P_J} = & \Sperp   \int_\tau^1\frac{\dd z}{z^2}\mathcal{J}_q^{(0)}(z)xq(x,\mu^2)  F(\qperp),
\\
\frac{\dd\sigma_{qq}^1}{\dd \eta\dd^{2}P_J} 
=&
\frac{\alpha_{s}}{2\pi}C_{F}\Sperp\int_\tau^1\frac{\dd z}{z^2}\mathcal{J}_q^{(0)}(z)
\int_{x}^{1}\dd\xi
 \frac{x}{\xi} q\left(\frac{x}{\xi},\mu^{2}\right)
\mathcal{P}_{qq}(\xi) 
\ln\frac{\Lambda^{2}}{\mu^{2}}F(\qperp),
\label{eq:sigqq1fir}
 \\
\frac{\dd\sigma_{qq}^2}{\dd \eta\dd^{2}P_J} 
=&
\frac{3}{2}\frac{\alphas}{2\pi} C_{F} \Sperp\int_\tau^1\frac{\dd z}{z^2}\mathcal{J}_q^{(0)}(z)
x q(x,\mu^{2})
\ln\frac{\qperp^{2}}{\Lambda^2}F(\qperp),
\\
\frac{\dd\sigma_{qq}^3}{\dd \eta\dd^{2}P_J}
=& \frac{\alpha_{s}}{2\pi^2} C_F \Sperp \int_\tau^1\frac{\dd z}{z^2}\mathcal{J}_q^{(0)}(z)\int_{x}^{1}\dd\xi\int\dd^{2}\qa\dd^{2}\qb
 \frac{x}{\xi} q\left(\frac{x}{\xi},\mu^{2}\right)
\frac{1+\xi^2}{(1-\xi)_+} \mathcal{T}_{qq}^{(1)}(\xi,\qa,\qb,\qperp), 
\\
\frac{\dd\sigma_{qq}^4}{\dd \eta\dd^{2}P_J}
=& 
- \frac{\alpha_{s}}{\pi} C_F \Sperp\int_\tau^1\frac{\dd z}{z^2}\mathcal{J}_q^{(0)}(z)
\int_{0}^{1}\dd\xi'\int\dd^{2}\qa 
 x q(x,\mu^{2})
\frac{1+{\xi'}^{2}}{\left(1-\xi'\right)_+}\ln\frac{(\qa-\xi'\qperp)^{2}}{\qperp^{2}}
F(\qa)F(\qperp),  \label{sigmaqq4}
\\
\frac{\dd\sigma_{qq}^5}{\dd \eta\dd^{2}P_J}
=& 
\frac{2\alpha_s}{\pi^2} C_F \Sperp\int_\tau^1\frac{\dd z}{z^2}\mathcal{J}_q^{(0)}(z) 
 \int \dd^2 \qa x q(x,\mu^2) 
\frac{1}{\qa^2} \ln \frac{\qperp^2}{\qa^2} [F(\qperp-\qa) - \theta(\qperp^2-\qa^2) F(\qperp)] 
\nonumber\\
+ & 
\frac{\alpha_s}{\pi} C_F \Sperp \int_\tau^1\frac{\dd z}{z^2}\mathcal{J}_q^{(0)}(z)
 \int \dd^2 \qa x q(x,\mu^2) 
F(\qa) F(\qperp) \ln^2 \frac{\qperp^2}{(\qperp - \qa)^2} \nonumber \\
- & \frac{2\alpha_s}{\pi^2} C_F \Sperp \int_\tau^1\frac{\dd z}{z^2}\mathcal{J}_q^{(0)}(z)
 \int \dd^2 \qa \int \dd^2 \qb x q(x,\mu^2) 
F(\qa) F(\qb) \ln \frac{\qperp^2}{(\qperp-\qa)^2} \nonumber\\
& \times \frac{(\qperp-\qa)\cdot (\qperp-\qb)}{(\qperp-\qa)^2 (\qperp-\qb)^2} ,  \label{sigmaqq5}\\
\frac{\dd\sigma_{qq}^6}{\dd \eta\dd^{2}P_J} 
=&\frac{\alpha_s}{2\pi} \Sperp C_F\left(6-\frac{4}{3}\pi^2\right) \int_\tau^1\frac{\dd z}{z^2}\mathcal{J}_q(z)xq(x,\mu^2)  F(\qperp),\\
\frac{\dd\sigma_{qq}^7}{\dd \eta\dd^{2}P_J} 
=&-\frac{\alpha_s}{2\pi}\Sperp C_F\int_\tau^1\frac{\dd z}{z^2}\mathcal{J}_q^{(0)}(z)\int_x^1d\xi \frac{x}{\xi}q\left(\frac{x}{\xi}\right)\left[\frac{\ln\frac{(1-\xi)^2}{\xi^2}}{1-\xi}\right]_+F(\qperp/\xi),\\
\frac{\dd\sigma_{qq}^8}{\dd \eta\dd^{2}P_J} 
=&
\frac{3}{2}\frac{\alphas}{2\pi} C_{F} \Sperp\int_\tau^1\frac{\dd z}{z^2}\mathcal{J}_q^{(0)}(z)
x q(x,\mu^{2})
\ln\frac{\qperp^{2}}{\Lambda^2}F(\qperp),\\
\frac{\dd\sigma_{qq}^{9}}{\dd \eta\dd^{2}P_J} 
=&\frac{\alpha_s}{2\pi}\Sperp C_F\int_\tau^1\frac{\dd z}{z^2}\mathcal{J}_q^{(0)}(z)\int_x^1d\xi \frac{x}{\xi}q\left(\frac{x}{\xi},\mu^2\right)\frac{1}{\xi^2}\mathcal{P}_{qq}(\xi)\ln\frac{\Lambda^2}{\qperp^2R^2}F(\qperp/\xi), \label{fstate}\\
\frac{\dd\sigma_{qq}^{10}}{\dd \eta\dd^{2}P_J} 
=&
\frac{\alpha_{s}}{2\pi}C_{F}\Sperp\int_\tau^1\frac{\dd z}{z^2}\mathcal{J}_q^{(0)}(z)
\int_{x}^{1}\dd\xi
 \frac{x}{\xi} q\left(\frac{x}{\xi},\mu^{2}\right)
\left( 1-\xi \right) F(\qperp),
 \\
\frac{\dd\sigma_{qq}^{11}}{\dd \eta\dd^{2}P_J} 
=&
-\frac{1}{2}\frac{\alphas}{2\pi} C_{F} \Sperp\int_\tau^1\frac{\dd z}{z^2}\mathcal{J}_q^{(0)}(z)
x q(x,\mu^{2})
F(\qperp),
\end{align}
where 
$\mathcal{T}_{qq}^{(1)}(\xi,\qa,\qb,\qperp)$ can be found in Ref.~\cite{Shi:2021hwx}. It is given by
\begin{align}
\mathcal{T}_{qq}^{(1)}(\xi,\qa,\qb,\qperp)= 
& \frac{(\qb-\qa/\xi)^{2}}{(\qperp+\qa)^2 (\qperp/\xi+\qb)^2} F(\qa)F(\qb)
\nonumber\\
& - \frac{1}{(\qperp+\qa)^2} \frac{\Lambda^{2}}{\Lambda^{2}+(\qperp+\qa)^2}
F(\qb)F(\qperp)
\nonumber \\
& 
- \frac{1}{(\qperp +\xi\qb)^2 } \frac{\Lambda^{2}}{\Lambda^{2}+(\qperp/\xi+\qb)^2}
F(\qperp/\xi)F(\qa).
\end{align}
By splitting the $\theta$-function inside $\sigma_{qq}^5$ into two terms, we can rewrite it as
\begin{align}
\frac{\dd\sigma_{qq}^5}{\dd \eta\dd^{2}P_J}
=
\frac{\dd\sigma_{qq}^{5a}}{\dd \eta\dd^{2}P_J}
+
\frac{\dd\sigma_{qq}^{5b}}{\dd \eta\dd^{2}P_J},
\end{align}
where,
\begin{align}
\frac{\dd\sigma_{qq}^{5a}}{\dd \eta\dd^{2}P_J}
=& - 
\frac{\alpha_s}{2\pi} C_F \Sperp\int_\tau^1\frac{\dd z}{z^2}\mathcal{J}_q^{(0)}(z) 
 x q(x,\mu^2)  F(\qperp)
\ln^2 \frac{\qperp^2}{\Lambda^2},
\\
\frac{\dd\sigma_{qq}^{5b}}{\dd \eta\dd^{2}P_J}
=& 
\frac{2\alpha_s}{\pi^2} C_F \Sperp \int_\tau^1\frac{\dd z}{z^2}\mathcal{J}_q^{(0)}(z)
\int \dd^2 \qa x q(x,\mu^2) 
\frac{1}{\qa^2} \ln \frac{\qperp^2}{\qa^2} [F(\qperp-\qa) -  \theta(\Lambda^2-\qa^2) F(\qperp)] 
\nonumber\\
- &
\frac{\alpha_s}{2\pi} C_F \Sperp \int_\tau^1\frac{\dd z}{z^2}\mathcal{J}_q^{(0)}(z)
x q(x,\mu^2)  F(\qperp)
\ln^2 \frac{\qperp^2}{\Lambda^2}  \nonumber\\
+ & 
\frac{\alpha_s}{\pi} C_F \Sperp \int_\tau^1\frac{\dd z}{z^2}\mathcal{J}_q^{(0)}(z)
 \int \dd^2 \qa x q(x,\mu^2) 
F(\qa) F(\qperp) \ln^2 \frac{\qperp^2}{(\qperp - \qa)^2} \nonumber \\
- & \frac{2\alpha_s}{\pi^2} C_F \Sperp\int_\tau^1\frac{\dd z}{z^2}\mathcal{J}_q^{(0)}(z) 
 \int \dd^2 \qa \int \dd^2 \qb x q(x,\mu^2) 
F(\qa) F(\qb) \ln \frac{\qperp^2}{(\qperp-\qa)^2} \nonumber\\
& \times \frac{(\qperp-\qa)\cdot (\qperp-\qb)}{(\qperp-\qa)^2 (\qperp-\qb)^2} .
\end{align}
This trick~\cite{Xiao:2018zxf,Shi:2021hwx} allows us to extract the Sudakov double logarithm from $\sigma_{qq}^5$.

\subsection{The \texorpdfstring{$g\to g$}{g->g} channel}
\label{section32}

Since we have done the calculations for the $q\to q$ channel in the previous section and established the procedure for the jet calculation in CGC formalism, the computation for the $g\to g$ channel then becomes straightforward. The momentum constraints in Eq.~(\ref{constraint1}) and Eq.~(\ref{constraint2}) remain the same, and the NJA can also be applied throughout the following calculations. Notice that $\xi$ here represents the longitudinal momentum fraction of the parent gluon carried by the observed gluon. The partonic cross-section of $g\to gg$ channel has been studied in Ref.~\cite{Dominguez:2011wm}. It can be written as
\begin{eqnarray}
\frac{\dd\sigma_{gA\to ggX}}{\dd^3l\dd^3k  }&=& \alpha_sN_c\delta(q^+-l^+-k^+)\int
\frac{\text{d}^{2}x_{\perp}}{(2\pi)^{2}}\frac{\text{d}^{2}x_{\perp}^{\prime }%
}{(2\pi )^{2}}\frac{\text{d}^{2}b_{\perp}}{(2\pi)^{2}} \frac{\text{d}^{2}b_{\perp}'}{(2\pi)^{2}}  \notag \\
&&\times e^{-ik_{\perp }\cdot(x_{\perp}-x^{\prime
}_{\perp})}e^{-il_{\perp }\cdot(b_{\perp}-b^{\prime
}_{\perp})}\sum_{\lambda\alpha\beta}
\psi^{\lambda\ast}_{gg\alpha\beta}(u^{\prime}_{\perp})\psi^\lambda_{gg\alpha%
\beta}(u_{\perp})  \notag \\
&&\times   \frac{1}{N_c(N_c^2-1)}   \left[\left\langle f_{ade}\left[W(x_\perp)W^\dagger(x'_\perp)\right]^{db}\left[W(b_\perp)W^\dagger(b_\perp')\right]^{ec}f_{abc}\right\rangle_Y\right.\notag\\
&&  \quad-\left\langle f_{ade}W^{db}(x_\perp)W^{ec}(b_\perp)f_{fbc}W^{fa}(v'_\perp)\right\rangle_Y\notag\\
&&\quad-\left\langle f_{ade}W^{db}(x'_\perp)W^{ec}(b_\perp')f_{fbc}W^{fa}(v_\perp)\right\rangle_Y\notag\\
&&\quad\left.+ N_c \left\langle\text{Tr}W(v_\perp)W^\dagger(v'_\perp)\right\rangle_Y\right]
,  \label{partgg}
\end{eqnarray}
with $f_{abc}$ being the antisymmetric structure constant. $k$ and $l$ are the momenta of the final state observed and unobserved gluons, respectively. 
\begin{figure}[ht]
\includegraphics[width=4cm]{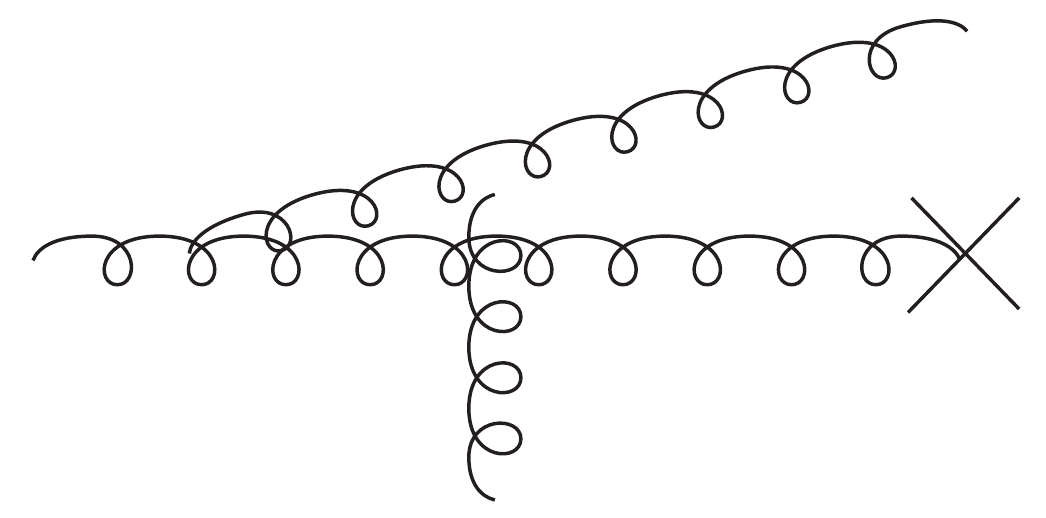}
\caption[*]{The Feynman diagram of the initial state gluon radiation for gluon jets.}
\label{ggginitial}
\end{figure}
The initial state gluon radiation of $g\to gg$ channel is depicted in Fig.~\ref{ggginitial}. The contribution which is proportional to $\left\langle f_{ade}\left[W(x_\perp)W^\dagger(x_\perp')\right]^{db}\left[W(b_\perp)W^\dagger(b_\perp')\right]^{ec}f_{abc}\right\rangle_Y$ in Eq.~(\ref{partgg}) stands for the initial state radiation. One can greatly simplify the multiple interaction factor when taking the large $N_c$ limit. Meanwhile, since we measure the jet which is initiated by the observed state gluon, one needs to integrate over the phase space of the unobserved gluon which leads to $b_\perp=b_\perp'$. Here $b_\perp$ and $b_\perp'$ are the transverse coordinates of the unobserved gluon in the amplitude and complex conjugate amplitude, respectively. Therefore, the multiple scattering factor $\left\langle f_{ade}\left[W(x_\perp)W^\dagger(x_\perp')\right]^{db}\left[W(b_\perp)W^\dagger(b_\perp')\right]^{ec}f_{abc}\right\rangle_Y$ is simplified to $ N_c^2 S^{(2)}_Y(x_\perp',x_\perp)S^{(2)}_Y(x_\perp,x_\perp')$. Then the contribution of the multiple interaction after the gluon splitting becomes
\begin{eqnarray}
\alpha_s
&&N_c\int^1_{x} \dd\xi \frac{x}{\xi}g\left(\frac{x}{\xi}\right)\int
\frac{\text{d}^{2}x_{\perp}}{(2\pi)^{2}}\frac{\text{d}^{2}x_{\perp}^{\prime }%
}{(2\pi )^{2}}\frac{\text{d}^{2}b_{\perp}}{(2\pi)^{2}}e^{-ik_{\perp }\cdot(x_{\perp}-x^{\prime
}_{\perp})}\sum_{\lambda\alpha\beta}
\psi^{\lambda\ast}_{gg\alpha\beta}(u^{\prime}_{\perp})\psi^\lambda_{gg\alpha%
\beta}(u_{\perp})S_Y^{(2)}(x_\perp,x_{\perp}^{\prime})S_Y^{(2)}(x_\perp^{%
\prime},x_{\perp}).
\end{eqnarray}
Note here the $g\to gg$ splitting kernel is found to be
\begin{equation}
\sum_{\lambda\alpha\beta} \psi^{\lambda\ast}_{gg\alpha\beta}(\xi,
u^{\prime}_{\perp})\psi^\lambda_{gg\alpha\beta}(\xi, u_{\perp})=4(2\pi)^2
\left[\frac{\xi}{1-\xi}+\frac{1-\xi}{\xi}+\xi(1-\xi)\right]\frac{1}{q^+}\frac{%
u_{\perp}^{\prime}\cdot u_{\perp}}{u_{\perp}^{\prime 2} u_{\perp}^{ 2}}.
\end{equation}
Fig.~\ref{gggfinal} shows the final state gluon radiation of  the $g\to gg$ channel. The correlator $ \left\langle\text{Tr}W(v_\perp)W^\dagger(v'_\perp)\right\rangle_Y$ in Eq.~(\ref{partgg}) arises from the fact that the multiple interaction takes place before the gluon splitting, and it can be simplified as $ N_c^2 S_Y^{(2)}(v_\perp,v_{\perp}^{\prime})S_Y^{(2)}(v_\perp^{\prime},v_{\perp})$ in the large $N_c$ limit.
\begin{figure}[ht]
\centering
\subfigure[]{\includegraphics[width=4cm]{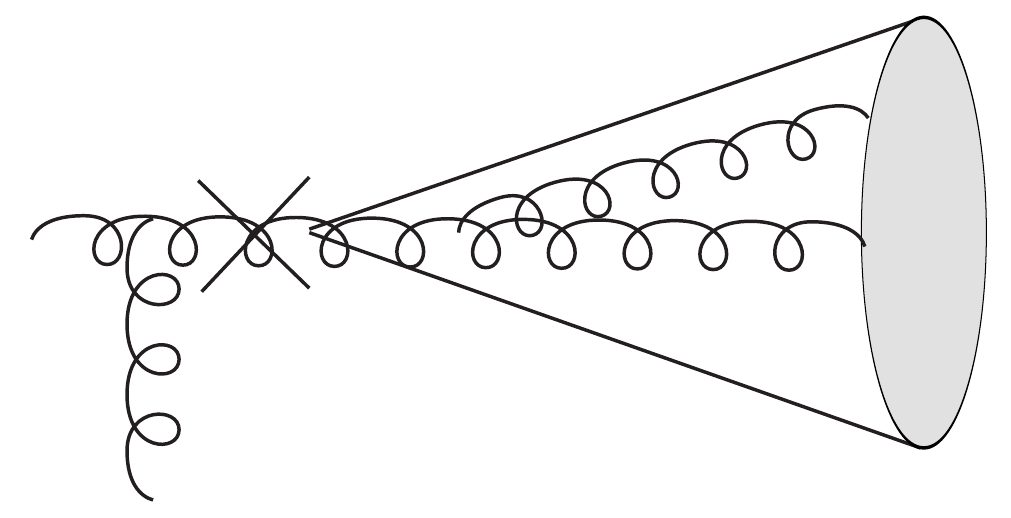}
}
~~\raisebox{28pt}[0pt][0pt]{--}~~
\subfigure[]{\includegraphics[width=4cm]{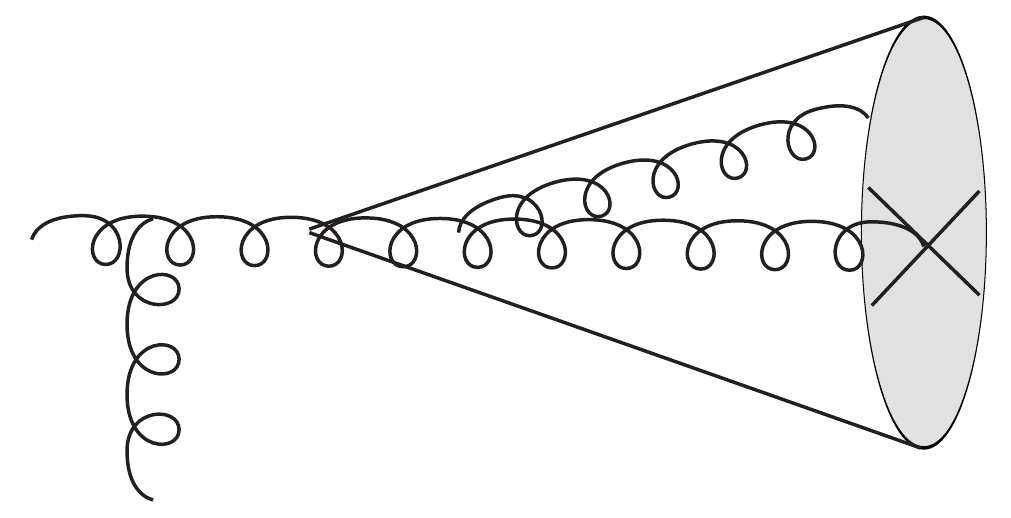}}
~~\raisebox{28pt}[0pt][0pt]{+}~~
\subfigure[]{\includegraphics[width=4cm]{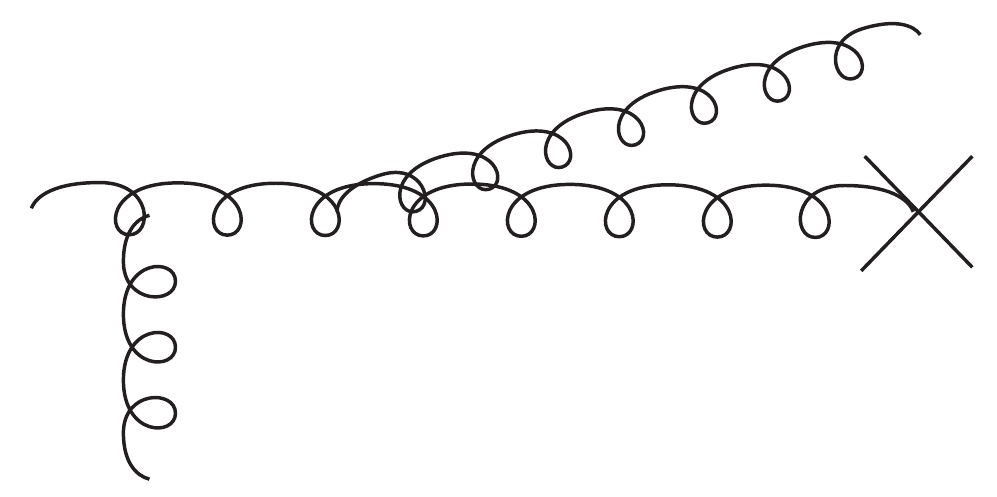}}
\caption[*]{The three real diagrams of gluon splittings contributing to the gluon jet.}
\label{gggfinal}
\end{figure}
Finally,  we get $\sigma_\text{a}$, $\sigma_\text{b}$ and $\sigma_\text{c}$ as follows
\begin{eqnarray}
\sigma_\text{a}&=&\frac{\alpha_s}{2\pi} N_cxg(x)\int
\frac{\text{d}^{2}v_{\perp}\text{d}^{2}v_{\perp}^{\prime }}{(2\pi)^{2}}e^{-iq_\perp\cdot(v_\perp-v_\perp')}\left[S_Y^{(2)}(v_\perp,v_\perp')\right]^2\notag\\
&&\times\left[\frac{1}{\epsilon^2}+\frac{11}{6\epsilon}-\frac{1}{\epsilon}\ln\frac{q_\perp^2R^2}{\mu^2}-\frac{11}{6}\ln\frac{q_\perp^2R^2}{\mu^2}+\frac{1}{2}\ln^2\frac{q_\perp^2R^2}{\mu^2}+\frac{67}{9}-\frac{3}{4}\pi^2   \right],\\
\sigma_\text{b}&=&\frac{\alpha_sN_c}{\pi}S_\perp\int\frac{\dd^2r_\perp}{(2\pi)^2}[S^{
(2)}(r_\perp)]^2\int_x^1\dd\xi\frac{[1-\xi(1-\xi)]^2}{\xi(1-\xi)_+}\frac{1}{\xi^{2}}\frac{x}{\xi}g\left(\frac{x}{\xi}\right)e^{-i\frac{q_\perp\cdot r_\perp}{\xi}}\left[-\frac{1}{\epsilon}-\ln\frac{\xi^2\mu^2}{q_\perp^2R^2}\right]\notag\\
&&+\frac{2\alpha_sN_c}{\pi}S_\perp\int\frac{\dd^2r_\perp}{(2\pi)^2}[S(r_\perp)]^2\int_x^1\dd\xi\left[\frac{\ln(1-\xi)}{(1-\xi)}\right]_+\frac{[1-\xi(1-\xi)]^2}{\xi^3}\frac{x}{\xi}g\left(\frac{x}{\xi}\right)e^{-i\frac{q_\perp\cdot r_\perp}{\xi}}\notag\\
&&+\frac{\alpha_sN_c}{2\pi}S_\perp xg(x)\int\frac{\dd^2r_\perp}{(2\pi)^2}[S^{
(2)}(r_\perp)]^2e^{-iq_\perp\cdot r_\perp}\left[\frac{1}{\epsilon^2}-\frac{1}{\epsilon}\ln\frac{q_\perp^2R^2}{\mu^2}-\frac{\pi^2}{12}+\frac{1}{2}\ln^2\frac{q_\perp^2R^2}{\mu^2}\right], \\
\sigma_\text{c}&=&\frac{\alpha_sN_C}{\pi}S_\perp\int\frac{\dd^2r_\perp}{(2\pi)^2}e^{-i\frac{q_\perp\cdot r_\perp}{\xi}}[S(r_\perp)]^2\int_x^1\dd\xi\frac{[1-\xi(1-\xi)]^2}{\xi^3(1-\xi)_+}\frac{x}{\xi}g\left(\frac{x}{\xi}\right)\left[-\frac{1}{\epsilon}+\ln\frac{c_0^2}{\mu^2r_\perp^2}\right].
\end{eqnarray}

Furthermore, we should also consider the $g\to q\bar{q}$ splitting when the final state particles are in the same jet cone. The in-cone contribution from the $g\to q\bar{q}$ channel is shown in Fig.~\ref{gqqfinal} and we label it as $\sigma_\text{d}$. The partonic cross-section of the $g\to q$ channel can be found in Sec.\ref{section34}. The second term in the  bracket of the Eq.~(\ref{xsgqqbar}) is the corresponding final state radiation contribution. Under the large $N_c$ limit approximation, the correlators of the multiple interaction before the gluon splitting can be expressed entirely in terms of 2-point functions $S^A_Y(v_\perp,v_\perp')\simeq S^{(2)}_Y(v_\perp,v_\perp')S^{(2)}_Y(v_\perp',v_\perp)$.
\begin{figure}[ht]
\centering
\subfigure{
\includegraphics[width=4.5cm]{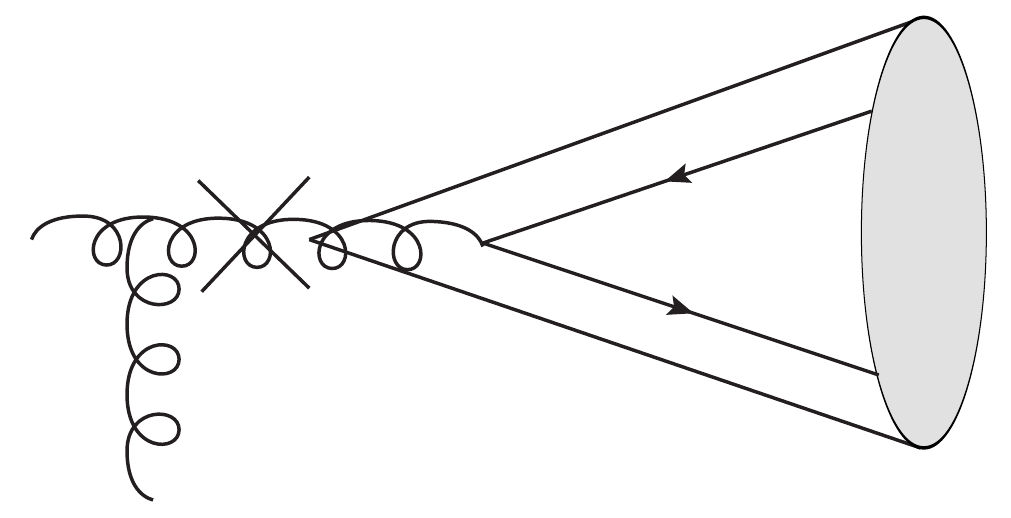}
}
\caption{The real diagram of the $g\to q\bar{q}$ splitting with $q$ and $\bar{q}$ being inside the jet cone. According to the jet definition and measurement, we categorize this contribution as a part of the gluon jet production.}
\label{gqqfinal}
\end{figure}
Then the cross-section as shown in Fig.~\ref{gqqfinal} becomes
\begin{eqnarray}
\sigma_\text{d}&=&\alpha_sT_R\delta(q^+-l^+-k^+)xg(x)\int\frac{\dd^2x_\perp}{(2\pi)^2}\frac{\dd^2x_\perp'}{(2\pi)^2}\frac{\dd^2b_\perp}{(2\pi)^2}\frac{\dd^2b_\perp'}{(2\pi)^2}e^{-ik_\perp\cdot(x_\perp-x_\perp')}e^{-il_\perp\cdot(b_\perp-b_\perp')}\notag\\
&&\times\sum_{\lambda\alpha\beta}\psi_{\alpha\beta}^{\lambda*}(u_\perp')\psi_{\alpha\beta}^\lambda(u_\perp)S_Y^{(2)}(v_\perp,v_\perp')S_Y^{(2)}(v_\perp',v_\perp),
\end{eqnarray} 
where $T_{R}=\frac{1}{2}$, and $k$ and $l$ are the momenta of quark and anti-quark, respectively. 
We have  adopted the $D$-dimensional splitting function of the $g\rightarrow q \bar q$ channel from Ref.~\cite{Catani:1998nv}, and modified it to the splitting kernel shown below
\begin{equation}
\sum_{\lambda \alpha \beta }\psi _{q\bar{q}\alpha \beta }^{\lambda \ast
}(q^{+},\xi ,u_{\perp })\psi _{q\bar{q}\alpha \beta }^{\lambda }(q^{+},\xi
,u_{\perp })=2(2\pi )^{2}\left[ \xi ^{2}+(1-\xi )^{2}  -2 \epsilon\xi (1-\xi) \right] \frac{1}{%
u_{\perp }^{2}}.
\end{equation}
This gives 
\begin{eqnarray}
\sigma_\text{d}=\frac{\alpha_s}{2\pi}N_fT_Rxg(x)\int
\frac{\text{d}^{2}v_{\perp}\text{d}^{2}v_{\perp}^{\prime }}{(2\pi)^{2}}e^{-iq_\perp\cdot(v_\perp-v_\perp')}\left[S_Y^{(2)}(v_\perp,v_\perp')\right]^2\left[-\frac{2}{3\epsilon}-\frac{26}{9}-\frac{2}{3}\ln\frac{\mu^2}{q_\perp^2R^2}  +\frac{1}{3} \right].
\end{eqnarray}
The last constant term ($\frac{1}{3}$) inside the square brackets in the above equation originates from the product of the $\epsilon$ term in the splitting kernel and the $\frac{1}{\epsilon}$ pole due to collinear divergence. There is a similar term from the virtual contribution of the $g\rightarrow q \bar q$ channel. These two term will cancel each other.

For the virtual contributions, we consider the virtual gluon loop diagrams in the Fig.~\ref{gluon-loop}. The partonic cross-section is given by
\begin{figure}[ht]
\centering
\subfigure[]{
\includegraphics[width=5cm]{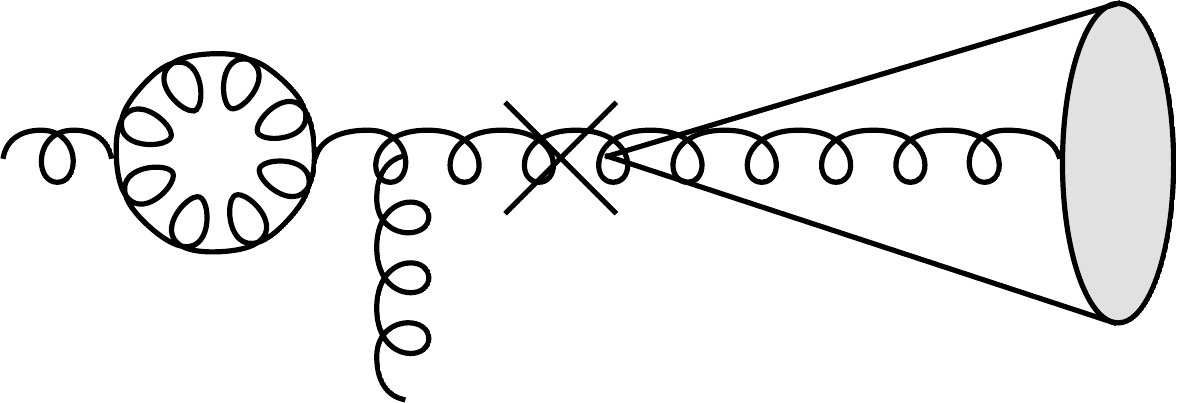}
}
\subfigure[]{
\includegraphics[width=5cm]{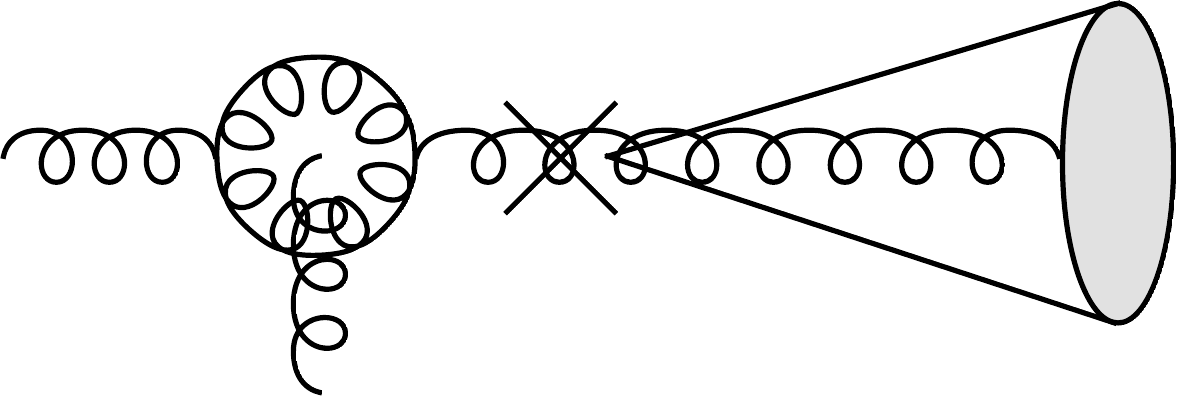}
}
\subfigure[]{
\includegraphics[width=5cm]{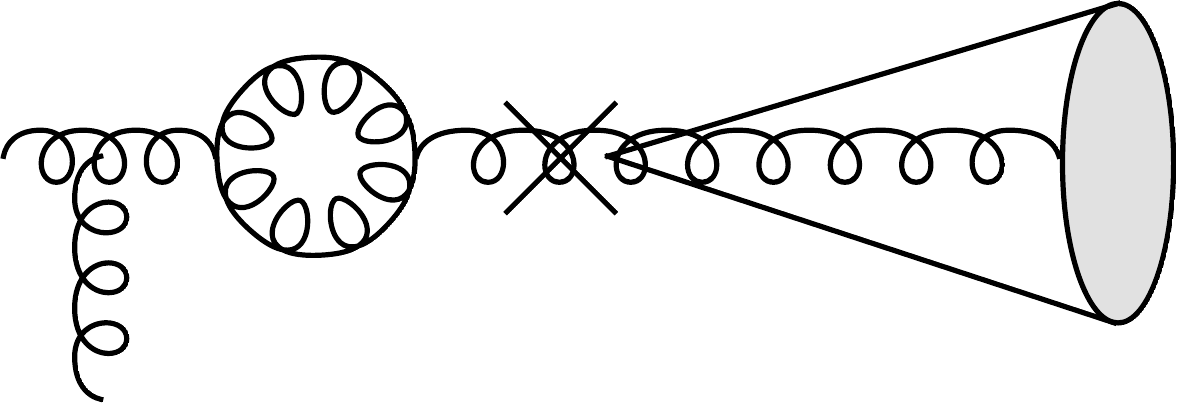}
}
\caption[*]{The virtual gluon loop diagrams of the $g\to g$ channel.}
\label{gluon-loop}
\end{figure}
\begin{eqnarray}
&&-\alpha _{s}N_{c}xg(x)\int_{0}^{1}\dd\xi \int \frac{\dd^{2}v_{\perp }}{(2\pi
)^{2}}\frac{\dd^{2}v_{\perp }^{\prime }}{(2\pi )^{2}}\frac{\dd^{2}u_{\perp }}{%
(2\pi )^{2}} e^{-ik_{\perp }\cdot (v_{\perp }-v_{\perp }^{\prime
})}\sum_{\lambda \alpha \beta }\psi _{gg\alpha \beta }^{\lambda \ast
}(q^{+},\xi ,u_{\perp })\psi _{gg\alpha \beta }^{\lambda }(q^{+},\xi
,u_{\perp }) \notag\\
&&\times \left[ S_Y^{(2)}(v_{\perp },v_{\perp }^{\prime })S_Y^{(2)}(v_{\perp
}^{\prime },v_{\perp })-S_Y^{(2)}(b_{\perp },x_{\perp })S_Y^{(2)}(x_{\perp },v_{\perp }^{\prime
})S_Y^{(2)}(v_{\perp }^{\prime },b_{\perp })%
\right] .\label{v2}
\end{eqnarray}%
With dimensional regularization and the $\overline{\text{MS}}$ scheme, we can perform the rest of the calculation directly and get
\begin{eqnarray}
\sigma_{gg}^{\text{virt}(\text{jet})}=\frac{11N_c}{6}\frac{\alpha_s}{2\pi}S_\perp xg(x)\int_\tau^1\frac{\dd z}{z^2}\mathcal{J}_g^{(0)}(z)\int\frac{\dd^2r_\perp}{(2\pi)^2}e^{-i q_\perp\cdot r_\perp }[S(r_\perp)]^2\left[-\frac{1}{\epsilon}+\ln\frac{q_\perp^2}{\mu^2}\right].
\end{eqnarray}
Furthermore, we  also compute the quark loop virtual contributions depicted in Fig.~\ref{quark-loop}, its contribution is given by 
\begin{figure}[ht]
\centering
\subfigure[]{
\includegraphics[width=5cm]{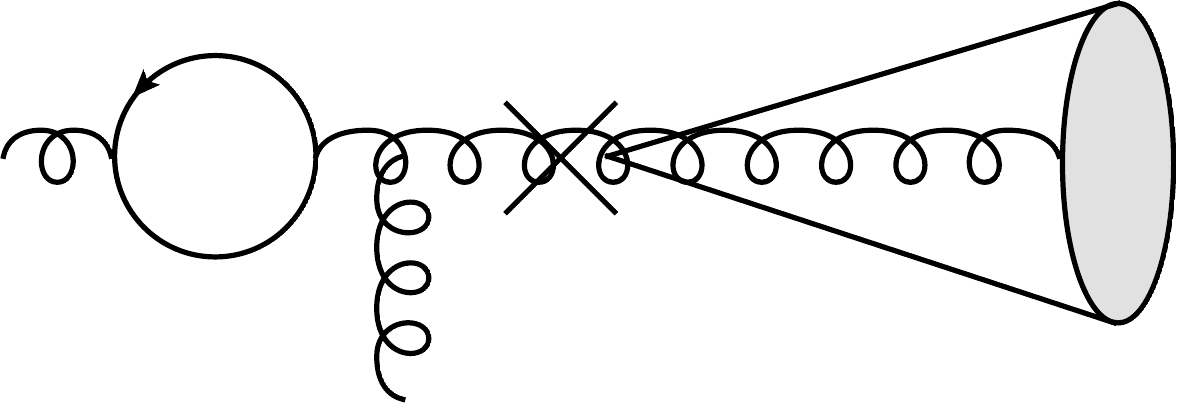}
}
\subfigure[]{
\includegraphics[width=5cm]{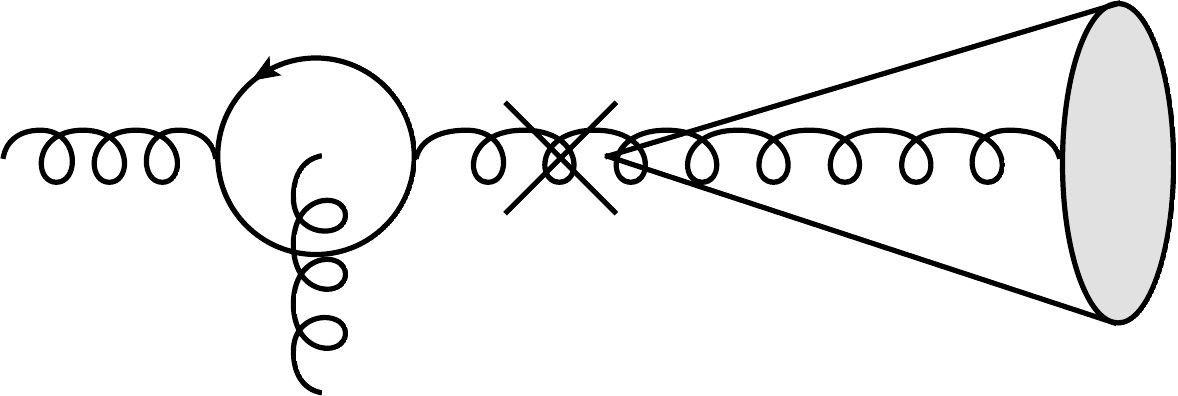}
}
\subfigure[]{
\includegraphics[width=5cm]{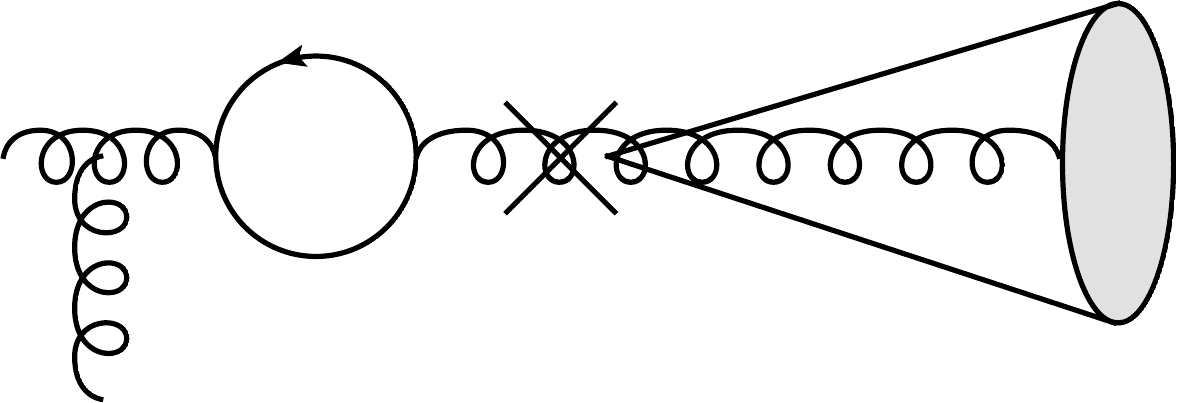}
}
\caption[*]{The virtual quark loop diagrams in the $g\to g$ channel.}
\label{quark-loop}
\end{figure}
\begin{eqnarray}
&&-2\alpha_sN_fT_Rxg(x)\int_0^1\dd\xi\int\frac{\dd^2u_\perp}{(2\pi)^2}\frac{\dd^2v_\perp}{(2\pi)^2}\frac{\dd^2v_\perp'}{(2\pi)^2}e^{-ik_\perp\cdot(v_\perp-v_\perp')}\notag\\
&&\times\sum_{\lambda\alpha\beta}\psi_{q\bar{q}\alpha\beta}^{\lambda*}(u_\perp)\psi_{q\bar{q}\alpha\beta}^\lambda(u_\perp)\left[S_Y^{(2)}(v_\perp,v_\perp')S_Y^{(2)}(v_\perp',v_\perp)-S_Y^{(2)}(x_\perp,v_\perp')S_Y^{(2)}(v_\perp',b_\perp)\right].
\end{eqnarray}
After the evaluation, we arrive at 
\begin{eqnarray}
\sigma^{\text{virt}(\text{jet})}_{q \bar q}=-\frac{2N_fT_R}{3}\frac{\alpha_s}{2\pi}S_\perp xg(x)\int_\tau^1\frac{\dd z}{z^2}\mathcal{J}_g^{(0)}(z)\int\frac{\dd^2r_\perp}{(2\pi)^2}e^{-i q_\perp\cdot r_\perp }[S(r_\perp)]^2\left[-\frac{1}{\epsilon}+\ln\frac{q_\perp^2}{\mu^2} +\frac{1}{2} \right].
\end{eqnarray}
Similar to the $q\to q$ channel, we only pick out part of virtual contributions to cancel the collinear singularities. The rest virtual contributions would be combined with other real diagrams. The final state contribution $\sigma_{gg}^{\text{final}}$ can be obtained by adding $\sigma_\text{a}$, $\sigma_\text{b}$, $\sigma_\text{c}$ , $\sigma_\text{d}$, $\sigma_{gg}^{\text{virt}(\text{jet})}$ and $\sigma_{q\bar{q}}^{\text{virt}(\text{jet})}$     together
\begin{eqnarray}
\sigma_{gg}^{\text{final}}=\sigma_\text{a}+(\sigma_\text{c}-\sigma_\text{b})+\sigma_\text{d}+\sigma_{gg}^{\text{virt}(\text{jet})}+\sigma_{q\bar{q}}^{\text{virt}(\text{jet})}. 
\end{eqnarray}
In the end, all the divergences cancel. With the CJF $\mathcal{J}_g(z)$, we get 
\begin{eqnarray}
\frac{\dd^3\sigma_{gg}^{\text{final}}}{\dd\eta \dd^2P_J}&=&-\frac{\alpha_sN_c}{2\pi}\int_x^1d\xi\frac{1}{\xi^2}\sigma_{\text{LO}}\left(\frac{x}{\xi},\frac{q_\perp}{\xi}\right)\left[2\left(\frac{\ln\frac{(1-\xi)^2}{\xi^2}}{(1-\xi)}\right)_+\frac{[1-\xi(1-\xi)]^2}{\xi}-2\frac{[1-\xi(1-\xi)]^2}{\xi(1-\xi)_+}\ln\frac{c_0^2}{q_\perp^2r_\perp^2}-\mathcal{P}_{gg}(\xi)\ln\frac{1}{R^2}\right]\notag\\
&&+ \frac{\alpha_s}{2\pi}\left[\left(\frac{67}{9}-\frac{4}{3}\pi^2  \right)N_c -  \frac{26}{9}N_f T_R\right]\sigma(x,q_\perp).
\end{eqnarray}
Note here we have defined $\sigma_{\text{LO}}\left(\frac{x}{\xi},\frac{q_\perp}{\xi}\right)=S_\perp\int_\tau^1\frac{\dd z}{z^2}\mathcal{J}_g^{(0)}(z)\frac{x}{\xi} g\left(\frac{x}{\xi}\right)  \int \frac{\dd^2\rperp}{(2\pi)^2} e^{-i\frac{\qperp\cdot \rperp}{\xi}} S^{(2)} (\rperp) S^{(2)} (\rperp)$. By combining the collinear singularities from both real and virtual diagrams of the initial state radiation, we find the coefficient of the collinear singularities becomes 
	\begin{equation}
	-N_c\frac{1}{2\pi}\frac{1}{\epsilon}\int_x^1\dd\xi\left[2\frac{[1-\xi(1-\xi)]^2}{\xi(1-\xi)_+}+\left(\frac{11}{6}-\frac{2N_fT_R}{3N_c}\right)\delta(1-\xi)\right]\frac{x}{\xi} g\left(\frac{x}{\xi}\right)=-N_c\frac{1}{2\pi}\frac{1}{\epsilon}\int_{x}^1\dd\xi\mathcal{P}_{gg}(\xi)\frac{x}{\xi} g\left(\frac{x}{\xi}\right),
	\end{equation}
where we have used the delta function again to change the second term in the left hand side. $\mathcal{P}_{gg}(\xi)$ is defined as
	\begin{equation}
		\mathcal{P}_{gg}(\xi)=2\frac{[1-\xi(1-\xi)]^2}{\xi(1-\xi)_+}+2\beta_0\delta(1-\xi),
	\end{equation} 
where $\beta_0=\frac{11}{12}-\frac{N_fT_R}{3N_c}$. As usual, we remove the collinear singularities by redefining the gluon distribution as follows
	\begin{equation}
		g(x,\mu)=g^{(0)}(x)-\frac{1}{\epsilon}\frac{\alpha_s(\mu)}{2\pi}\int_x^1\frac{\dd\xi}{\xi}N_c\mathcal{P}_{gg}(\xi)g\Big(\frac{x}{\xi}\Big).
	\end{equation}
Once we remove all the divergences, the final cross-section is finite. Similarly, the next-to-leading order cross-section for the $g\to g$ channel in the coordinate space can be expressed as 
\begin{align}
\frac{\dd \sigma_{gg}}{\dd \eta \dd^2 P_J}
= 
\frac{\dd \sigma_{gg}^{\rm LO}}{\dd \eta \dd^2 P_J}
+
\frac{\dd \sigma_{gg}^{\rm NLO}}{\dd \eta \dd^2 P_J}
= 
\frac{\dd \sigma_{gg}^{\rm LO}}{\dd \eta \dd^2 P_J}
+
\sum_{i=a}^n
\frac{\dd \sigma_{gg}^{i}}{\dd \eta \dd^2 P_J},
\end{align}
where the LO and NLO cross-section are given by

\begin{align}
\frac{\dd \sigma_{gg}^{\rm LO}}{\dd \eta \dd^2 P_J}
= &
\Sperp \int_\tau^1\frac{\dd z}{z^2}\mathcal{J}_g^{(0)}(z)x g(x,\mu^2)  \int \frac{\dd^2\rperp}{(2\pi)^2} e^{-i\qperp\cdot \rperp} S^{(2)} (\rperp) S^{(2)} (\rperp),
\\
\frac{\dd \sigma_{gg}^a}{\dd \eta \dd^2 P_J}
= &
\frac{\alpha_s}{2\pi} \Sperp N_c \int_\tau^1\frac{\dd z}{z^2}\mathcal{J}_g^{(0)}(z) \int_{x}^1 \dd\xi \frac{x}{\xi} g\left(\frac{x}{\xi},\mu^2\right)  \mathcal{P}_{gg} (\xi) \int \frac{\dd^2\rperp}{(2\pi)^2} S^{(2)} (\rperp) S^{(2)} (\rperp)
 e^{-i\qperp\cdot \rperp}\ln \frac{c_0^2}{\rperp^2\mu^2},
\\
\frac{\dd \sigma_{gg}^b}{\dd \eta \dd^2 P_J}
= &
-  2\beta_0\frac{\alpha_s}{2\pi}  \Sperp N_c \int_\tau^1\frac{\dd z}{z^2}\mathcal{J}_g^{(0)}(z) x g(x,\mu^2) \int \frac{\dd^2\rperp}{(2\pi)^2} e^{-i\qperp\cdot \rperp} S^{(2)} (\rperp) S^{(2)} (\rperp)
\ln \frac{c_0^2}{\rperp^2 \qperp^2} ,
\\
\frac{\dd \sigma_{gg}^c}{\dd \eta \dd^2 P_J}
= &
8\pi \Sperp N_f T_R \frac{\alpha_s}{2\pi}  \int_\tau^1\frac{\dd z}{z^2}\mathcal{J}_g^{(0)}(z)x g(x,\mu^2) \int \frac{\dd^2\uperp\dd^2\vperp}{(2\pi)^4} e^{-i\qperp\cdot\vperp} S^{(2)} (\uperp) S^{(2)} (\vperp)
\nonumber\\
& \times \int_0^1 \dd\xi' [\xi'^2 + (1-\xi')^2] \left[ \frac{e^{-i\xi' \qperp\cdot (\uperp-\vperp)}}{(\uperp-\vperp)^2} -\delta^2 (\uperp-\vperp) \int \dd^2 \rperp' \frac{e^{i\qperp\cdot \rperp' }}{\rperp'^2 } \right],
\\
\frac{\dd \sigma_{gg}^d}{\dd \eta \dd^2 P_J}
= & 
-16\pi \Sperp N_c \frac{\alpha_s}{2\pi} \int_\tau^1\frac{\dd z}{z^2}\mathcal{J}_g^{(0)}(z)
\int_{x}^1 \dd\xi\frac{x}{\xi} g\left(\frac{x}{\xi},\mu^2\right)  \int\frac{\dd^2\uperp\dd^2\vperp}{(2\pi)^4} 
e^{-i\qperp\cdot(\uperp-\vperp)} e^{-i\frac{\qperp\cdot\vperp}{\xi}} 
\nonumber\\
&
\times S^{(2)} (\uperp) S^{(2)} (\vperp) S^{(2)} (\uperp-\vperp) \frac{[1-\xi(1-\xi)]^2}{(1-\xi)_+}
\frac{1}{\xi^2} \frac{(\uperp-\vperp) \cdot\vperp}{(\uperp-\vperp)^2\vperp^2},
\\
\frac{\dd \sigma_{gg}^e}{\dd \eta \dd^2 P_J}
= & 
16\pi \Sperp N_c \frac{\alpha_s}{2\pi}  \int_\tau^1\frac{\dd z}{z^2}\mathcal{J}_g^{(0)}(z)x g(x,\mu^2)  \int\frac{\dd^2\uperp\dd^2\vperp}{(2\pi)^4} S^{(2)} (\uperp) S^{(2)} (\vperp) S^{(2)} (\uperp-\vperp)
\nonumber\\
&
\times e^{-i\qperp\cdot(\uperp-\vperp)} \int_0^1 \dd\xi' \left[ \frac{\xi'}{(1-\xi')_+} + \frac{\xi' (1-\xi')}{2} \right]
\left[ \frac{e^{-i\xi'\qperp\cdot\vperp}}{\vperp^2} - \delta^2 (\vperp) \int \dd^2 \rperp' \frac{e^{i\qperp\cdot \rperp' }}{\rperp'^2 }  \right],
\\
\frac{\dd \sigma_{gg}^f}{\dd \eta \dd^2 P_J}
= & 
\frac{\alpha_s}{\pi^2} \Sperp N_c  \int_\tau^1\frac{\dd z}{z^2}\mathcal{J}_g^{(0)}(z)x g(x) \int \frac{\dd^2\uperp\dd^2\vperp}{(2\pi)^2} e^{-i\qperp\cdot(\uperp-\vperp)} [S^{(2)}(\uperp)S^{(2)}(\vperp)-S^{(2)}(\uperp-\vperp)] 
\nonumber\\
& 
\times S^{(2)}(\uperp-\vperp) 
\Bigg[ 
  \frac{1}{\uperp^2} \ln \frac{\qperp^2\uperp^2}{c_0^2} 
+ \frac{1}{\vperp^2} \ln \frac{\qperp^2\vperp^2}{c_0^2}- \frac{2\uperp\cdot \vperp}{\uperp^2 \vperp^2} \ln \frac{\qperp^2 |\uperp||\vperp|}{c_0^2}
\Bigg],\\
 \frac{\dd \sigma_{gg}^{g}}{\dd \eta \dd^2 P_J}
= &
 \frac{\alpha_s}{2\pi}\Sperp N_c\left(\frac{67}{9}-\frac{4}{3}\pi^2  \right) \int_\tau^1\frac{\dd z}{z^2}\mathcal{J}_g(z)x g(x)  \int \frac{\dd^2\rperp}{(2\pi)^2} e^{-i\qperp\cdot \rperp} S^{(2)} (\rperp) S^{(2)} (\rperp)\nonumber \\
 & - \frac{\alpha_s}{2\pi}\Sperp N_f T_R \frac{26}{9}  \int_\tau^1\frac{\dd z}{z^2}\mathcal{J}_g^{(0)}(z)x g(x)  \int \frac{\dd^2\rperp}{(2\pi)^2} e^{-i\qperp\cdot \rperp} S^{(2)} (\rperp) S^{(2)} (\rperp),
\\
\frac{\dd \sigma_{gg}^h}{\dd \eta \dd^2 P_J}
= &
-\frac{\alpha_s}{2\pi} \Sperp N_c  \int_\tau^1\frac{\dd z}{z^2}\mathcal{J}_g^{(0)}(z)\int_{x}^1 \dd\xi \frac{x}{\xi} g\left(\frac{x}{\xi}\right)  \left[\frac{\ln\frac{(1-\xi)^2}{\xi^2}}{1-\xi}\right]_+ \frac{2[1-\xi(1-\xi)]^2}{\xi}\frac{1}{\xi^2}\int \frac{\dd^2\rperp}{(2\pi)^2} S^{(2)} (\rperp) S^{(2)} (\rperp)
  e^{-i\frac{\qperp\cdot \rperp}{\xi} },\\
 \frac{\dd \sigma_{gg}^{m}}{\dd \eta \dd^2 P_J}
= &
-2\beta_0\frac{\alpha_s}{2\pi} \Sperp N_c\int_\tau^1\frac{\dd z}{z^2}\mathcal{J}_g^{(0)}(z)x g(x)  \int \frac{\dd^2\rperp}{(2\pi)^2} e^{-i \qperp \cdot \rperp} S^{(2)} (\rperp) S^{(2)} (\rperp)\ln \frac{c_0^2}{\rperp^2\qperp^2},\\
\frac{\dd \sigma_{gg}^n}{\dd \eta \dd^2 \qperp}
= &
\frac{\alpha_s}{2\pi} \Sperp N_c  \int_\tau^1\frac{\dd z}{z^2}\mathcal{J}_g^{(0)}(z)\int_{x}^1 \dd\xi \frac{x}{\xi} g\left(\frac{x}{\xi}\right)  \mathcal{P}_{gg} (\xi) \int \frac{\dd^2\rperp}{(2\pi)^2} S^{(2)} (\rperp) S^{(2)} (\rperp)
 \frac{1}{\xi^2} e^{-i\frac{\qperp\cdot \rperp}{\xi}  }\ln\frac{c_0^2}{\rperp^2\qperp^2R^2},\\
 \frac{\dd \sigma_{gg}^o}{\dd \eta \dd^2 P_J}
= &
-\frac{1}{3} \frac{\alpha_s}{2\pi} N_f T_R \Sperp  \int_\tau^1\frac{\dd z}{z^2}\mathcal{J}_g^{(0)}(z) x g(x,\mu^2) \int \frac{\dd^2\rperp}{(2\pi)^2} e^{-i\qperp\cdot \rperp} S^{(2)} (\rperp) S^{(2)} (\rperp).
\end{align}
Due to the reasons discussed in the last subsection, we also need to Fourier transform the above equations to the momentum space analytically. The cross-section in the momentum space is given by
\begin{equation}
\frac{\dd \sigma_{gg}}{\dd \eta\dd^{2}P_J}
=
\frac{\dd \sigma^\text{LO}_{gg}}{\dd \eta\dd^{2}P_J}
+
\frac{\dd\sigma^\text{NLO}_{gg}}{\dd \eta\dd^{2}P_J}
=
\frac{\dd \sigma^\text{LO}_{gg}}{\dd \eta\dd^{2}P_J}
+
\sum_{i=1}^{11}
\frac{\dd\sigma_{gg}^i}{\dd \eta\dd^{2}P_J},
\end{equation}
with 
\begin{align}
\frac{\dd \sigma^\text{LO}_{gg}}{\dd \eta\dd^{2}P_J}
= &
\Sperp   \int_\tau^1\frac{\dd z}{z^2}\mathcal{J}_g^{(0)}(z)x g(x,\mu^{2})  \int \dd^2 \qa F(\qa) F(\qperp-\qa),
\\
\frac{\dd\sigma_{gg}^1}{\dd \eta\dd^{2}P_J}
=
&
\frac{\alpha_s}{2\pi} N_c \Sperp  \int_\tau^1\frac{\dd z}{z^2}\mathcal{J}_g^{(0)}(z)\int_{x}^1 d\xi \int \dd^2 \qa \frac{x}{\xi} g\left(\frac{x}{\xi},\mu^2\right)  \mathcal{P}_{gg} (\xi) 
\ln\frac{\Lambda^2}{\mu^2} 
F(\qperp-\qa)F(\qa),\label{ggistate}
\\
\frac{\dd\sigma_{gg}^2}{\dd \eta\dd^{2}P_J}
=
&
2\beta_0\frac{\alpha_s}{2\pi} N_c \Sperp \int_\tau^1\frac{\dd z}{z^2}\mathcal{J}_g^{(0)}(z)x g\left(x,\mu^2\right)
 \int \dd^2 \qa   \
F(\qperp-\qa) F(\qa)
 \ln \frac{\qperp^2}{\Lambda^2},
\\
\frac{\dd\sigma_{gg}^3}{\dd \eta\dd^{2}P_J}
=
&
-\frac{1}{3} \frac{\alpha_s}{2\pi} N_f T_R \Sperp  \int_\tau^1\frac{\dd z}{z^2}\mathcal{J}_g^{(0)}(z)x g\left(x,\mu^2\right)
 \int \dd^2 \qa   \
F(\qperp-\qa) F(\qa),
\\
\frac{\dd\sigma_{gg}^4}{\dd \eta\dd^{2}P_J}
=
&
\frac{\alpha_s}{\pi^2} N_c \Sperp\int_\tau^1\frac{\dd z}{z^2}\mathcal{J}_g^{(0)}(z)
\int_{x}^1 \dd \xi \frac{x}{\xi} g\left(\frac{x}{\xi},\mu^2\right) \frac{[1-\xi(1-\xi)]^2}{\xi(1-\xi)_+}\int \dd^2 \qa \dd^2 \qb \dd^2 \qc  
 \mathcal{T}_{gg}^{(1)} (\xi,\qa,\qb,\qc,\qperp),
\\
\frac{\dd\sigma_{gg}^5}{\dd \eta\dd^{2}P_J}
=
& -
2N_f T_R \frac{\alpha_s}{2\pi} \Sperp\int_\tau^1\frac{\dd z}{z^2}\mathcal{J}_g^{(0)}(z) x g(x,\mu^2) \int_0^1 \dd\xi' \dd^2 \qa  [\xi'^2 + (1-\xi')^2] 
F(\qa) F(\qperp-\qa) \ln \frac{(\qa-\xi'\qperp)^2}{\qperp^2},
\\
\frac{\dd\sigma_{gg}^6}{\dd \eta\dd^{2}P_J}
=
&
- 4N_c \frac{\alpha_s}{2\pi} \Sperp \int_\tau^1\frac{\dd z}{z^2}\mathcal{J}_g^{(0)}(z)x g(x,\mu^2)\int_0^1 \dd\xi' \dd^2\qa \dd^2\qb   \left[ \frac{\xi'}{(1-\xi')_+} + \frac{1}{2} \xi' (1-\xi') \right] 
\nonumber\\
& 
\times F(\qa) F(\qb) F(\qperp-\qa) \ln \frac{(\qa+\qb-\xi'\qperp)^2}{\qperp^2}, 
\\
\frac{\dd\sigma_{gg}^7}{\dd \eta\dd^{2}P_J}
=& 
\frac{2\alpha_s}{\pi^2} N_c \Sperp \int_\tau^1\frac{\dd z}{z^2}\mathcal{J}_g^{(0)}(z)x g(x,\mu^2) 
\int \dd^2 \qa  \dd^2 \qb 
\frac{1}{\qb^2} \ln\frac{\qperp^2}{\qb^2} F(\qperp-\qa) 
 [F(\qa+\qb) - \theta(\qperp^2 - \qb^2) F(\qa)] 
\nonumber\\
+& 
\frac{\alpha_s}{\pi} N_c \Sperp\int_\tau^1\frac{\dd z}{z^2}\mathcal{J}_g^{(0)}(z) x g(x,\mu^2) 
\int \dd^2 \qa  \dd^2 \qb
F(\qa) F(\qb) F(\qperp-\qb) \ln^2 \frac{\qperp^2}{(\qa+\qb-\qperp)^2}
\nonumber \\
- & 
\frac{2\alpha_s}{\pi^2} N_c \Sperp\int_\tau^1\frac{\dd z}{z^2}\mathcal{J}_g^{(0)}(z) x g(x,\mu^2) 
 \int \dd^2 \qa  \dd^2 \qb  \dd^2 \qc  F(\qa) F(\qb) F(\qc)
\nonumber \\
& \times 
\frac{(\qperp - \qa + \qc) \cdot (\qperp - \qb + \qc)}{(\qperp - \qa + \qc)^2 (\qperp - \qb + \qc)^2}
\ln \frac{\qperp^2}{(\qperp - \qa + \qc)^2},
\\
\frac{\dd\sigma_{gg}^{8}}{\dd \eta\dd^{2}P_J}
=&\frac{\alpha_s}{2\pi}\Sperp N_c\left(\frac{67}{9}-\frac{4}{3}\pi^2\right) \int_\tau^1\frac{\dd z}{z^2}\mathcal{J}_g(z)x g(x)\int \dd^2 \qa F(\qa) F(\qperp-\qa)\nonumber \\
 & - \frac{\alpha_s}{2\pi}\Sperp N_f T_R \frac{26}{9}  \int_\tau^1\frac{\dd z}{z^2}\mathcal{J}_g^{(0)}(z)x g(x) \int \dd^2 \qa F(\qa) F(\qperp-\qa),
\\
\frac{\dd\sigma_{gg}^{9}}{\dd \eta\dd^{2}P_J}
=
&-\frac{\alpha_s}{2\pi} \Sperp N_c \int_\tau^1\frac{\dd z}{z^2}\mathcal{J}_g^{(0)}(z) \int_{x}^1 \dd\xi \frac{x}{\xi} g\left(\frac{x}{\xi}\right)  \left[\frac{\ln\frac{(1-\xi)^2}{\xi^2}}{1-\xi}\right]_+ \frac{2[1-\xi(1-\xi)]^2}{\xi^3}\int \dd^2 \qa F(\qa) F(\qperp/\xi-\qa),\\
\frac{\dd\sigma_{gg}^{10}}{\dd \eta\dd^{2}P_J}
=
&2\beta_0\frac{\alpha_s}{2\pi} N_c \Sperp \int_\tau^1\frac{\dd z}{z^2}\mathcal{J}_g^{(0)}(z)
\int \dd^2 \qa x g\left(x\right) 
F(\qperp-\qa) F(\qa)
\ln \frac{\qperp^2}{\Lambda^2},\\
\frac{\dd\sigma_{gg}^{11}}{\dd \eta\dd^{2}P_J}
=
&
\frac{\alpha_s}{2\pi} N_c \Sperp  \int_\tau^1\frac{\dd z}{z^2}\mathcal{J}_g^{(0)}(z) \int_{x}^1 \dd\xi \frac{x}{\xi} g\left(\frac{x}{\xi}\right)\frac{1}{\xi^2} \mathcal{P}_{gg} (\xi) \int \dd^2 \qa 
 F(\qperp/\xi-\qa)F(\qa)\ln\frac{\Lambda^2}{q_\perp^2 R^2}.\label{ggfstate}
\end{align}
Again $\mathcal{T}_{gg}^{(1)} (\xi,\qa,\qb,\qc,\qperp)$ can also be found in Ref.~\cite{Shi:2021hwx} which is
\begin{align}
\mathcal{T}_{gg}^{(1)} (\xi,\qa,\qb,\qc,\qperp) =
&
\frac{1}{\xi^2} \frac{[(1-\xi)\qa +\qc - \xi\qb]^2}{(\qa+\qc-\qperp)^2 (\qa + \qb - \qperp/\xi)^2} F(\qa) F(\qb) F(\qc)
\nonumber\\
- & 
\frac{1}{(\qa+\qc-\qperp)^2} \frac{\Lambda^2}{\Lambda^2 + (\qa+\qc-\qperp)^2} F(\qperp-\qa) F(\qb) F(\qc) 
\nonumber \\
- &
\frac{1}{\xi^2} \frac{1}{(\qa + \qb - \qperp/\xi)^2} \frac{\Lambda^2}{\Lambda^2 + (\qa + \qb - \qperp/\xi)^2} F(\qperp/\xi-\qb) F(\qb) F(\qc).
\end{align}
Similarly, we can extract the Sudakov double logarithm from $\sigma_{gg}^{7}$~\cite{Xiao:2018zxf,Shi:2021hwx}. Then we get
\begin{align}
\frac{\dd\sigma_{gg}^7}{\dd \eta\dd^{2}P_J}
=& 
\frac{\dd\sigma_{gg}^{7a}}{\dd \eta\dd^{2}P_J}
+
\frac{\dd\sigma_{gg}^{7b}}{\dd \eta\dd^{2}P_J},
\end{align}
where
\begin{align}
\frac{\dd\sigma_{gg}^{7a}}{\dd \eta\dd^{2}P_J}
=& 
- \frac{\alpha_s}{2\pi} N_c \Sperp \int_\tau^1\frac{\dd z}{z^2}\mathcal{J}_g^{(0)}(z)
 x g(x,\mu^2) 
\ln^2 \frac{\qperp^2}{\Lambda^2} \int \dd^2 \qa F(\qperp-\qa)F(\qa),
\\
\frac{\dd\sigma_{gg}^{7b}}{\dd \eta\dd^{2}P_J}
=& 
\frac{2\alpha_s}{\pi^2} N_c \Sperp \int_\tau^1\frac{\dd z}{z^2}\mathcal{J}_g^{(0)}(z)x g(x,\mu^2) 
\int \dd^2 \qa  \dd^2 \qb 
\frac{1}{\qb^2} \ln\frac{\qperp^2}{\qb^2} F(\qperp-\qa) 
\nonumber\\
&
\times [F(\qa+\qb) - \theta(\Lambda^2 - \qb^2) F(\qa)] 
\nonumber\\
- & \frac{\alpha_s}{2\pi} N_c \Sperp \int_\tau^1\frac{\dd z}{z^2}\mathcal{J}_g^{(0)}(z)
x g(x,\mu^2) 
\ln^2 \frac{\qperp^2}{\Lambda^2} \int \dd^2 \qa F(\qperp-\qa)F(\qa) \nonumber\\
+& 
\frac{\alpha_s}{\pi} N_c \Sperp\int_\tau^1\frac{\dd z}{z^2}\mathcal{J}_g^{(0)}(z)x g(x,\mu^2) 
 \int \dd^2 \qa  \dd^2 \qb 
F(\qa) F(\qb) F(\qperp-\qb)
\nonumber\\
&
\times \ln^2 \frac{\qperp^2}{(\qa+\qb-\qperp)^2}
\nonumber \\
- & 
\frac{2\alpha_s}{\pi^2} N_c \Sperp \int_\tau^1\frac{\dd z}{z^2}\mathcal{J}_g^{(0)}(z)xg(x,\mu^2)
 \int \dd^2 \qa  \dd^2 \qb  \dd^2 \qc   F(\qa) F(\qb) F(\qc)
\nonumber \\
& \times 
\frac{(\qperp - \qa + \qc) \cdot (\qperp - \qb + \qc)}{(\qperp - \qa + \qc)^2 (\qperp - \qb + \qc)^2}
\ln \frac{\qperp^2}{(\qperp - \qa + \qc)^2}.
\end{align}

\subsection{The \texorpdfstring{$q\to g$}{q->g} channel}
\label{section33}

Before we start the calculation for the off-diagonal channels, let us comment on the similarity and difference as compared to that for the diagonal ones. Note that the partonic cross-section of this channel is the same as the $q\to q $ channel~\cite{Dominguez:2011wm}, which is given by Eq. (\ref{partqqg}). For the splitting after the multiple scattering case, we have already computed the in-cone contribution in which the final state quark and the radiated gluon are within the same jet cone. Therefore, it is not necessary to consider in-cone contribution anymore. We do not have virtual contribution either since we observe the gluon jet in this channel. Here, the variable $\xi$ is defined as the longitudinal momentum fraction of the initial state quark carried by the radiated gluon, and this $\xi$ is different from what was defined in the $q\to q$ channel. In fact, we always use $\xi$ to denote the momentum fraction of the produced particle. Note that the transverse momentum of the measured jet is equal to the momentum of the radiated gluon which is $l_\perp=P_J$, while in the $q\to q $ channel we have $k_\perp=P_J$. 

\begin{figure}[ht]
\raisebox{28pt}[0pt][0pt]{--}~~
\subfigure[]{\includegraphics[width=4cm]{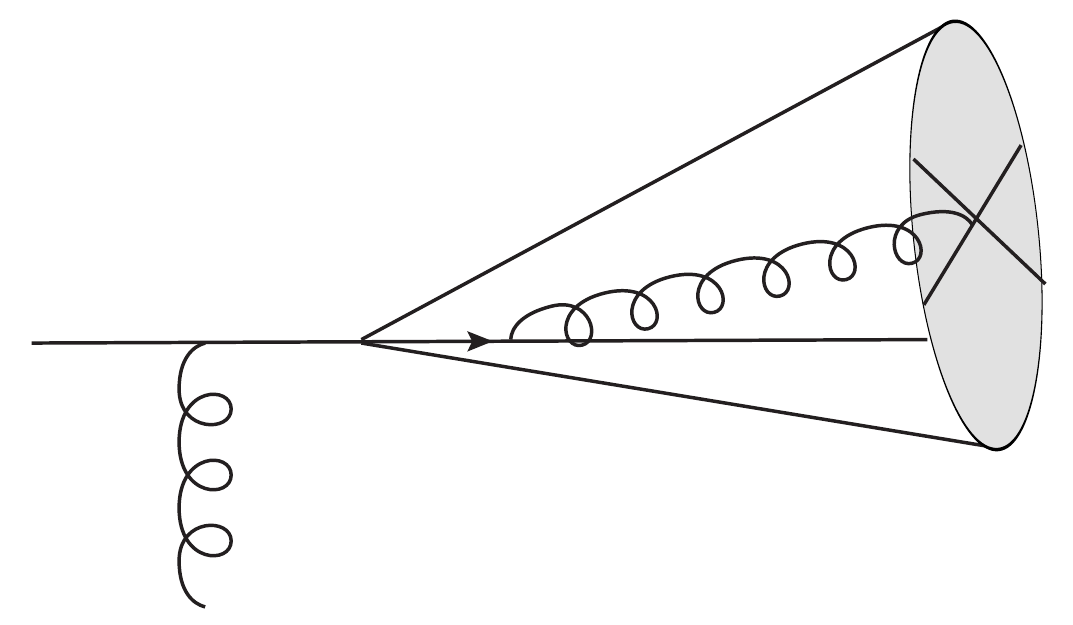}}
\raisebox{28pt}[0pt][0pt]{+}~~
\subfigure[]{\includegraphics[width=4cm]{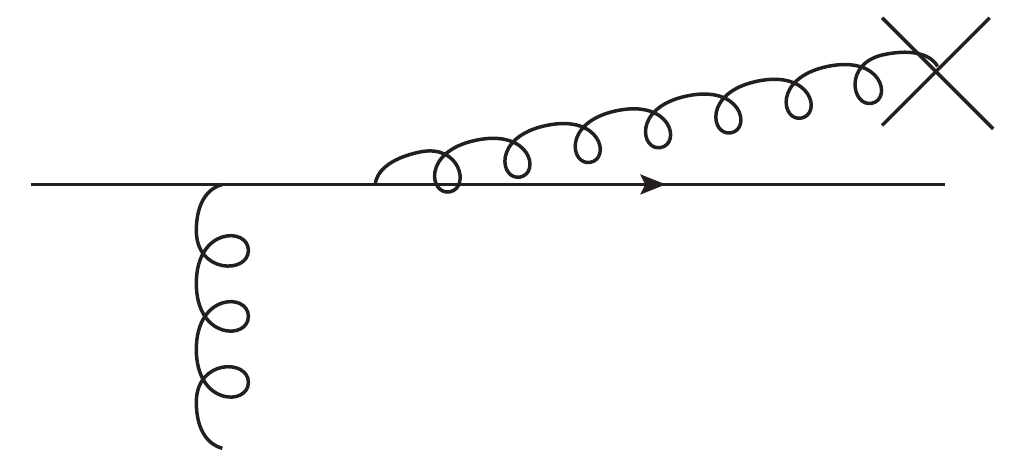}}
\caption[*]{The final state radiation contribution from the $q\to g$ channel.}
\label{qtog1}
\end{figure}

To compute the jet cross-section in the $q\to g$ channel, let us consider the diagrams shown in Fig.~\ref{qtog1}. Fig.~\ref{qtog1} (a) stands for the false identification of a gluon jet, while the final state gluon and quark reside in the same jet cone. This is similar to the case shown in Fig.~\ref{qqfinal} (b), where the quark is falsely identified as the jet. Fig.~\ref{qtog1} (b) represents the $q\to g$ contribution without any constraint. Hence the correct out-cone contribution for the $q\to g$ channel can be obtained by taking the difference of the contributions from these two diagrams.

\begin{figure}[ht]
\subfigure{
\includegraphics[width=4cm]{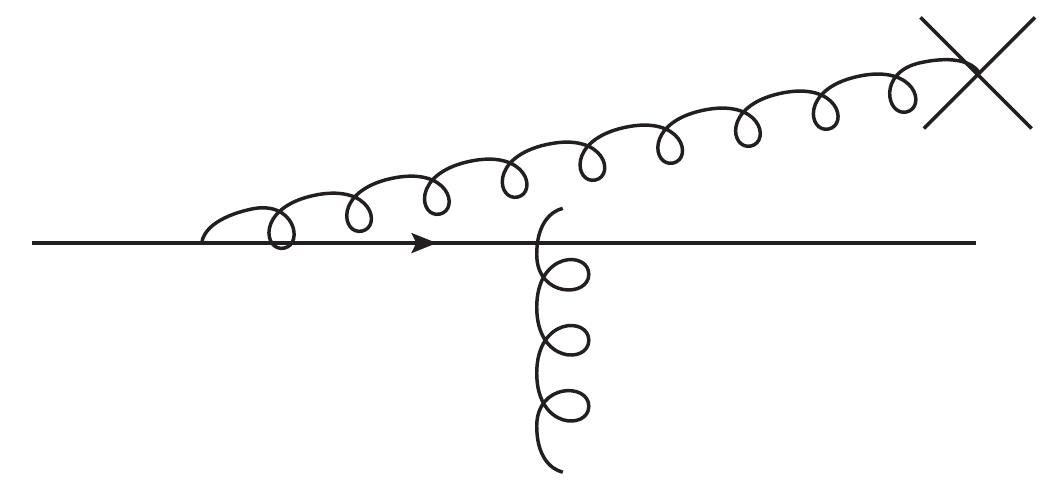}
}
\caption[*]{The initial state radiation contribution from the $q\to g$ channel.}
\label{qtog2}
\end{figure}
For the multiple interactions that take place before the gluon emission, as shown in Fig.~\ref{qtog2}, the corresponding correlator is $S^{(6)}_Y(b_\perp,x_\perp,b_\perp',x_\perp')$. Once we integrate over the momentum of the final state quark, we set $b_\perp=b_\perp'$. Therefore, we simplify $S^{(6)}_Y(b_\perp,x_\perp,b_\perp',x_\perp')$ to $S^{(2)}_Y(x_\perp',x_\perp)S^{(2)}_Y(x_\perp,x_\perp')$. Then the following evaluation is straightforward. Again, the usual dimensional regularization and $\overline{\text{MS}}$ subtraction scheme are applied to perform the rest of the calculations. Following the same calculation as that in the $q\to q$ channel, we remove the collinear singularity by redefining the gluon distribution as follows
\begin{equation}
g(x,\mu)=g^{(0)}(x)-\frac{1}{\epsilon}\frac{\alpha_s(\mu)}{2\pi}\int_x^1\frac{\dd\xi}{\xi}C_F\mathcal{P}_{gq}(\xi)q\Big(\frac{x}{\xi}\Big),
\end{equation}
where $\mathcal{P}_{gq}(\xi)=\frac{1+(1-\xi)^2}{\xi}$. At the end of the day, the final cross-section is found to be finite. In the $q \to g$ channel, there is no LO contribution. We write the cross-section in the coordinate space as
\begin{align}
\frac{\dd\sigma_{gq}}{\dd \eta\dd^{2}P_J}
=
\frac{\dd\sigma^\text{NLO}_{gq}}{\dd \eta\dd^{2}P_J}
= 
\sum_{i=a}^c
\frac{\dd\sigma_{gq}^i}{\dd \eta\dd^{2}P_J},
\end{align}
where,
\begin{align}
\frac{\dd\sigma_{gq}^a}{\dd \eta\dd^{2}P_J}
= &
\frac{\alpha_s}{2\pi} \Sperp C_F\int_\tau^1\frac{\dd z}{z^2}\mathcal{J}_g^{(0)}(z)  \int_{x}^1 \dd\xi \frac{x}{\xi} q\left(\frac{x}{\xi},\mu^2\right)  
\int \frac{\dd^2\rperp}{(2\pi)^2} e^{-i\qperp\cdot\rperp} S^{(2)} (\rperp) S^{(2)} (\rperp) 
\left[
\mathcal{P}_{gq} (\xi) \ln \frac{c_0^2}{\rperp^2\mu^2} +\xi
\right],  \label{xiterm}
\\
\frac{\dd\sigma_{gq}^b}{\dd \eta\dd^{2}P_J}
= &
8\pi \Sperp C_F \frac{\alpha_s}{2\pi}\int_\tau^1\frac{\dd z}{z^2}\mathcal{J}_g^{(0)}(z)
 \int_{x}^1 \dd\xi  \frac{x}{\xi} q\left(\frac{x}{\xi},\mu^2\right) 
\int\frac{\dd^2\uperp\dd^2\vperp}{(2\pi)^4} 
e^{-i\frac{\qperp}{\xi}\cdot(\uperp-\vperp)} e^{-i\qperp\cdot\vperp}
\nonumber\\
&\times
S^{(2)} (\uperp) S^{(2)} (\vperp) \mathcal{P}_{gq} (\xi) \frac{1}{\xi} \frac{(\uperp-\vperp) \cdot \vperp}{(\uperp-\vperp)^2 \vperp^2},\\
\frac{\dd\sigma_{gq}^c}{\dd y\dd^{2}P_J}
= &
\frac{\alpha_s}{2\pi} \Sperp C_F\int_\tau^1\frac{\dd z}{z^2}\mathcal{J}_g^{(0)}(z)\int_{x}^1 \dd\xi\frac{x}{\xi} q\left(\frac{x}{\xi}\right) 
\int \frac{\dd^2\rperp}{(2\pi)^2} e^{-i\frac{\qperp}{\xi} \cdot\rperp} S^{(2)} (\rperp) \notag\\
&\times
\frac{1}{\xi^2} \mathcal{P}_{gq} (\xi) \left[ 
\ln\frac{c_0^2}{\rperp^2\qperp^2R^2} - \ln\frac{(1-\xi)^2}{\xi^2}
\right].
\end{align}
Similar to the $q\to q$ channel, the extra $\xi$ term inside the square brackets of Eq.~(\ref{xiterm}) arises from the additional $-\epsilon \xi$ correction in the $q\to g$ splitting function in $4-2\epsilon$ dimension. After the Fourier transform, we obtain the cross-section in the momentum space as follows 
\begin{align}
\frac{\dd\sigma_{gq}}{\dd \eta\dd^{2}P_J}
=
\frac{\dd\sigma^\text{NLO}_{gq}}{\dd y\dd^{2}P_J}
= 
\sum_{i=1}^5
\frac{\dd\sigma_{gq}^i}{\dd \eta\dd^{2}P_J},
\end{align}
where
\begin{align}
\frac{\dd\sigma_{gq}^1}{\dd \eta\dd^{2}P_J} = 
& 
\frac{\alpha_s}{2\pi} C_F \Sperp\int_\tau^1\frac{\dd z}{z^2}\mathcal{J}_g^{(0)}(z)
\int_{x}^1 \dd \xi \int \dd^2 \qa
\frac{x}{\xi} q\left(\frac{x}{\xi},\mu^2\right)  \mathcal{P}_{gq} (\xi) \ln \frac{\Lambda^2}{\mu^2}
F(\qa) F(\qperp-\qa),\label{qgistate}
\\
\frac{\dd\sigma_{gq}^2}{\dd \eta\dd^{2}P_J} = 
& 
\frac{\alpha_s}{2\pi} C_F \Sperp\int_\tau^1\frac{\dd z}{z^2}\mathcal{J}_g^{(0)}(z)
\int_{x}^1 \dd \xi   \int \dd^2 \qa
\frac{x}{\xi}  q\left(\frac{x}{\xi},\mu^2\right)
\xi F(\qa) F(\qperp-\qa),\\
\frac{\dd\sigma_{gq}^3}{\dd \eta\dd^{2}P_J} = 
&
\frac{\alpha_s}{2\pi^2} C_F \Sperp\int_\tau^1\frac{\dd z}{z^2}\mathcal{J}_g^{(0)}(z)
 \int_{x}^1 \dd \xi \int \dd^2 \qa \int \dd^2\qb
x q(x,\mu^2) \mathcal{P}_{gq} (\xi) 
\mathcal{T}_{gq}^{(1)} (\xi, \qa,\qb,\qperp),\\
\frac{\dd\sigma_{gq}^4}{\dd \eta\dd^{2}P_J} = 
& 
\frac{\alpha_s}{2\pi} C_F \Sperp\int_\tau^1\frac{\dd z}{z^2}\mathcal{J}_g^{(0)}(z)
 \int_{x}^1 \dd \xi
\frac{x}{\xi} q\left(\frac{x}{\xi}\right) \frac{1}{\xi^2}\mathcal{P}_{gq} (\xi) \ln \frac{\Lambda^2}{\qperp^2R^2}
 F(\qperp/\xi),\label{qgfstate}\\
\frac{\dd\sigma_{gq}^5}{\dd \eta\dd^{2}P_J} = 
& -\frac{\alpha_s}{2\pi} C_F \Sperp\int_\tau^1\frac{\dd z}{z^2}\mathcal{J}_g^{(0)}(z)
 \int_{x}^1 \dd \xi
\frac{x}{\xi} q\left(\frac{x}{\xi}\right) \mathcal{P}_{gq} (\xi) \frac{1}{\xi^2} F(\qperp/\xi) \ln\frac{(1-\xi)^2}{\xi^2},
\end{align}
with
\begin{align}
\mathcal{T}_{gq}^{(1)} (\xi, \qa,\qb,\qperp) 
= 
&
\left( \frac{\qperp -\qa - \qb}{(\qperp -\qa - \qb)^2} - \frac{\qperp - \xi \qb}{(\qperp - \xi\qb)^2} \right)^2
F(\qa) F(\qb)
\nonumber\\
&
- \frac{\Lambda^2}{\Lambda^2 + (\qperp -\qa - \qb)^2} \frac{1}{(\qperp -\qa - \qb)^2} 
F(\qb) F(\qperp-\qb)
\nonumber\\
& 
- \frac{\Lambda^2}{\Lambda^2 + (\qperp/\xi-\qb)^2} \frac{1}{(\qperp-\xi\qb)^2} F(\qa) F(\qperp/\xi).
\end{align}

\subsection{The \texorpdfstring{$g\to q$}{g->q} channel}
\label{section34}

The partonic cross-section for the gluon splitting into a quark (with the momentum $l$) and an anti-quark (with the momentum $k$) can be written as~\cite{Dominguez:2011wm} 
\begin{eqnarray}
\frac{\dd\sigma _{gA\rightarrow q\bar{q}X}}{\dd^3l\dd^3k}&=&\alpha _S\delta(q^+-l^+-k^+)T_R \int \frac{\text{d}^{2}x_\perp}{(2\pi)^{2}}\frac{\text{d}^{2}x_\perp^{\prime }}{(2\pi )^{2}}\frac{\text{d}^{2}b_\perp}{(2\pi)^{2}}\frac{\text{d}^{2}b_\perp'}{(2\pi )^{2}} \notag \\
&&\times e^{-ik_{\perp }\cdot(x_\perp-x_\perp^{\prime })} e^{-il_\perp\cdot (b_\perp-b_\perp^{\prime })} \sum_{\lambda\alpha\beta} \psi_{\alpha\beta}^{T\lambda*}(u_\perp') \psi_{\alpha\beta}^{T\lambda}(u_\perp) \notag \\
&&\times \left[C_Y(x_\perp,b_\perp,x'_\perp,b_\perp)+S^A_Y(\xi x_\perp+(1-\xi)b_\perp,\xi x'_\perp+(1-\xi)b_\perp')\right.\notag\\
&&\quad\left.-S^{(3)}_Y(x_\perp,\xi x'_\perp+(1-\xi)b_\perp',b_\perp)-S^{(3)}_Y(b_\perp',\xi x_\perp+(1-\xi)b_\perp,x_\perp')\right] \ ,\label{xsgqqbar}
\end{eqnarray}
with
\begin{equation}
 \sum_{\lambda\alpha\beta} \psi_{\alpha\beta}^{T\lambda*}(u_\perp') \psi_{\alpha\beta}^{T\lambda}(u_\perp)
 =\frac{2(2\pi^2)}{p^+}  \frac{u_\perp ^\prime \cdot u_\perp}{{{u_\perp ^\prime }^2 u_\perp^2 }}
 \left[  \xi^2 (1-\xi)^2 -2\epsilon\xi(1-\xi)  \right]~.
\end{equation}
The correlator $S^{(3)}_Y$ is given in the previous section and the other two correlators read
\begin{align}
C_Y(x_\perp,b_\perp,x_\perp',b_\perp')&=\frac{1}{C_FN_c}\left\langle\text{Tr}\left(U^\dagger(b_\perp)T^cU(x_\perp)U^\dagger(x_\perp')T^cU(b_\perp')\right)\right\rangle_Y,\\
S^A_Y(v_\perp,v_\perp')&=\frac{1}{N_c^2-1}\left\langle\text{Tr}W(v_\perp)W^\dagger(v_\perp')\right\rangle_Y.
\end{align}

\begin{figure}[ht]
\raisebox{27pt}[0pt][0pt]{--}~~
\subfigure[]{\begin{minipage}[t]{0.25\linewidth}
\includegraphics[width=4cm]{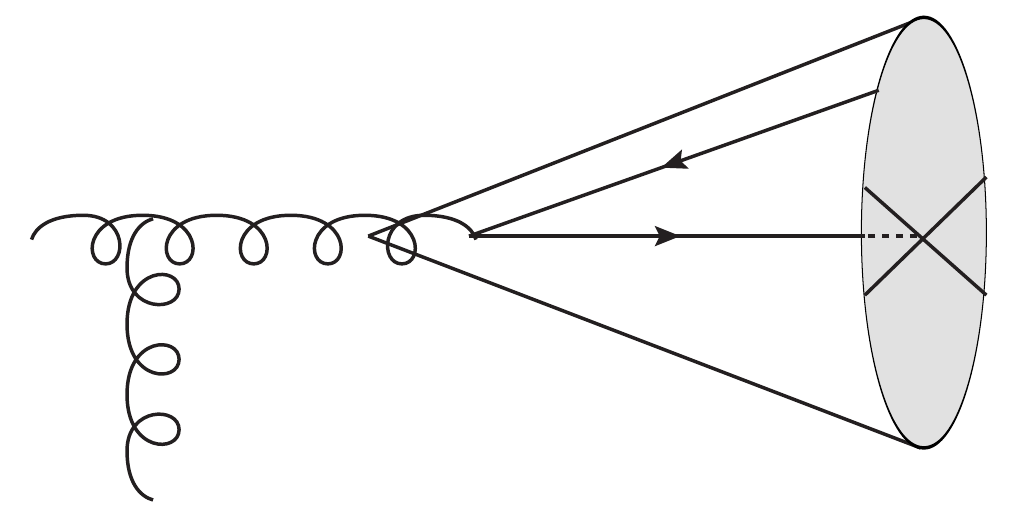}
\end{minipage}
}
~~\raisebox{27pt}[0pt][0pt]{+}~~
\subfigure[]{
\begin{minipage}[t]{0.25\linewidth}
\centering
\includegraphics[width=4cm]{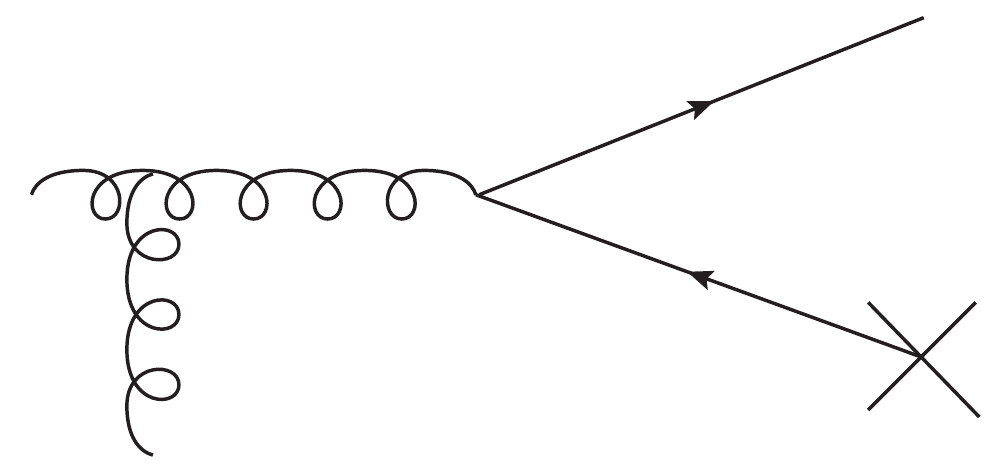}
\end{minipage}
}
\caption{The contribution of the pair production after the multiple scattering of the $g\to q$ channel.}
\label{gqqbefore}
\end{figure}
Fig.~\ref{gqqbefore} indicates that the multiple scatterings take place before the pair production. For the interaction before the gluon splitting, we only need to consider the out-cone contribution since we have already considered in-cone contribution in the Sec.~\ref{section32}. Once we take the large-$N_c$ limit approximation, the correlator $S_Y^A(v_\perp,v_\perp')$ in Eq.(\ref{xsgqqbar}) can be expressed entirely in terms of the two-point function $S^{(2)}_Y(v_\perp,v_\perp')S^{(2)}_Y(v_\perp',v_\perp)$.

\begin{figure}[ht]
\includegraphics[width=4cm]{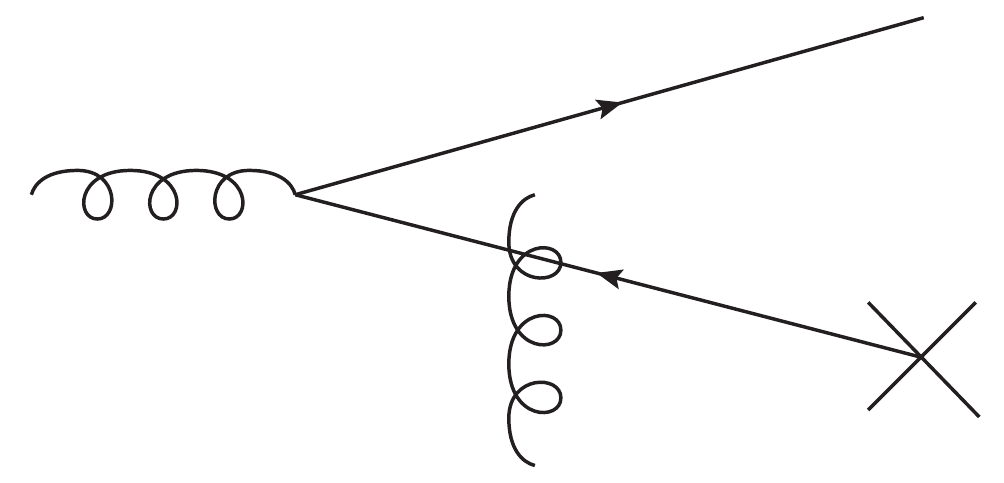}
\caption{The contribution of the gluon splitting after the multiple scattering of the $g\to q$ channel.}
\label{gqqafter}
\end{figure}

For the interaction after the gluon splitting as shown in Fig.~\ref{gqqafter}, it is easy to write down the cross-section after integrating the momentum of the final state quark in Eq.(\ref{xsgqqbar}). Similarly, the multiple interaction correlator $C_Y(x_\perp,b_\perp,x'_\perp,b'_\perp)$ can be expressed in terms of $S^{(2)}_Y(x_\perp,x'_\perp)S^{(2)}_Y(b'_\perp,b_\perp)$ in the large-$N_c$ limit. After integrating the momentum of the anti-quark, we identify $b_\perp=b_\perp'$, which gives $C_Y(x_\perp,b_\perp,x'_\perp,b'_\perp)\simeq S^{(2)}_Y(x_\perp,x'_\perp)$. Following the same calculation as previous channels, we remove the collinear singularities by redefining the quark distribution as follows
	\begin{equation}
		q(x,\mu)=q^{(0)}(x)-\frac{1}{\epsilon}\frac{\alpha_s(\mu)}{2\pi}\int_x^1\frac{\dd\xi}{\xi}C_F\mathcal{P}_{qg}(\xi)g\Big(\frac{x}{\xi}\Big),
	\end{equation}
where $\mathcal{P}_{qg}(\xi)=(1-\xi)^2+\xi^2$. In addition, we also have interference contribution which is the same as hadron production. At the end of the day, we get the cross-section of the $g \to q$ channel in the coordinate space as follows
\begin{align}
\frac{\dd\sigma_{qg}}{\dd \eta\dd^2 P_J} = \sum_{i=a}^c \frac{\dd\sigma_{qg}^i}{\dd y\dd^2 P_J},
\end{align}
where
\begin{align}
\frac{\dd\sigma_{qg}^a}{\dd \eta\dd^2 P_J}
= &
\frac{\alpha_s}{2\pi} \Sperp T_R\int_\tau^1\frac{\dd z}{z^2}\mathcal{J}_q^{(0)}(z) \int_{x}^1 \dd \xi \frac{x}{\xi} g\left(\frac{x}{\xi},\mu^2\right) 
\int \frac{\dd^2\rperp}{(2\pi)^2} e^{-i\qperp\cdot\rperp} S^{(2)} (\rperp) 
\left[ 
\mathcal{P}_{qg} (\xi) 
\ln\frac{c_0^2}{\rperp^2\mu^2} + 2\xi(1-\xi)
\right],
\\
\frac{\dd\sigma_{qg}^b}{\dd \eta\dd^2 P_J}
= &
8\pi \Sperp T_R \frac{\alpha_s}{2\pi}\int_\tau^1\frac{\dd z}{z^2}\mathcal{J}_q^{(0)}(z)   \int_{x}^1 \dd \xi \frac{x}{\xi} g\left(\frac{x}{\xi},\mu^2\right) 
\int \frac{\dd^2\uperp\dd^2\vperp}{(2\pi)^4} e^{-i\qperp\cdot (\uperp-\vperp) -i\frac{\qperp\cdot\vperp}{\xi} }
\nonumber\\
& \times
\frac{1}{\xi} \mathcal{P}_{qg} (\xi) S^{(2)} (\uperp) S^{(2)} (\vperp) 
\frac{(\uperp-\vperp) \cdot \vperp}{(\uperp-\vperp)^2 \vperp^2},\\
\frac{\dd\sigma_{qg}^c}{\dd \eta\dd^2 P_J}
= &
\frac{\alpha_s}{2\pi} \Sperp T_R \int_\tau^1\frac{\dd z}{z^2}\mathcal{J}_q^{(0)}(z) \int_{x}^1 \dd \xi \frac{x}{\xi} g\left(\frac{x}{\xi},\mu^2\right) 
\int \frac{\dd^2\rperp}{(2\pi)^2} e^{-i\frac{\qperp}{\xi}\cdot\rperp} S^{(2)} (\rperp) S^{(2)} (\rperp) \nonumber\\
& \times
\frac{1}{\xi^2} \mathcal{P}_{qg} (\xi) 
\left[ 
\ln\frac{c_0^2}{\rperp^2\qperp^2R^2}- \ln\frac{(1-\xi)^2}{\xi^2}
\right].
\end{align}
The cross-section in the momentum space reads
\begin{align}
\frac{\dd\sigma_{qg}}{\dd \eta\dd^{2}P_J}
= 
\sum_{i=1}^5
\frac{\dd\sigma_{qg}^i}{\dd \eta\dd^{2}P_J},
\end{align}
with
\begin{align}
\frac{\dd\sigma_{qg}^1}{\dd \eta\dd^{2}P_J}
=
&
\frac{\alpha_s}{2\pi} \Sperp T_R\int_\tau^1\frac{\dd z}{z^2}\mathcal{J}_q^{(0)}(z)
 \int_{x}^1 \dd \xi 
\frac{x}{\xi}
g\left(\frac{x}{\xi},\mu^2\right) \mathcal{P}_{qg} (\xi) \ln \frac{\Lambda^2}{\mu^2} 
F(\qperp),\label{gqistate}
\\
\frac{\dd\sigma_{qg}^2}{\dd \eta\dd^{2}P_J}
=
&
2\frac{\alpha_s}{2\pi} \Sperp T_R\int_\tau^1\frac{\dd z}{z^2}\mathcal{J}_q^{(0)}(z)
 \int_{x}^1 \dd \xi 
\frac{x}{\xi}
g\left(\frac{x}{\xi},\mu^2\right) \xi(1-\xi)
F(\qperp),
\\
\frac{\dd\sigma_{qg}^3}{\dd \eta\dd^{2}P_J}
= 
&
\frac{\alpha_s}{2\pi^2} \Sperp T_R\int_\tau^1\frac{\dd z}{z^2}\mathcal{J}_q^{(0)}(z)
 \int_{x}^1 \dd \xi \int \dd^2 \qa \int \dd^2 \qb\frac{x}{\xi}
g\left(\frac{x}{\xi},\mu^2\right)  \mathcal{P}_{qg} (\xi)
\mathcal{T}_{qg}^{(1)} (\xi, \qa,\qb,\qperp),\\
\frac{\dd\sigma_{qg}^4}{\dd \eta\dd^{2}P_J}
=
&
\frac{\alpha_s}{2\pi} \Sperp T_R\int_\tau^1\frac{\dd z}{z^2}\mathcal{J}_q^{(0)}(z)
 \int_{x}^1 \dd \xi \int \dd^2 \qa
\frac{x}{\xi}
g\left(\frac{x}{\xi},\mu^2\right) \frac{1}{\xi^2} \mathcal{P}_{qg} (\xi) \ln \frac{\Lambda^2}{\qperp^2R^2} 
F(\qa) F(\qperp/\xi-\qa),\label{gqfstate}\\
\frac{\dd\sigma_{qg}^5}{\dd \eta\dd^{2}P_J}
=
&
-\frac{\alpha_s}{2\pi} \Sperp T_R\int_\tau^1\frac{\dd z}{z^2}\mathcal{J}_q^{(0)}(z)
 \int_{x}^1 \dd \xi \int \dd^2 \qa
\frac{x}{\xi}
g\left(\frac{x}{\xi},\mu^2\right) \frac{1}{\xi^2} \mathcal{P}_{qg} (\xi) 
F(\qa) F(\qperp/\xi-\qa) \ln\frac{(1-\xi)^2}{\xi^2},
\end{align}
where 
\begin{align}
\mathcal{T}_{qg}^{(1)} (\xi, \qa,\qb,\qperp)
=
& 
\left( 
\frac{\qperp- \xi\qa -\xi\qb}{(\qperp - \xi\qa -\xi\qb)^2} 
- \frac{\qperp-\qb}{(\qperp-\qb)^2}
\right)^2
F(\qa) F(\qb) 
\nonumber\\
& 
- \frac{1}{(\qperp - \xi\qa -\xi\qb)^2}\frac{\Lambda^2}{\Lambda^2 + (\qperp/\xi - \qa - \qb)^2}
F(\qb) F(\qperp/\xi-\qb)
\nonumber\\
&
- \frac{1}{(\qperp-\qb)^2} \frac{\Lambda^2}{\Lambda^2 + (\qperp-\qb)^2}
F(\qa) F(\qperp).
\end{align}

So far, we have achieved the full NLO results of single inclusive jet productions in $pA$ collisions. And the cross-section is consistent with Refs.~\cite{Chirilli:2012jd,Shi:2021hwx}. Furthermore, we have also extracted all the large logarithms in the momentum space. In the next section, we focus on how to deal with these large logarithms in detail.

\section{Reummation of large logarithms}
\label{section4}

As discussed before, to improve the accuracy of the theoretical prediction and numerical implementation we need to resum all the large logarithms arising near the threshold boundary. 
Therefore, this section is devoted to resum such large logarithms in the previous calculations. The resummation strategy is similar to that used in the hadron productions. 

Before providing the details of the threshold resummation, let us comment on one technical issue. The resummation strategy used here is analogous that in Ref.~\cite{Shi:2021hwx} except the terms which are proportional to $\left(\frac{\ln^n(1-\xi)}{1-\xi}\right)_+$ (e.g., $\dd\sigma_{qq}^g$ and $\dd\sigma_{gg}^h$).  As a matter of fact, those terms which are proportional to $\left(\frac{\ln^n(1-\xi)}{1-\xi}\right)_+$ stem from the final state gluon radiations. When $\xi\to 1$ ($\tau\to 1$) near the threshold limit, these terms would give us $\ln^2N$ contribution in the Mellin space. Indeed, this term has a sign difference comparing with the result  from Cantani and Trentadue for the Drell-Yan process~\cite{Catani:1989ne}. It is well known that this kind of double logs is notoriously difficult to deal with when one performs inverse Mellin transform because of the so-called Landau pole problem. 

We first note that there are two kinds of sources that contribute to these double logs, where one originates from $\dd\sigma_{qq}^g$ and $\dd\sigma_{gg}^h$, the other from part of the kinematic constraint correction. From Eqs.~(\ref{sigmaa}) and (\ref{sigmab}), we can show that these double logs from the jet contribution cancel. It means that there is no Sudakov factor associated with the final state radiation.
This case is different from the hadron production case where the double logs can contribute from both initial and final state.

Another intuitive way to think about this question is from the physical point of view. By tracing the source, we found that the plus function  $\left(\frac{\ln^n(1-\xi)}{1-\xi}\right)_+$ comes from the false identification, where we treat the final state quark as a jet when the radiated gluon is inside the jet cone. Once the radiated gluon is inside the jet cone, it can not be real soft since the phase space of the gluon emission is small but not zero. As long as the jet cone $R$ is large enough, there is no soft divergence anymore. This means that there are no double logs for the final state radiations in the end.

With the above arguments in mind, we believe that there are no double logarithmic divergences in the final state gluon radiations of forward single inclusive jet production in $pA$ collisions. Therefore, we expect that there is a cancellation between the terms proportional to the plus function $\left(\frac{\ln^n(1-\xi)}{1-\xi}\right)_+$ and the double logs from final state kinematic constraint corrections. The combined results are expected to be small. Thus we do not resum them, and put them together with other terms in the NLO hard factor.
The remaining double logs terms that we need to resum now come from initial state radiation and the left over (initial) part of the kinematic constraint correction. Furthermore, there is no single logarithmic divergences for the final state gluon radiation either. We put those log terms in the NLO hard factor.
 
Before we resum all the large logarithmic terms, let us specify these logarithms. As can be seen from previous calculation results, there are collinear logarithms ($\ln\frac{\Lambda^2}{\mu^2}$ and $\ln\frac{\Lambda^2}{\mu_J^2}$) and Sudakov logarithms ($\ln\frac{\qperp^2}{\Lambda^2}$, $\ln^2\frac{\qperp^2}{\Lambda^2}$). More discussions of the collinear logarithms can be found in the Sec.~\ref{collinear1}. We list all the large logarithms in the Table~\ref{table:1}. 

\begin{table}[ht]
\begin{tabular}{c|c|c|c|c}
\hline
\hline
process &  collinear log(initial) & ~~ single log ~~ & ~~ double log~~ &  collinear log(final)\\
\hline
$q\to q$ &  $\mathcal{P}_{qq}(\xi)\ln\frac{\Lambda^2}{\mu^2}$  &  $\ln\frac{\qperp^2}{\Lambda^2}$ & $\ln^2\frac{\qperp^2}{\Lambda^2}$  & $\frac{1}{\xi^2}\mathcal{P}_{qq}(\xi)\ln\frac{\Lambda^2}{\mu_J^2}$  \\
\hline
$g\to g$ &  $\mathcal{P}_{gg}(\xi)\ln\frac{\Lambda^2}{\mu^2}$  &  $\ln\frac{\qperp^2}{\Lambda^2}$ & $\ln^2\frac{\qperp^2}{\Lambda^2}$  & $\frac{1}{\xi^2}\mathcal{P}_{gg}(\xi)\ln\frac{\Lambda^2}{\mu_J^2}$    \\ 
\hline
$q\to g$ &  $\mathcal{P}_{gq}(\xi)\ln\frac{\Lambda^2}{\mu^2}$  &  $/$ & $/$&   $\frac{1}{\xi^2}\mathcal{P}_{gq}(\xi)\ln\frac{\Lambda^2}{\mu_J^2}$\\
\hline
$g\to q$ &  $\mathcal{P}_{qg}(\xi)\ln\frac{\Lambda^2}{\mu^2}$  &  $/$ & $/$ &   $\frac{1}{\xi^2}\mathcal{P}_{qg}(\xi)\ln\frac{\Lambda^2}{\mu_J^2}$\\
\hline
\hline
\end{tabular}
\caption{List of collinear and Sudakov logarithms in different channels.} 
\label{table:1}
\end{table}

In subsection~\ref{collinear1} and~\ref{collinear2}, we demonstrate two different approaches developed by Ref.~\cite{Shi:2021hwx} to resum the collinear logarithms. 
The resummation of soft logarithms is shown in section~\ref{soft}.
Because the resummation of the collinear and soft logarithms for jet productions differ only slightly from hadron production, we will show only the main results in the latter subsections, and more detailed discussions can be found in the supplemental material of the Ref.~\cite{Shi:2021hwx}.

\subsection{Resummation via the DGLAP evolution equation}
\label{collinear1}

As discussed previously, there are two types of collinear logarithms. One originates from the initial state radiation, and the other one from the final state emission. Their resummation corresponds to the scale evolution of PDFs and CJFs, respectively. For the diagonal channels, we know that the differential cross-section is proportional to the plus functions as follows
\begin{equation}
\int_\tau^1\dd\xi\frac{f(\xi)}{(1-\xi)_+}\propto \ln (1-\tau).  
\end{equation} 
When the gluon emission is near the boundary of the allowed phase space, the integration over the plus function then becomes divergent in the limit $\tau\rightarrow 1$. 
More discussion can be found in Ref.~\cite{Shi:2021hwx}. Therefore, one needs to resum such collinear logarithms associated with the plus function. Basically, there are two approaches to resum the collinear logarithms. Motivated by the works in Ref.~\cite{Xiao:2018zxf,Shi:2021hwx}, the collinear logarithms~\cite{Becher:2006qw, Becher:2006nr,Becher:2006mr} can intuitively be resumed with the help of the DGLAP evolution equations. This methods is called the reverse-evolution method in Ref.~\cite{Shi:2021hwx}. 

For the collinear logarithms $\ln\frac{\Lambda^2}{\mu^2}$ associated with initial state gluon emissions as in Eqs.(\ref{eq:sigqq1fir}, \ref{ggistate}, \ref{qgistate}, \ref{gqistate}), once we evolve the factorization scale $\mu$ to the auxiliary scale $\Lambda$, the resummation of the collinear part can be achieved automatically. As demonstrated in Ref.~\cite{Shi:2021hwx}, we can apply the following replacement
\begin{align}
\left[
\begin{array}{c}
q\left(x,\mu \right) \\
g\left(x,\mu \right)
\end{array}
\right] 
+ \frac{\alpha_s}{2\pi} \ln\frac{\Lambda^2}{\mu^2} \int_{x}^1 \frac{\dd\xi}{\xi}
\left[
\begin{array}{cc}
C_F \mathcal{P}_{qq} (\xi) & T_R \mathcal{P}_{qg} (\xi) \\
C_F \mathcal{P}_{gq} (\xi) & N_C \mathcal{P}_{gg} (\xi)
\end{array}
\right]
\left[
\begin{array}{c}
q\left(x/\xi,\mu \right) \\
g\left(x/\xi,\mu \right)
\end{array}
\right]
\Rightarrow
\left[
\begin{array}{c}
q\left(x,\Lambda \right) \\
g\left(x,\Lambda \right)
\end{array}
\right]. 
\label{PDFs}
\end{align}

In order to resum the collinear logarithm ($\ln\frac{\Lambda^2}{\mu_J^2}$ with $\mu_J=P_J R$) arising from the final state radiations, our strategy is to redefine the CJFs. Similar to the FFs in the hadron production case~\cite{Chirilli:2012jd}, we take $q\to q$ channel as an example and start with Eq.~(\ref{fstate}), which can be cast into  
\begin{align}
&\frac{\alpha_s}{2\pi}\int_\tau^1\frac{\dd z}{z^2}\mathcal{J}_q^{(0)}(z)\ln\frac{\Lambda^2z^2}{P_J^2R^2}\int_{\tau/z}^1\dd\xi C_F\mathcal{P}_{qq}(\xi)\frac{x}{\xi}q\left(\frac{x}{\xi}\right)\frac{1}{\xi^2}F\left(\frac{P_J}{z\xi}\right)\Big\vert_{x=\frac{\tau}{z}}\notag\\
=&\frac{\alpha_s}{2\pi}\int_\tau^1\frac{\dd z'}{z'^2}x q\left(x \right)F\left(\frac{P_J}{z'}\right)\ln\frac{\Lambda^2z'^2}{P_J^2R^2\xi^2}\int_{z'}^1\frac{\dd\xi}{\xi} C_F\mathcal{P}_{qq}(\xi)\mathcal{J}_q^{(0)}\left(\frac{z'}{\xi}\right)\Big\vert_{x=\frac{\tau}{z'},z'=z\xi}\notag\\
=&\frac{\alpha_s}{2\pi}\int_\tau^1\frac{\dd z}{z^2}x q\left(x \right)F\left(\frac{P_J}{z}\right)\ln\frac{\Lambda^2}{P_J^2R^2}\int_{z}^1\frac{\dd\xi}{\xi} C_F\mathcal{P}_{qq}(\xi)\mathcal{J}_q^{(0)}\left(\frac{z}{\xi}\right),
\end{align}
where we have used $\int_z'^1\dd z'\ln\frac{z'^2}{\xi^2}\mathcal{J}_q^{(0)}\left(\frac{z'}{\xi}\right)=\int_z'^1\dd z'\ln\frac{z'^2}{\xi^2}\delta\left(1-\frac{z'}{\xi}\right)=0$ in the last line, and changed the integration variable $z'\to z$. By combining the LO,  $q\to q$ channel and  $q\to g$ channel contributions together, we redefine the collinear quark jet function as follows
\begin{align}
\mathcal{J}_q(z, \Lambda)=\mathcal{J}^{(0)}_q(z)+\frac{\alpha_s}{2\pi}\ln\frac{\Lambda^2}{P_J^2R^2}\int_{z}^1\frac{\dd\xi}{\xi} \left[C_F\mathcal{P}_{qq}(\xi)\mathcal{J}_q^{(0)}\left(\frac{z}{\xi}\right)+C_F\mathcal{P}_{gq}(\xi)\mathcal{J}_g^{(0)}\left(\frac{z}{\xi}\right)\right].
\label{quarkjetdefinition}
\end{align}
By differentiating Eq.~(\ref{quarkjetdefinition}) with respect to $\ln\Lambda^2$, we can obtain
 \begin{align}
\frac{\partial\mathcal{J}_q(z, \Lambda)}{\partial\ln\Lambda^2}=\frac{\alpha_s}{2\pi}\int_{z}^1\frac{\dd\xi}{\xi} \left[C_F\mathcal{P}_{qq}(\xi)\mathcal{J}_q^{(0)}\left(\frac{z}{\xi}\right)+C_F\mathcal{P}_{gq}(\xi)\mathcal{J}_g^{(0)}\left(\frac{z}{\xi}\right)\right].
\label{quarkjetDGLAP}
\end{align}
At first, Eq.~(\ref{quarkjetDGLAP}) is not a closed equation. However, by taking higher loop contributions into account, we can promote $\mathcal{J}^{(0)}_q(z)$ to $\mathcal{J}_q(z, \Lambda)$ and thus arrive at the closed evolution equation as follows
\begin{align}
\frac{\partial\mathcal{J}_q(z, \Lambda)}{\partial\ln\Lambda^2}=\frac{\alpha_s}{2\pi}\int_{z}^1\frac{\dd\xi}{\xi} \left[C_F\mathcal{P}_{qq}(\xi)\mathcal{J}_q\left(\frac{z}{\xi}, \Lambda\right)+C_F\mathcal{P}_{gq}(\xi)\mathcal{J}_g\left(\frac{z}{\xi}, \Lambda\right)\right].
\end{align}
It is obvious that the differential equation for the collinear quark jet function is identical to the DGLAP evolution equation. The initial condition for this equation is given by the $\mathcal{J}_q^{(0)}(z)=\delta (1-z)$ at scale $\mu_J=P_J R$. We should also consider running coupling solution when we perform the numerical calculations. Following the same procedure, we can obtain the evolution equation of the collinear gluon jet function
\begin{align}
\frac{\partial\mathcal{J}_g(z, \Lambda)}{\partial\ln\Lambda^2}=\frac{\alpha_s}{2\pi}\int_{z}^1\frac{\dd\xi}{\xi} \left[T_R\mathcal{P}_{qg}(\xi)\mathcal{J}_q\left(\frac{z}{\xi}, \Lambda\right)+N_C\mathcal{P}_{gg}(\xi)\mathcal{J}_g\left(\frac{z}{\xi}, \Lambda\right)\right],
\end{align}
with initial condition $\mathcal{J}_g^{(0)}(z)=\delta (1-z)$ at scale $\mu_J=P_J R$. Now we can resum the final state collinear logarithms $\ln \frac{\Lambda^2}{\mu_J^2}$ as in Eqs.(\ref{fstate}, \ref{ggfstate}, \ref{qgfstate}, \ref{gqfstate}) through the DGLAP equation with the following replacement
\begin{align}
\left[
\begin{array}{c}
\mathcal{J}_q(z, \mu_J) \\
\mathcal{J}_g(z, \mu_J)
\end{array}
\right] 
+ \frac{\alpha_s}{2\pi} \ln\frac{\Lambda^2}{\mu_J^2} \int_{z}^1 \frac{\dd\xi}{\xi}
\left[
\begin{array}{cc}
C_F \mathcal{P}_{qq} (\xi) & C_F \mathcal{P}_{gq} (\xi) \\
T_R \mathcal{P}_{qg} (\xi) & N_C \mathcal{P}_{gg} (\xi)
\end{array}
\right]
\left[
\begin{array}{c}
\mathcal{J}_q(z/\xi, \mu_J) \\
\mathcal{J}_g(z/\xi, \mu_J)
\end{array}
\right]
\Rightarrow
\left[
\begin{array}{c}
\mathcal{J}_q(z, \Lambda) \\
\mathcal{J}_g(z, \Lambda)
\end{array}
\right].
\label{CJFs}
\end{align}
In practice, we require $\mu_J\gg \Lambda_{QCD}$. This requirement allows us to perform perturbative QCD calculations.

The quantitative prescription for the choice of $\Lambda^2$ is the same as 
hadron production case. The detailed derivation which determines the proper value of the auxiliary scale $\Lambda^2$ can be found in Ref.~\cite{Shi:2021hwx}. Particularly, we can identify the dominant contribution for the NLO correction via the saddle point approximation and we find the natural choice for the semi-hard scale $\Lambda^2$ in the $q\to q$ channel 
\begin{align}
\Lambda^2 \approx \max \left\{ \Lambda_{\rm QCD}^2 \left[ \frac{\qperp^2 (1-\xi)}{\Lambda^2_{\rm QCD}} \right]^{\frac{C_F}{C_F+N_c\beta_0}}, Q_s^2\right\}. \label{lambdavalue2}
\end{align}
Once we replace the color factor $C_F$ with $N_c$ in the above equation and change $Q_s$ to the adjoint representation, we can get the result for the $g\to g$ channel.

\subsection{The resummation of the collinear logarithm in the Mellin space}
\label{collinear2}

Alternatively, we can resum the collinear logarithms in the Mellin space~\cite{Sterman:1986aj,Catani:1989ne, Catani:1996yz, deFlorian:2008wt}. This type of threshold resummation was first introduced for the DIS process~\cite{Bosch:2004th,Becher:2006qw, Becher:2006nr,Becher:2006mr} within the soft collinear effective theory framework. Our strategy used here is based on our previous work~\cite{Shi:2021hwx}. Due to the plus functions and delta functions in $\mathcal{P}_{qq}(\xi)$ and $\mathcal{P}_{gg}(\xi)$, there are endpoint singularities in the $\xi\to 1$ limit. This lmit corresponds to the large $N$ limit in the Mellin moment space. Therefore, the dominant contributions arise from these endpoint singularities and they are from diagonal channels. In contrast, the off-diagonal channels have no plus functions or delta functions. Therefore, we expect that the threshold effects from the off-diagonal terms are small. We simply deal with the diagonal channels in this subsection and keep the off-diagonal channels unchanged.

We first Mellin transform the cross-section of the diagonal channels into the Mellin space. 
In the Mellin space, the convolution of the differential cross-section can be factorized into an independent integral product and the integration over $\xi$. One can exponentiate the corresponding large logarithms in the Mellin space under the large-$N$ limit. 
At the end of the day, we need to perform the cross-section back to the momentum space with the help of the inverse Mellin transform. Since the calculation is straightforward, we only list the main results here and more details can be found in our previous work \cite{Shi:2021hwx}.

We first take the $q\to q $ channel as an example and show what is going on, then the $g\to g$ channel can be done similarly. Utilizing Mellin transform and inverse Mellin transform, we resum the collinear logarithms associated with parton distribution functions (PDFs) and CJFs separately. For PDFs and CJFs, we write
\begin{align}
\int_0^1 \dd x x^{N-1} \int_{x}^1 \frac{\dd\xi}{\xi} q\left(\frac{x}{\xi}\right) {\cal P}_{qq} (\xi)= {\cal P}_{qq}(N) q(N),\\
\int_0^1 \dd z z^{N-1} \int_z^1 \frac{\dd\xi}{\xi} \mathcal{J}_q\left(\frac{z}{\xi}\right) {\cal P}_{qq} (\xi) = {\cal P}_{qq}(N) \mathcal{J}_q(N),
\end{align}
where $q(N) \equiv \int_0^1 \dd x x^{N-1} q(x)$ and $\mathcal{P}_{qq}(N) \equiv \int_0^1\dd\xi\xi^{N-1}\mathcal{P}_{qq}(\xi)$. The resummed quark distributions and CJFs in the Mellin space can be cast into
\begin{align}
& q^{\rm res} (N) = q (N) \exp \left[ -\frac{\alpha_s}{\pi} C_F \ln\frac{\Lambda^2}{\mu^2} (\gamma_E-\frac{3}{4} +\ln N) \right], \\
& \mathcal{J}_q^{\rm res} (N) =  \mathcal{J}_q (N)\exp \left[ -\frac{\alpha_s}{\pi} C_F \ln\frac{\Lambda^2}{\mu_J^2} (\gamma_E-\frac{3}{4} +\ln N) \right].
\end{align} 
In arriving at the above expressions, we have taken the large $N$ limit and exponentiated the collinear logarithms. Next, we perform the inverse Mellin transform with respect to $q^{\rm res} (N)$ and obtain    
\begin{eqnarray}
q^{\rm res} (x,\Lambda ^2, \mu ^2   ) & = &  \int  _{  \mathcal C }  \frac{\dd N}{ 2\pi i}  x ^ {-N}  q ( N) \exp \left[ -\frac{\alpha_s}{\pi} C_F \ln\frac{\Lambda^2}{\mu^2} (\gamma_E-\frac{3}{4} +\ln N) \right] \nonumber \\
& =& 
\exp \left[ -\frac{\alpha_s}{\pi} C_F \ln\frac{\Lambda^2}{\mu^2} (\gamma_E-\frac{3}{4}) \right] \int ^1_0 \frac{\dd x'  }{x'} q(x', \mu ^2) \int _{\mathcal C}   \frac{\dd N}{2\pi i} \left(\frac{x'}{x}\right)^{N}  \exp \left[ -\frac{\alpha _s}{\pi} C_F \ln\frac{\Lambda^2}{\mu^2} \ln N \right].
\end{eqnarray}
After integrating over $N$, we arrive at the resummed expression of quark distribution for the $q\to q$ channel. It is given by
\begin{align}
&   q^{\rm res} (x,\Lambda ^2, \mu ^2   ) 
  =  
   \frac{  e^{ - \gamma^q_{\Lambda, \mu} (\gamma_E-\frac{3}{4}) }  }{\Gamma( \gamma^q_{\Lambda, \mu}   )   }   
 \int ^1_{x} \frac{\dd x'  }{x'} q(x', \mu ^2)    
  \left( \ln \frac{x' }{x}  \right)  ^{ \gamma^q_{\Lambda, \mu} -1  },
&& \text{Re}\left[  \gamma^q_{\Lambda, \mu}  \right]  >0,
\end{align}
where $\gamma^q_{\Lambda, \mu} = \frac{\alpha _s}{\pi} C_F \ln\frac{\Lambda^2}{\mu^2}$. Similarly, with the same strategy, for the quark jet function, we have
\begin{align}
&  \mathcal{J}_q^{\rm res} (z,\Lambda ^2, \mu_J^2   ) 
  =  
   \frac{  e^{ - \gamma^q_{\Lambda, \mu_J} (\gamma_E-\frac{3}{4}) }  }{\Gamma( \gamma^q_{\Lambda, \mu_J}   )   }   
 \int ^1_{z} \frac{\dd z'  }{z'}\mathcal{J}_q (z', \mu_J ^2)    
  \left( \ln \frac{z'}{z}  \right)  ^{ \gamma^q_{\Lambda, \mu_J} -1  },
&& \text{Re}\left[  \gamma^q_{\Lambda, \mu_J}  \right]  >0.
\end{align}
Note that the above anomalous dimensions $\gamma^q_{\Lambda, \mu}$ and $\gamma^q_{\Lambda, \mu_J}$ are formulated in the fixed coupling case. In the running coupling scenario, they read as follows
\begin{align}
\gamma^q_{\Lambda, \mu} = C_F \int ^{\Lambda^2} _{\mu^2} \frac{\dd {\mu'}^2}{{\mu'}^2} \frac{\alpha_s ( {\mu'}^2 )}{\pi},\\
\gamma^q_{\Lambda, \mu_J} = C_F \int ^{\Lambda^2} _{\mu_J^2} \frac{\dd {\mu'}^2}{{\mu'}^2} \frac{\alpha_s ( {\mu'}^2 )}{\pi}.
\end{align}

For the $g\to g$ channel, one just need to replace the splitting function $\mathcal{P}_{qq}(\xi)$ to $\mathcal{P}_{gg}(\xi)$ and color factor $C_F$ to $N_c$. The Mellin transform of ${\cal P}_{gg} (\xi)$ is given by ${\cal P}_{gg} (N) \equiv \int_0^1 \dd\xi \xi^{N-1} {\cal P}_{gg} (\xi)$.
\if{false}
\begin{align}
{\cal P}_{gg} (N) \equiv \int_0^1 \dd\xi \xi^{N-1} {\cal P}_{gg} (\xi) = 
- 2 \left[ \gamma_E + \psi(N) - \beta_0 - \frac{2}{N(N^2-1)} + \frac{1}{N+2} \right]
= - 2 \left[ \gamma_E - \beta_0 + \ln N \right] + {\cal O}(\frac{1}{N}),
\end{align}
where we have taken large-$N$ limit in the last step. 
\fi
Therefore, for the gluon case, we obtain the following expressions for the resummed gluon PDFs and CJFs
\begin{align}
g^{\rm res} (x, \Lambda^2, \mu^2) &= 
\frac{e^{-\gamma^g_{\Lambda,\mu} (\gamma_E-\beta_0)}}{\Gamma(\gamma^g_{\Lambda,\mu})} \int_{x}^1 \frac{\dd x'}{x'} g(x',\mu^2)
\left(\ln \frac{x'}{x} \right)^{\gamma^g_{\Lambda,\mu} - 1},
&& \text{Re}\left[  \gamma^g_{\Lambda, \mu}  \right]  >0, 
\\
\mathcal{J}_g^{\rm res} (z, \Lambda^2, \mu_J^2) &= 
\frac{e^{-\gamma^g_{\Lambda,\mu_J} (\gamma_E-\beta_0)}}{\Gamma(\gamma^g_{\Lambda,\mu_J})} \int_z^1 \frac{\dd z'}{z'} \mathcal{J}_g (z', \mu_J^2)
\left( \ln \frac{z'}{z} \right)^{\gamma^g_{\Lambda,\mu_J} - 1},
&& \text{Re}\left[  \gamma^g_{\Lambda, \mu_J}  \right]  >0, 
\end{align}
where the gluon channel anomalous dimensions read
\begin{align}
\gamma^g_{\Lambda,\mu}=N_c \int_{\mu^2}^{\Lambda^2} \frac{\dd \mu'^2}{\mu'^2} \frac{\alpha_s (\mu'^2)}{\pi},\\
\gamma^g_{\Lambda,\mu_J}=N_c \int_{\mu_J^2}^{\Lambda^2} \frac{\dd \mu'^2}{\mu'^2} \frac{\alpha_s (\mu'^2)}{\pi}.
\end{align}
One should note that the above resummed results are only applicable in the region $\text{Re}\left[\gamma_{\Lambda,\mu/\mu_J}^{q/g}\right]>0$. Therefore, we need to do the analytic continuation to extend to the whole space. Inspired by the analytic continuation of the gamma function, the resummed PDFs and CJFs can be rewritten with the help of the star distribution~\cite{Becher:2006nr,Becher:2006mr,Bosch:2004th}. With the star distribution, they are given by
\begin{align}
   q^{\rm res} (x,\Lambda ^2, \mu ^2) 
&  =  
   \frac{e^{ - \gamma^q_{\Lambda, \mu} (\gamma_E-\frac{3}{4})}}{\Gamma(\gamma^q_{\Lambda, \mu})}   
 \int ^1_{x} \frac{\dd x'}{x'} q(x', \mu ^2)    
  \left( \ln \frac{x' }{x}  \right)_*^{ \gamma^q_{\Lambda, \mu} -1  },
\\
 g^{\rm res} (x, \Lambda^2, \mu^2) &= 
\frac{e^{-\gamma^g_{\Lambda,\mu} (\gamma_E-\beta_0)}}{\Gamma(\gamma^g_{\Lambda,\mu})} \int_{x}^1 \frac{\dd x'}{x'} g(x',\mu^2)
\left(\ln \frac{x'}{x} \right)_*^{\gamma^g_{\Lambda,\mu} - 1},
\\
 \mathcal{J}_q^{\rm res} (z,\Lambda ^2, \mu_J^2) 
&  =  
   \frac{  e^{ - \gamma^q_{\Lambda, \mu_J} (\gamma_E-\frac{3}{4}) }  }{\Gamma( \gamma^q_{\Lambda, \mu_J}   )   }   
 \int ^1_{z} \frac{\dd z'  }{z'}\mathcal{J}_q (z', \mu_J ^2)    
  \left( \ln \frac{z'}{z}  \right)_*^{ \gamma^q_{\Lambda, \mu_J} -1  },
\\
 \mathcal{J}_g^{\rm res} (z, \Lambda^2, \mu_J^2) &= 
\frac{e^{-\gamma^g_{\Lambda,\mu_J} (\gamma_E-\beta_0)}}{\Gamma(\gamma^g_{\Lambda,\mu_J})} \int_z^1 \frac{\dd z'}{z'} \mathcal{J}_g(z', \mu_J^2)
\left( \ln \frac{z'}{z} \right)_*^{\gamma^g_{\Lambda,\mu_J} - 1}, 
\end{align}
where the detailed prescription of the star distribution can also be found in Sec. III3 in the supplemental material of Ref.~\cite{Shi:2021hwx}. 

\subsection{Resummation of the soft logarithms}
\label{soft}

The resummation procedures for both single and double Sudakov logarithms are almost the same as those in the hadron production case, except for the final state radiation. As we have discussed at the beginning of this section, the counting rule for the double logarithmic contribution is different between hadron and jet production.
We have only initial state contribution for the jet production. Therefore, we identify the following Sudakov logarithms for the $q\to q$ channel and similarly for the $g\to g$ channel
\begin{align}
-\frac{\alpha_s}{2\pi} \frac{C_F}{2} \ln^2 \frac{\qperp^2}{\Lambda^2} &+ \frac{\alpha_s}{2\pi} \frac{3}{2} C_F \ln \frac{\qperp^2}{\Lambda^2},\\
-\frac{\alpha_s}{2\pi} \frac{N_c}{2} \ln^2 \frac{\qperp^2}{\Lambda^2} &+ \frac{\alpha_s}{2\pi} 2\beta_0N_C \ln \frac{\qperp^2}{\Lambda^2}.
\end{align}
In addition, we can extend the above expression by considering the running of the coupling as follows
\begin{align}
S_{\rm Sud}^{qq} 
=& \frac{C_F}{2} \int_{\Lambda^2}^{\qperp^2} \frac{\dd \mu^2}{\mu^2} \frac{\alpha_s(\mu^2)}{\pi} \ln \frac{\qperp^2}{\mu^2}
- \frac{3}{2} C_F \int_{\Lambda^2}^{\qperp^2} \frac{\dd \mu^2}{\mu^2} \frac{\alpha_s(\mu^2)}{2\pi},\label{eq:rs1}\\
 S_{\rm Sud}^{gg}
=& \frac{N_c}{2} \int_{\Lambda^2}^{\qperp^2} \frac{\dd \mu^2}{\mu^2} \frac{\alpha_s(\mu^2)}{\pi} \ln \frac{\qperp^2}{\mu^2}
- 2\beta_0N_c\int_{\Lambda^2}^{\qperp^2} \frac{\dd \mu^2}{\mu^2} \frac{\alpha_s(\mu^2)}{2\pi}.\label{eq:rs2}
\end{align}
The resummation of the soft logarithms can be achieved by the exponentiating the above Sudakov factor. Following Ref.~\cite{Shi:2021hwx},
to treat the NLO correction, the Sudakov matching term is defined as follows
\begin{align}
\frac{\dd\sigma_{\rm Sud~matching}}{\dd \eta\dd^2P_J}
= & 
\Sperp  \int_\tau^1 \frac{\dd z}{z ^2} x q(x,\mu^2)\mathcal{J}_q(z,\mu_J^2) F(\qperp) \left\{S_{\rm Sud}^{qq} - \left[ C_F\frac{\alpha_s}{2\pi}\left(\frac{1}{2}\ln^2 \frac{\qperp^2}{\Lambda^2}-\frac{3}{2}\ln \frac{\qperp^2}{\Lambda^2}\right)\right]\right\}
\nonumber\\
+ & 
\Sperp  \int_\tau^1 \frac{\dd z}{z ^2} x g(x,\mu^2)\mathcal{J}_g(z,\mu_J^2) \int \dd^2 \qa F(\qa) F(\qperp-\qa) 
\nonumber\\
& \phantom{xx} 
\times
\left\{ S_{\rm Sud}^{gg} - \left[ N_c\frac{\alpha_s}{2\pi}\left(\frac{1}{2}\ln^2 \frac{\qperp^2}{\Lambda^2}-2\beta_0\ln \frac{\qperp^2}{\Lambda^2}\right) \right] \right\}. \label{eq:sudakov-mismatch}
\end{align}
 
\subsection{The full resummation results}

By using the DGLAP evolution equations, we resum the initial state collinear logarithms in $\sigma_{qq}^1$, $\sigma_{gg}^{1}$, $\sigma_{gq}^1$ and $\sigma_{qg}^{1}$ by setting the factorization scale $\mu^2$ to $\Lambda^2$ in Eq.~\eqref{PDFs}, then resum the final state collinear logarithms associated with the jet in $\sigma_{qq}^9$, $\sigma_{gg}^{11}$, $\sigma_{gq}^4$ and $\sigma_{qg}^4$ by replacing $P_J^2 R^2$ by $\Lambda^2$ in Eq.~\eqref{CJFs}.

As discussed previously, only initial state radiations contain the genuine Sudakov logarithms, thus we extract the corresponding initial state Sudakov logarithms in $\sigma_{qq}^2$, $\frac{1}{2}\sigma_{qq}^{5a}$, $\sigma_{gg}^{2}$ and $\frac{1}{2}\sigma_{gg}^{7a}$ and resum them by exponentiating the Sudakov factor in Eq.~(\ref{eq:rs1}) and Eq.~(\ref{eq:rs2}). 
In contrast, the remaining logarithms $\frac{1}{2}\sigma_{qq}^{5a}$ and $\frac{1}{2}\sigma_{gg}^{7a}$ are associated with the final state gluon radiation from the jet. We treat them as normal NLO corrections along with other NLO terms in the hard factor. In the NLO hard matching term, there are nine terms  ($\sigma_{qq}^3$, $\sigma_{qq}^4$,$\frac{1}{2}\sigma_{qq}^{5a}$, $\sigma_{qq}^{5b}$, $\sigma_{qq}^6$, $\sigma_{qq}^7$, $\sigma_{qq}^8$, $\sigma_{qq}^{10}$, $\sigma_{qq}^{11}$) in the $q\to q$ channel after removing the large logarithms ($\sigma_{qq}^1$, $\sigma_{qq}^2$, $\frac{1}{2}\sigma_{qq}^{5a}$ ). In the $g\to g$ channel, there are also nine terms ($\sigma_{gg}^{3}$, $\sigma_{gg}^{4}$, $\sigma_{gg}^{5}$, $\sigma_{gg}^{6}$, $\frac{1}{2}\sigma_{gg}^{7a}$, $\sigma_{gg}^{7b}$, $\sigma_{gg}^{8}$, $\sigma_{gg}^{9}$, $\sigma_{gg}^{10}$) left after removing ($\sigma_{gg}^{1}$, $\sigma_{gg}^{2}$ and $\frac{1}{2}\sigma_{gg}^{7a}$ ). Besides, there are three terms ($\sigma_{gq}^2$,  $\sigma_{gq}^3$, $\sigma_{gq}^5$) and  ($\sigma_{qg}^2$,  $\sigma_{qg}^3$ and  $\sigma_{qg}^5$) in the $q\to g$ and $g\to q$ channels, respectively. We put all these remaining small terms in the NLO hard factor which is referred to as the ``NLO~matching'' contribution.

The fully resummed result can be derived by collecting Eq.~\eqref{PDFs}, Eq.~\eqref{CJFs}, Eq.~\eqref{eq:rs1} and Eq.~\eqref{eq:rs2} together, the detailed derivation can be found in Ref.~\cite{Shi:2021hwx} and we present the final ``Resummed'' result here
\begin{align}
\frac{\dd\sigma_{\rm resummed}}{\dd \eta\dd^2P_J}
= & 
\Sperp  \int_\tau^1 \frac{\dd z}{z ^2} x q(x,\Lambda^2)\mathcal{J}_q(z,\Lambda^2) F(\qperp) e^{-S_{\rm Sud}^{qq}}
\nonumber\\
+ &
\Sperp  \int_\tau^1 \frac{\dd z}{z ^2} x g(x,\Lambda^2)\mathcal{J}_g(z,\Lambda^2) \int \dd^2 \qa F(\qa) F(\qperp-\qa) e^{-S_{\rm Sud}^{gg}}. \label{eq:resed}
\end{align}
At the end of the day, the resummation improved NLO cross-section is then given by
\begin{align}
\frac{\dd\sigma}{\dd \eta\dd^2P_J} =
\frac{\dd\sigma_{\rm resummed}}{\dd \eta\dd^2P_J} + \frac{\dd\sigma_{\rm NLO~matching}}{\dd \eta\dd^2P_J}
+ \frac{\dd\sigma_{\rm Sud~matching}}{\dd \eta\dd^2P_J},
\label{eq:final-prescription}
\end{align}
where 
\begin{align}
\frac{\dd\sigma_{\rm NLO~matching}}{\dd \eta\dd^2P_J}
= &
\sum_{i=3,4,6,7,8,10,11} \frac{\dd\sigma_{qq}^i}{\dd \eta\dd^2P_J} + \frac{1}{2}\frac{\dd\sigma_{qq}^{5a}}{\dd \eta\dd^2P_J}+  \frac{\dd\sigma_{qq}^{5b}}{\dd \eta\dd^2P_J}+ \sum_{i=2,3,5}\frac{\dd\sigma_{qg}^i}{\dd \eta\dd^2P_J} \notag\\
+ & \sum_{i=3,4,5,6,8,9, 10} \frac{\dd\sigma_{gg}^i}{\dd \eta\dd^2P_J} + \frac{1}{2}\frac{\dd\sigma_{gg}^{7a}}{\dd \eta\dd^2P_J} + \frac{\dd\sigma_{gg}^{7b}}{\dd \eta\dd^2P_J}+  \sum_{i=2,3,5}\frac{\dd\sigma_{gq}^i}{\dd \eta\dd^2P_J},
\end{align}
and ``$\rm Sud~matching$" is given by Eq.~(\ref{eq:sudakov-mismatch}).
Due to the resummation of the threshold collinear logarithms, the scale $\mu$ for PDFs (or $\mu_J$ for CJFs) in $\sigma_{\text{resummed}}$ becomes $\Lambda$. Meanwhile, the scales remain unchanged in $\sigma_{\text{NLO matching}}$. After all the resummations, we believe that all the large logarithms have been taken care of and the remaining NLO hard factors are numerically small. Therefore, the resummation improved results allow us to obtain reliable predictions for forward jet productions. 
\end{widetext}

\section{conclusion}
\label{section5}

In summary, we have systematically calculated the complete NLO cross-section for single inclusive jet production in $pA$ collisions at forward rapidity region within the small-$x$ framework. 
As shown above, the narrow jet approximation allows us to neglect the small contribution from the kinematic region where the radiated gluon is located inside the jet cone. Therefore, the calculation for the initial state radiation becomes identical to the single hadron production case.
The collinear divergences associated with the initial state gluon radiation can also be factorized into the splittings of the PDFs of the incoming nucleon. 
Thanks to the jet algorithm, complete cancellations occur for final state gluon radiations as expected. The residual contribution after the cancellation is proportional to $\ln\frac{1}{R^2}$, which is only divergent in the small cone limit($R \to 0$). It is the signature of final state collinear divergence, and corresponds to the collinear singularity for FFs in the hadron production case. By employing proper subtractions of both rapidity and collinear divergences, we obtain the NLO hard coefficients which can be numerically evaluated for future phenomenological studies. The one-loop results obtained in this study are consistent with the results in Ref.~\cite{Liu:2022ijp}. However, our resummation strategies for the collinear and Sudakov logarithms from the initial state radiations and the jet cone logarithms from the final state radiations are new. 

Furthermore, by applying the threshold resummation technique in the CGC formalism, we can improve the theoretical calculation precision by resumming threshold logarithms. In addition, the resummation of the collinear logarithms can be achieved automatically through evolving the scale $\mu$ for PDFs (or $\mu_J$ for CJFs) to the auxiliary scale $\Lambda$. The results provide another channel at the NLO level for the study of the onset of the gluon saturation phenomenon in high energy collisions. The numerical evaluation of the NLO forward jet production is underway, and it will be presented in a separate work. 

At last, the calculation presented in this paper can be extended to the NLO computation of the well-known Mueller-Navelet jet~\cite{Mueller:1986ey} process in proton-proton collisions in which two jets with a large rapidity gap are produced. The Mueller-Navelet jet offers a unique channel for us to understand the BFKL dynamics. By choosing the Coulomb gauge for this process, one can separate the gluon radiation off the upper jet from the gluon emission from the bottom one. Thus, similar techniques used in this paper can be applied to both the forward and backward rapidity regions. We will leave this study for future work.

\section*{acknowledgments}
This work is partly supported by the Natural Science Foundation of China (NSFC) under Grant Nos. 11575070 and by the university development fund of CUHK-Shenzhen under Grant No. UDF01001859. S.Y.W. is also supported by the Tanshan fellowship of Shandong Province for junior scientists.

\end{document}